\documentclass[iop]{emulateapj}

\newcommand{\tnm}{\tablenotemark}
\newcommand{\tnt}{\tablenotetext}

\bibpunct{(}{)}{;}{a}{}{,}

\slugcomment{Accepted to the ApJ}

\shorttitle{Lens Models for {\it Herschel} SMGs}
\shortauthors{Bussmann et al.}

\begin{document}

\title{Gravitational Lens Models Based on Submillimeter Array$^{\dagger}$
Imaging of {\it Herschel}$^{\dagger\dagger}$-selected Strongly Lensed
Sub-millimeter Galaxies at {${\textit z}>1.5$}}


\author{R.~S.~Bussmann\altaffilmark{1},
I.~P{\'e}rez-Fournon\altaffilmark{2,3},
S.~Amber\altaffilmark{4},
J.~Calanog\altaffilmark{5},
M.~A.~Gurwell\altaffilmark{1},
H.~Dannerbauer\altaffilmark{6},
F.~De Bernardis\altaffilmark{5},
Hai~Fu\altaffilmark{7},
A.~I.~Harris\altaffilmark{8},
M.~Krips\altaffilmark{9},
A.~Lapi\altaffilmark{10},
R.~Maiolino\altaffilmark{11,12},
A.~Omont\altaffilmark{13},
D.~Riechers\altaffilmark{14},
J.~Wardlow\altaffilmark{5},
A.~J.~Baker\altaffilmark{15},
M.~Birkinshaw\altaffilmark{16},
J.~Bock\altaffilmark{17,18},
N.~Bourne\altaffilmark{19},
D.~L.~Clements\altaffilmark{20},
A.~Cooray\altaffilmark{5,17},
G.~De Zotti\altaffilmark{21},
L.~Dunne\altaffilmark{22},
S.~Dye\altaffilmark{19},
S.~Eales\altaffilmark{23},
D.~Farrah\altaffilmark{24},
R.~Gavazzi\altaffilmark{13},
J.~Gonz\'alez~Nuevo\altaffilmark{25},
R.~Hopwood\altaffilmark{20},
E.~Ibar\altaffilmark{26},
R.~J.~Ivison\altaffilmark{27,28},
N.~Laporte\altaffilmark{2,3},
S.~Maddox\altaffilmark{22},
P.~Mart\'inez-Navajas\altaffilmark{2,3},
M.~Michalowski\altaffilmark{28},
M.~Negrello\altaffilmark{21},
S.~J.~Oliver\altaffilmark{29},
I.~G.~Roseboom\altaffilmark{29,28},
Douglas~Scott\altaffilmark{30},
S.~Serjeant\altaffilmark{4},
A.~J.~Smith\altaffilmark{29},
Matthew~Smith\altaffilmark{23},
A.~Streblyanska\altaffilmark{2,3},
E.~Valiante\altaffilmark{30},
P.~van der Werf\altaffilmark{31},
A.~Verma\altaffilmark{32},
J.~D.~Vieira\altaffilmark{17},
L.~Wang\altaffilmark{29},
D.~Wilner\altaffilmark{1}}

\altaffiltext{$\dagger$}{The Submillimeter Array is a joint project between the
Smithsonian Astrophysical Observatory and the Academia Sinica Institute of
Astronomy and Astrophysics and is funded by the Smithsonian Institution and the
Academia Sinica.}

\altaffiltext{$\dagger\dagger$}{{\it Herschel} is an ESA space observatory with
science instruments provided by European-led Principal Investigator consortia
and with important participation from NASA.}

\altaffiltext{1}{Harvard-Smithsonian Center for Astrophysics, 60 Garden Street, Cambridge, MA 02138}
\altaffiltext{2}{Instituto de Astrof{\'\i}sica de Canarias (IAC), E-38200 La Laguna, Tenerife, Spain}
\altaffiltext{3}{Departamento de Astrof{\'\i}sica, Universidad de La Laguna (ULL), E-38205 La Laguna, Tenerife, Spain}
\altaffiltext{4}{Department of Physical Sciences, The Open University, Milton Keynes MK7 6AA, UK}
\altaffiltext{5}{Dept. of Physics \& Astronomy, University of California, Irvine, CA 92697}
\altaffiltext{6}{Laboratoire AIM-Paris-Saclay, CEA/DSM/Irfu - CNRS - Universit\'e Paris Diderot, CE-Saclay, pt courrier 131, F-91191 Gif-sur-Yvette, France}
\altaffiltext{7}{Department of Physics and Astronomy, The University of Iowa, 203 Van Allen Hall, Iowa City, IA 52242}
\altaffiltext{8}{Department of Astronomy, University of Maryland, College Park, MD 20742-2421}
\altaffiltext{9}{Institut de RadioAstronomie Millim\'etrique, 300 Rue de la Piscine, Domaine Universitaire, 38406 Saint Martin d'H\`eres, France}
\altaffiltext{10}{Dip. Fisica, Univ. Tor Vergata, Via Ricerca Scientifica 1, 00133 Rome, Italy and SISSA, Via Bonomea 265, 34136 Trieste, Italy}
\altaffiltext{11}{Cavendish Laboratory, University of Cambridge, 19 J.J. Thomson Ave., Cambridge CB3 OHE, UK}
\altaffiltext{12}{Kavli Institute for Cosmology, University of Cambridge, Madingley Road, Cambridge CB3 OHA, UK}
\altaffiltext{13}{Institut d'Astrophysique de Paris, UMR 7095, CNRS, UPMC Univ. Paris 06, 98bis boulevard Arago, F-75014 Paris, France}
\altaffiltext{14}{Department of Astronomy, Space Science Building, Cornell University, Ithaca, NY, 14853-6801}
\altaffiltext{15}{Department of Physics and Astronomy, Rutgers, The State University of New Jersey, 136 Frelinghuysen Rd, Piscataway, NJ 08854}
\altaffiltext{16}{HH Wills Physics Laboratory, University of Bristol, Tyndall Avenue, Bristol BS8 1TL, U.K.}
\altaffiltext{17}{California Institute of Technology, 1200 E. California Blvd., Pasadena, CA 91125}
\altaffiltext{18}{Jet Propulsion Laboratory, 4800 Oak Grove Drive, Pasadena, CA 91109}
\altaffiltext{19}{School of Physics and Astronomy, University of Nottingham, NG7 2RD, UK}
\altaffiltext{20}{Astrophysics Group, Imperial College London, Blackett Laboratory, Prince Consort Road, London SW7 2AZ, UK}
\altaffiltext{21}{INAF - Osservatorio Astronomico di Padova, Vicolo dell'Osservatorio 5, I-35122 Padova, Italy.}
\altaffiltext{22}{Department of Physics and Astronomy, University of Canterbury, Private Bag 4800, Christchurch, 8140, New Zealand}
\altaffiltext{23}{School of Physics and Astronomy, Cardiff University, Queens Buildings, The Parade, Cardiff CF24 3AA, UK}
\altaffiltext{24}{Department of Physics, Virginia Tech, Blacksburg, VA 24061}
\altaffiltext{25}{Instituto de F\'i sica de Cantabria (CSIC-UC), Av. los Castros s/n, 39005 Santander, Spain}
\altaffiltext{26}{Instituto de Astrof\'isica. Facultad de F\'isica. Pontificia Universidad Cat\'olica de Chile. Casilla 306, Santiago 22, Chile}
\altaffiltext{27}{UK Astronomy Technology Centre, Royal Observatory, Blackford Hill, Edinburgh EH9 3HJ, UK}
\altaffiltext{28}{Institute for Astronomy, University of Edinburgh, Royal Observatory, Blackford Hill, Edinburgh EH9 3HJ, UK}
\altaffiltext{29}{Astronomy Centre, Dept. of Physics \& Astronomy, University of Sussex, Brighton BN1 9QH, UK}
\altaffiltext{30}{Department of Physics \& Astronomy, University of British Columbia, 6224 Agricultural Road, Vancouver, BC V6T~1Z1, Canada}
\altaffiltext{31}{Leiden Observatory, P.O. Box 9513, NL - 2300 RA Leiden, The Netherlands}
\altaffiltext{32}{Department of Astrophysics, Denys Wilkinson Building, University of Oxford, Keble Road, Oxford OX1 3RH, UK}


\begin{abstract}

Strong gravitational lenses are now being routinely discovered in wide-field surveys at (sub-)millimeter wavelengths.  We present Submillimeter Array (SMA) high-spatial resolution imaging and Gemini-South and Multiple Mirror Telescope optical spectroscopy of strong lens candidates discovered in the two widest extragalactic surveys conducted by the {\it Herschel Space Observatory}: the {\it Herschel}-Astrophysical Terahertz Large Area Survey (H-ATLAS) and the {\it Herschel} Multi-tiered Extragalactic Survey (HerMES).  From a sample of 30 {\it Herschel} sources with $S_{500}>100\,$mJy, 21 are strongly lensed (i.e., multiply imaged), 4 are moderately lensed (i.e., singly imaged), and the remainder require additional data to determine their lensing status.  We apply a visibility-plane lens modeling technique to the SMA data to recover information about the masses of the lenses as well as the intrinsic (i.e., unlensed) sizes ($r_{\rm half}$) and far-infrared luminosities ($L_{\rm FIR}$) of the lensed submillimeter galaxies (SMGs).  The sample of lenses comprises primarily isolated massive galaxies, but includes some groups and clusters as well.  Several of the lenses are located at $z_{\rm lens}>0.7$, a redshift regime that is inaccessible to lens searches based on Sloan Digital Sky Survey spectroscopy.  The lensed SMGs are amplified by factors that are significantly below statistical model predictions given the 500$\,\mu$m flux densities of our sample.  We speculate that this may reflect a deficiency in our understanding of the intrinsic sizes and luminosities of the brightest SMGs.  The lensed SMGs span nearly one decade in $L_{\rm FIR}$ (median $L_{\rm FIR}=7.9\times10^{12}\,$L$_\sun$) and two decades in FIR luminosity surface density (median $\Sigma_{\rm FIR}=6.0\times10^{11}\,$L$_\sun\;$kpc$^{-2}$).  The strong lenses in this sample and others identified via (sub-)mm surveys will provide a wealth of information regarding the astrophysics of galaxy formation and evolution over a wide range in redshift.

\end{abstract}



\section{Introduction} \label{sec:intro} 

Strong gravitational lensing by massive galaxies provides one of the most
striking visual confirmations of Einstein's theory of General Relativity.  In
the case of galaxy-galaxy lensing, the chance alignment of two galaxies along
the line of sight provides information about both the lens and the source that
cannot be obtained in any other way.  The angular separation of multiple images
of a lensed galaxy is typically parameterized in terms of the angular Einstein
radius (here, denoted $\theta_{\rm E}$) and provides an unambiguous measurement
of the total mass of the lens (baryonic plus non-baryonic) inside $\theta_{\rm
E}$, as long as the distances to the lens and source are known
\citep{Schneider:1992fk}.  At the same time, lensing increases the apparent
size of the background source and conserves surface brightness in the process.
A spatially unresolved measurement of the flux density from a lensed source is
therefore a factor of order $\mu$ (the magnification factor, defined in detail
in Section~\ref{sec:lensmethod}) brighter than for an unlensed source, while
spatially resolved measurements can provide a factor of $\sim \sqrt{\mu}$
higher resolution \citep{Schneider:1992fk}.

Given the benefits of studying strong lenses, it is no surprise that
significant efforts have been devoted to the assembly of large samples of them.
The earliest of these efforts focused on surveys in the radio with the Jodrell
Bank Very Large Array gravitational lens survey \citep[JVAS;][]{King:1996fk}
and the Cosmic Lens All-Sky Survey \citep[CLASS;][]{Myers:2003lr} or on {\it
Hubble Space Telescope} ({\it HST}) follow-up of known strong lenses as part of
the Center for Astrophysics Arizona Space Telescope Lens Survey
\citep[CASTLeS;][]{Munoz:1998mz}.  Together these surveys have created a sample
of $\approx 80$ strong lenses \citep{Schneider:2006ab}.  More recently, surveys
based on {\it HST} Advanced Camera for Surveys (ACS) follow-up of candidates
selected from Sloan Digital Sky Survey (SDSS) spectroscopy as part of the Sloan
Lens ACS Survey \citep[SLACS;][]{Bolton:2008wd} and the SDSS Quasar Lens Search
\citep[SQLS;][]{Inada:2012lr}, as well as candidates selected from the Canada
France Hawaii Telescope Legacy Survey as part of the Strong Lensing in the
Legacy Survey (SL2S) project \citep[e.g.,][]{Sonnenfeld:2013fj} have more than
doubled this number.  More recent upgrades associated with the SDSS-III Baryon
Oscillation Spectroscopic Survey \citep[BOSS;][]{Eisenstein:2011fr} that
promise to increase the sample size of SDSS-selected lenses by a factor of
several \citep[the BOSS Emission-Line Lens Survey, or
BELLS;][]{Brownstein:2012rt}.  Finally, a new method of finding lenses has come
to sudden prominence with the launch of the {\it Herschel Space Observatory}
\citep[{\it Herschel};][]{Pilbratt:2010fk} and the advent of the South Pole
Telescope \citep[SPT;][]{Carlstrom:2011qy} and the Atacama Cosmology Telescope
\citep{Swetz:2011qy}: wide-field surveys at submillimeter (submm) and
millimeter (mm) wavelengths.

Surveys at (sub-)mm wavelengths are ideal tools for discovering lenses, in part
because the observed-frame (sub-)mm flux density of a dusty galaxy at a given
luminosity is approximately independent of redshift for $z > 1$
\citep{Blain:1993lr} and in part because the number counts of unlensed submm
sources (SMGs) fall off very steeply at high flux densities compared to
optically selected galaxies \citep[e.g.,][]{Barger:1999rt, Coppin:2006lr,
2010A&A...518L..21O, Clements:2010fk}.  Strong lensing events are rare, so the
key requirement for identifying them is wide-area coverage.  This is now being
provided by the {\it Herschel} Astrophysical Terahertz Large Area Survey
\citep[H-ATLAS;][]{2010PASP..122..499E}, the {\it Herschel} Multi-Tiered
Extragalactic Survey
\citep[HerMES;][]{Oliver:2012lr}\footnote{http://hermes.sussex.ac.uk}, the SPT
\citep{Vieira:2010rr, Mocanu:2013fk}, and ACT
\citep{Marsden:2013lr}.  In this paper, we focus on strong lens candidates
selected from the two {\it Herschel} surveys.

Studies based on the H-ATLAS Science Demonstration Phase field (covering
14.4$\,$deg$^2$) and on HerMES (covering 94.8$\,$deg$^2$) have found that a
simple selection at 500$\,\mu$m of $S_{500} > 100\,$mJy finds
lenses with an efficiency of $70-100\%$ \citep{Negrello:2010fk,
Wardlow:2013lr}.  This single selection criterion also yields low-$z$ spiral
galaxies \citep{2005MNRAS.356..192S} and blazars with synchrotron emission
spectra in the {\it Herschel} bands \citep{2005A&A...431..893D}, but these are
easily identified and removed using shallow optical imaging (SDSS is
sufficient) and shallow radio imaging from the National Radio Astronomy
Observatory Very Large Array Sky Survey \citep[NVSS;][]{Condon:1998uq}.

A world-wide, multi-wavelength follow-up effort is now underway to study strong
lens candidates from H-ATLAS and HerMES that are selected to have $S_{500} >
100\,$mJy.  The main goals of this effort are to: (1) confirm that the
candidates are indeed strong lenses; (2) use the lenses to study massive galaxy
evolution over $0.2 < z < 1.3$; and (3) use the lensed SMGs to learn about the
nature of star-formation and galaxy evolution in luminous, dusty galaxies at
$z \gtrsim 1.5$.  

There are three key steps that must be taken to confirm the lensing hypothesis
and study a member of the {\it Herschel} sample in detail: a redshift
measurement for the lens (typically from optical spectroscopy); a distinct
redshift for the background source (typically from radio or (sub-)mm wave
spectroscopy); and spatially-resolved imaging of the source that is consistent
with strong lensing.  In this paper, we present data that mark significant
progress on two of these three fronts.  First, we give results from a large,
multi-semester program with the Submillimeter Array \citep[SMA;][]{Ho:2004lr}
that comprises over 160~hours of on-source integration time and provides
sub-arcsecond, spatially-resolved 880$\,\mu$m images of 30 {\it Herschel} lens
candidates (some of the SMA data have been published previously and are known
lenses---we highlight where this is the case and provide references in
section~\ref{sec:objectbyobject}).  Second, we provide redshifts from optical
spectroscopy of 8 lens candidates obtained with the Multiple Mirror Telescope
(MMT), Gemini-South telescope (Gemini-S), William Herschel Telescope (WHT), and
Very Large Telescope (VLT).
Redshift measurements for the background sources for nearly all of the objects
in this paper are already available from a wide range of facilities (see
section~\ref{sec:select}).

We use the SMA imaging, optical spectroscopy, and radio or (sub-)mm wave
spectroscopy to determine which of the {\it Herschel}-selected lens candidates
are indeed strongly lensed ($\mu > 2$, with multiple images of the lensed
source), which are only moderately lensed ($1 < \mu < 2$, with only a single
image of the lensed source), and which are inconclusive.  We then develop and
apply lens models in the visibility plane ---as is appropriate for
interferometers like the SMA---for all of the objects that show convincing
evidence of moderate or strong lensing.  This provides measurements of the
total (bayonic and non-baryonic) masses within $\theta_{\rm E}$ ($M_{\rm E}$),
the magnification factors of the background sources at 880$\, \mu$m ($\mu_{880}$)
and the sizes of the background sources ($r_{\rm half}$).  These are fundamental
parameters needed to understand the physics of galaxy evolution at intermediate
redshift ($0.2 < z < 1.3$) and high redshift ($z > 1.5$).

In section~\ref{sec:obs}, we describe our selection technique and present the
SMA, MMT, Gemini-S, WHT, and VLT data (highlighting which datasets are
new to this paper and which have been published previously).  We also show {\it
HST} or Keck adaptive optics (AO) imaging for comparison with the SMA imaging
and reference the future papers that will present the {\it HST} and Keck data.
Section~\ref{sec:lensmodels} contains a description of our lens modeling
methodology and a detailed description of each object in the sample.  We
discuss the implications of the lens modeling for the population of foreground
galaxies discovered by {\it Herschel} and compare to existing surveys for
lenses in section~\ref{sec:lensingresults}.  The lens model implications for
the lensed SMGs are discussed in section~\ref{sec:lensedresults}, with an
emphasis on the size-scale of star-formation in SMGs at $1.5 < z < 4.5$.
Finally, we present our conclusions in section~\ref{sec:conclusions}.

Throughout this paper, we assume $H_0=$71~km~s$^{-1}$~Mpc$^{-1}$, $\Omega_{\rm
m_0} = 0.27$, and $\Omega_{\Lambda_0} = 0.73$.

\section{Data}\label{sec:obs}

In this section, we describe how candidate strongly lensed SMGs are selected
from wide-field {\it Herschel} surveys and present SMA high-spatial resolution
imaging of the dust continuum emission from these candidate lensed SMGs as well
as optical spectroscopy obtained with the MMT, Gemini-S, and WHT.  We also
highlight ancillary optical and near-IR imaging that is used to determine the
position of the lensing galaxy or galaxies and reference the papers that fully
present and analyze those data.

\subsection{Selection of Candidate Lensed SMGs}\label{sec:select}

The first suggestion that wide-field surveys (i.e., covering $\gtrsim
100\;$deg$^2$) at submm or mm wavelengths would efficiently identify strongly
lensed galaxies was made nearly two decades ago \citep{1996MNRAS.283.1340B},
but it is only in the past few years, with the advent of {\it Herschel} and the
SPT, that such surveys have reached the requisite survey area and sensitivity
to discover them in large numbers.  We select candidate strongly lensed
galaxies from the two widest {\it Herschel} extra-galactic surveys: H-ATLAS and
HerMES.  The total area considered for the candidate selection is $\approx300
\,$deg$^2$ in H-ATLAS (comprising the full equatorial fields and $\approx75$\%
of the northern galactic pole field) and 94.8$\,$deg$^2$ in nine independent
fields in HerMES \citep[for details of the fields, see][]{Oliver:2012lr}.  The
total area surveyed by {\it Herschel} as part of H-ATLAS and HerMES is $\sim
1000\,$deg$^2$ (i.e., roughly a three-fold increase over the area considered
for this paper).

An important aspect of candidate lens selection is source extraction and
photometry for the {\it Herschel} Spectral and Photometric Imaging REceiver
\citep[SPIRE;][]{2010A&A...518L...3G} data.  We summarize the relevant aspects
of the methodology here and provide references where appropriate.  

In the HerMES fields, source detection is achieved by applying the {\sc
StarFinder} code \citep{Diolaiti:2000qy} to the 250$\, \mu$m images.
Photometry is then computed using the HerMES XID pipeline
\citep{Roseboom:2010lk}, which allocates flux density based on the 250$\, \mu$m
position priors obtained with {\sc StarFinder}.  

In the H-ATLAS fields, sources are identified and flux densities are measured
using the Multi-band Algorithm for source eXtraction (MADX; Maddox et al., in
prep.). MADX first subtracts a smooth background, and then filters with the
point spread function (PSF) appropriate for each band. Next, $>2.5\sigma$ peaks
are identified in the 250$\,\mu$m map, and `first-pass' flux density estimates
are obtained from the pixel values at these positions in each band. Sub-pixel
positions are estimated by fitting to the 250$\,\mu$m peaks, and more accurate
flux-densities are estimated using bi-cubic interpolation to the accurate
250$\,\mu$m position. In each band, the sources are sorted in order of
decreasing flux density using the first-pass pixel values, and a scaled PSF is
subtracted from the map before estimation of flux densities for any fainter
sources. This step prevents faint source flux densities from being
overestimated when they lie near brighter sources.  Finally candidate sources
are retained in the catalog if their flux densities are more than 5$\sigma$ in
any of the three bands. The 5$\sigma$ flux density limits in H-ATLAS, including
confusion noise \citep[typically $\approx 6\,$mJy in all three SPIRE
bands][]{Nguyen:2010fk}, are 32~mJy at 250$\,\mu$m, 36~mJy at 350$\,\mu$m, and
45~mJy at 500$\,\mu$m \citep{2011MNRAS.415..911P, 2011MNRAS.415.2336R}.  

Although the HerMES and H-ATLAS teams use different methods to extract
photometry, the sources that are the subject of this paper are all high S/N,
point sources as seen by {\it Herschel}.  In this regime, we expect that the
different methods should provide consistent flux density measurements.

The essence of the candidate lens selection technique is to identify objects
that are bright at 500$\, \mu$m.  A complete description of the selection
technique for the HerMES lens candidates is given by \citet{Wardlow:2013lr}.
This paper includes the objects tabulated in \citet{Wardlow:2013lr} as well as
objects identified in H-ATLAS.  We select objects that satisfy $S_{500} >
100~$mJy, a regime that has been shown from previous studies of smaller areas
to have relatively little contamination from unlensed SMGs
\citep{Negrello:2010fk, Wardlow:2013lr}.  The primary contaminant is local
Universe galaxies ($z < 0.1$).  These galaxies are spatially resolved in SDSS
imaging and therefore trivial to remove.  There is also a small contamination
from blazars, which are non-thermal emitters and are easily removed using data
from the NVSS or the Very Large Array Faint Images of the Radio Sky at
Twenty-Centimeters survey \citep[FIRST;][]{Becker:1995fj}.

A total of 13 objects in HerMES satisfy $S_{500} > 100$~mJy and are not local
galaxies or blazars \citep{Wardlow:2013lr}.  In the H-ATLAS fields, there are
91 objects that satisfy these criteria.  Considering the combination of the two
surveys, this results in a surface density on the sky of $\approx
0.26$~deg$^{-2}$.  This value lies between the values of 0.32~deg$^{-2}$ from
\citet{Negrello:2010fk} and $0.14 \pm 0.04$~deg$^{-2}$ from
\citet{Wardlow:2013lr}, as expected since it represents a combination and
extension of these previous efforts.  Cosmic variance likely explains the
difference in the surface densities of lenses between HerMES and H-ATLAS.  A
detailed calculation of this effect requires taking into account the cosmic
variance of both the lensed SMGs and the lenses themselves and is beyond the
scope of this paper.

Efforts are on-going to obtain a complete database of follow-up observations
for this sample of 104 candidate lensed SMGs.  The present paper focuses on a
subset of 30 candidates with superb existing follow-up observations (hereafter,
we refer to this as the ``SMA subsample'').  These targets were initially
selected on the basis of strong 1.2$\,$mm detections from the Max Planck
Millimeter Bolometer (MAMBO) array \citep{Kreysa:1998uq} at the Institut de
Radioastronomie Millim\'etrique (IRAM) 30$\,$m telescope (Dannerbauer et al. in
prep.).  Subsequent follow-up efforts have now provided high-spatial resolution
880$\, \mu$m imaging with the SMA, spectroscopic redshifts of the lensed SMGs
obtained with GBT, CSO, CARMA, PdBI, and {\it Herschel} \citep[][Riechers et
al., in prep.; Krips et al., in prep., George et al., in prep.]{Cox:2011fk,
Harris:2012fr, Lupu:2012ly}, and spectroscopic redshifts to the lenses obtained
with the MMT, Gemini-S, or WHT.  In addition, Keck-II Near InfraRed Camera 2
(NIRC2) laser guide star adaptive optics (LGSAO) imaging has been obtained for
nearly half of the candidate lensed SMG sample \citep[][Calanog et al., in
prep.]{Wardlow:2013lr}.  These datasets provide the information needed to
confirm the lensing hypothesis and begin analysis of the source and lens
properties.  Table~\ref{tab:position} provides basic positional data for the
SMA subsample, including the International Astronomical Union names, short
names to aid comparison with previous publications, positions measured from
the SMA 880$\, \mu$m image (see section~\ref{sec:smaobs}), and redshift
measurements for the lens(es) (see section~\ref{sec:mmtobs},
section~\ref{sec:geminiobs}, section~\ref{sec:whtobs},
section~\ref{sec:vltobs}, and references in the
Table) and background sources (references given in the Table), where available.

\begin{deluxetable*}{llccccccc}[!tbp]
\tabletypesize{\scriptsize} 
\tablecolumns{9}
\tablewidth{0in}
\tablecaption{Positions and redshifts of SMA candidate strong gravitational lens sample.
Definition of lens grades: A = Obvious strong lensing morphology in SMA map and
distinct lens and source redshifts; B = Obvious strong lensing morphology in
SMA map, but only a single redshift measurement (either lens or source); C =
Evidence for moderate lensing from SMA map and distinct lens and source
redshifts; X = SMA imaging and spectroscopic redshifts do not provide
conclusive evidence of lensing.\\
Reference key: G05 = \citet{Gladders:2005qy}; B06 = \citet{Borys:2006lr}; N10 =
\citet{Negrello:2010fk}; S11 = \citet{2011ApJ...733...29S}; F11 = \citet{2011ApJ...726L..22F}; C11 =
\citet{Cox:2011fk}; H12 = \citet{Harris:2012fr}; B12 = \citet{Bussmann:2012lr};
L12 = \citet{Lupu:2012ly}; W13 = \citet{Wardlow:2013lr}; I13 =
\citet{Ivison:2013fk}; G13 = George et al. (in prep.); R13 = Riechers et al.
(in prep.); K13 = Krips et al. (in prep.); L13 = Lupu et al. (in prep.); H13 = Harris et al. (in prep.).}
\tablehead{
\colhead{} & 
\colhead{} & 
\colhead{RA$_{880}$} &
\colhead{Dec$_{880}$} &
\colhead{} &
\colhead{} &
\colhead{} &
\colhead{} &
\colhead{Lens}
\\
\colhead{IAU name} & 
\colhead{Short name} & 
\colhead{(J2000)} &
\colhead{(J2000)} &
\colhead{$z_{\rm lens}$} &
\colhead{Ref.} &
\colhead{$z_{\rm source}$} &
\colhead{Ref.} &
\colhead{Grade}
}
\startdata
1HerMES S250 J021830.5$-$053124  & HXMM02     & 02:18:30.679 & $-$05:31:31.60 & $1.35  \pm 0.01$  & W13        & $3.39\pm0.01$      & R13 & A \\
1HerMES S250 J022016.5$-$060143  & HXMM01     & 02:20:16.603 & $-$06:01:43.20 & $0.654 ~ 0.502$\tnm{a} & W13  & $2.307\pm0.001$    & F13 & C \\
H-ATLAS J083051.0$+$013224       & G09v1.97   & 08:30:51.156 & $+$01:32:24.35 & $0.626 ~ 1.002$\tnm{a} & New  & $3.634 \pm0.001$   & R13 & A \\
H-ATLAS J084933.4$+$021443       & G09v1.124  & 08:49:33.362 & $+$02:14:42.30 & $0.3478\pm0.0001$ & I13        & $2.410 \pm0.003$   & H12 & C \\
H-ATLAS J085358.9$+$015537       & G09v1.40   & 08:53:58.862 & $+$01:55:37.70 &   ---             & ---        & $2.0894\pm0.0009$  & L13 & B \\
H-ATLAS J090302.9$-$014127       & SDP17      & 09:03:03.031 & $-$01:41:27.11 & $0.9435\pm0.0009$ & N10        & $2.3051\pm0.0002$  & L12 & A \\
H-ATLAS J090311.6$+$003906       & SDP81      & 09:03:11.568 & $+$00:39:06.43 & $0.2999\pm0.0002$ & SDSS       & $3.042 \pm0.001$   & F11 & A \\
H-ATLAS J090740.0$-$004200       & SDP9       & 09:07:40.022 & $-$00:41:59.80 & $0.6129\pm0.0005$ & New        & $1.577 \pm0.008$   & L12 & A \\
H-ATLAS J091043.1$-$000321       & SDP11      & 09:10:43.061 & $-$00:03:22.76 & $0.7932\pm0.0012$ & N10        & $1.786 \pm0.005$   & L12 & A \\
H-ATLAS J091305.0$-$005343       & SDP130     & 09:13:05.107 & $-$00:53:43.05 & $0.220\pm0.002$   & N10        & $2.6256\pm0.0005$  & F11 & A \\
H-ATLAS J091840.8$+$023047       & G09v1.326  & 09:18:40.927 & $+$02:30:45.90 &   ---             & ---        & $2.5811\pm0.0012$  & H12 & X \\
1HerMES S250 J103826.6$+$581542  & HLock04    & 10:38:26.611 & $+$58:15:42.47 & $0.61  \pm0.02 $  & W13        &  ---               & --- & B \\
1HerMES S250 J105712.2$+$565457  & HLock03    & 10:57:12.262 & $+$56:54:58.70 &   ---             & W13        & $2.771\pm0.001$    & R13 & X \\
1HerMES S250 J105750.9$+$573026  & HLock01    & 10:57:51.022 & $+$57:30:26.80 & $0.60  \pm0.04 $  & W13        & $2.957\pm0.001$    & S11 & A \\
H-ATLAS J113526.3$-$014605       & G12v2.43   & 11:35:26.273 & $-$01:46:06.55 &   ---             & ---        & $3.1276\pm0.0005$  & H12 & X \\
H-ATLAS J114637.9$-$001132       & G12v2.30   & 11:46:37.980 & $-$00:11:31.80 & $1.2247\pm0.0001$ & New        & $3.2592\pm0.0010$  & H12 & A \\
H-ATLAS J125135.4$+$261457       & NCv1.268   & 12:51:35.412 & $+$26:14:58.63 & ---               & ---        & $3.675 \pm0.001$   & K13 & B \\
H-ATLAS J125632.7$+$233625       & NCv1.143   & 12:56:32.544 & $+$23:36:27.63 & $0.2551\pm0.0001$ & New        & $3.565 \pm0.001$   & R13 & A \\
H-ATLAS J132427.0$+$284452       & NBv1.43    & 13:24:27.206 & $+$28:44:49.40 & $0.997 \pm0.017 $ & G05        & $1.676 \pm0.001$   & G13 & C \\
H-ATLAS J132630.1$+$334410       & NAv1.195   & 13:26:30.216 & $+$33:44:07.60 & $0.7856\pm0.0003$ & New        & $2.951 \pm0.001$   & H13 & A \\
H-ATLAS J132859.3$+$292317       & NAv1.177   & 13:28:59.246 & $+$29:23:26.13 &   ---             & ---        & $2.778 \pm001$     & K13 & X \\
H-ATLAS J133008.4$+$245900       & NBv1.78    & 13:30:08.520 & $+$24:58:59.17 & $0.4276\pm0.0003$ & New        & $3.1112\pm0.0001$  & R13 & A \\
H-ATLAS J133649.9$+$291801       & NAv1.144   & 13:36:49.985 & $+$29:17:59.77 &   ---             & ---        & $2.2024\pm0.0002$  & H12 & B \\
H-ATLAS J134429.4$+$303036       & NAv1.56    & 13:44:29.518 & $+$30:30:34.05 & $0.6721\pm0.0004$ & New        & $2.3010\pm0.0009$  & H12 & A \\
H-ATLAS J141351.9$-$000026       & G15v2.235  & 14:13:52.092 & $-$00:00:24.43 & $0.5470\pm0.0003$ & New        & $2.4782\pm0.0005$  & H12 & C \\
H-ATLAS J142413.9$+$022303       & G15v2.779  & 14:24:13.975 & $+$02:23:03.60 & $0.595\pm0.005$   & B12        & $4.243 \pm0.001$   & C11 & A \\
1HerMES S250 J142823.9$+$352619  & HBootes03  & 14:28:24.074 & $+$35:26:19.35 & $1.034\pm0.001$   & B06        & $1.325  \pm0.001$  & B06 & C \\
1HerMES S250 J142825.5$+$345547  & HBootes02  & 14:28:25.476 & $+$34:55:47.10 & $0.414\pm0.001$   & W13        & $2.804  \pm0.001$  & R13 & A \\
1HerMES S250 J143330.8$+$345439  & HBootes01  & 14:33:30.826 & $+$34:54:39.75 & $0.59  \pm0.08 $  & W13        & $3.274 \pm 0.001$  & R13 & A \\
H-ATLAS J144556.1$-$004853       & G15v2.481  & 14:45:56.297 & $-$00:48:51.70 &   ---             & ---        & ---                & --- & X \\
\enddata
\label{tab:position}
\tnt{a}{Multiple lens redshifts have been measured for these targets.  The redshift uncertainty is 0.001 in all cases.}
%
\end{deluxetable*}

\begin{figure}[!tbp] 
\includegraphics[width=\linewidth]{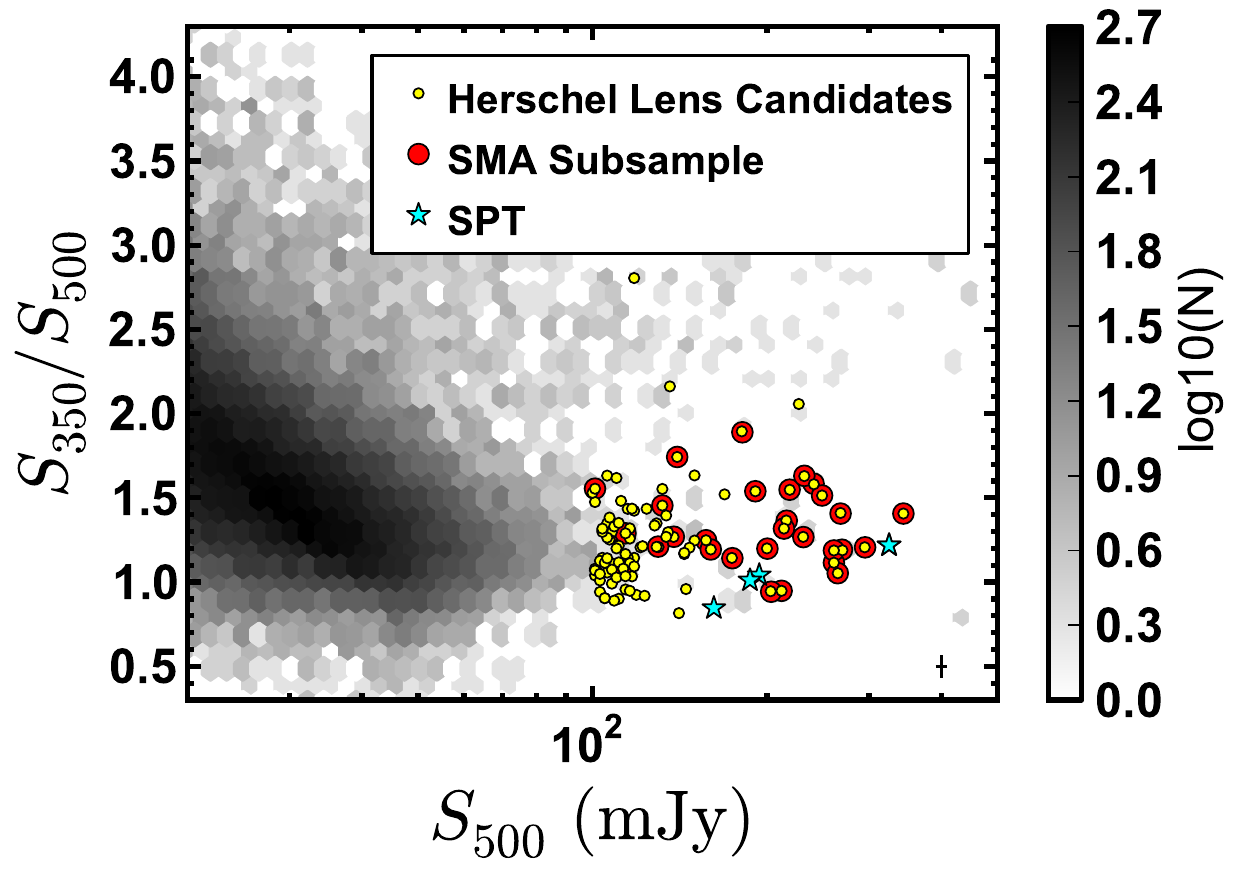}

\caption{ {\it Herschel}/SPIRE photometry of all galaxies in the H-ATLAS
phase~I catalog with S/N$ > 3$ at 250$\,\mu$m, 350$\,\mu$m, and 500$\,\mu$m
(grayscale).  The sample of H-ATLAS and HerMES sources that satisfy the
selection criteria used to select lens candidates are overplotted with yellow
filled squares.  The targets presented in this paper are represented by red
filled circles (``SMA subsample''), and a comparison sample of lensed SMGs
discovered by the SPT that have published lens models are represented by cyan
stars.  Representative error bars are shown in the lower right corner.  The SMA
subsample is biased towards higher 500$\,\mu$m flux densities but has similar
$S_{350}/S_{500}$ and $S_{250}/S_{350}$ colors (not shown).} \label{fig:sample}

\end{figure}

Figure~\ref{fig:sample} shows the $S_{350}/S_{500}$ SPIRE colors as a function
of $S_{500}$ flux density for all galaxies in the H-ATLAS phase~I catalog with
S/N$ > 3$ in all SPIRE bands (grayscale, logarithmic scaling), the full sample
of 104 candidate {\it Herschel} lensed SMGs in HerMES and H-ATLAS (cyan
squares), the SMA subsample with superb follow-up data that is the focus of
this paper, and a sample of objects selected from the SPT survey with published
lens models \citep{Hezaveh:2013fk}.  The SMA subsample is biased to higher
$S_{500}$ values than the full sample.  We therefore expect the lensing rate to
be higher than in the full sample.  The SPIRE colors ($S_{350}/S_{500}$ and
$S_{250}/S_{350}$) are comparable between the full sample and the SMA
subsample.  The SMA subsample contains nearly all galaxies in the parent sample
with $S_{500} > 170\,$mJy (the
lone exception is H-ATLAS~J1429$-$002, which has the highest $S_{350}/S_{500}$
ratio in the full lens candidate sample and is the subject of a study based on
data from the Atacama Large Millimeter/submillimeter Array (ALMA) and other facilities;
Messias et al., in prep.).

\subsection{SMA Imaging}\label{sec:smaobs}

SMA data were obtained over a period of multiple semesters from 2010
March through 2013 May with a total of 162~hours of
on-source integration time.  Each target was typically observed
in multiple array configurations with $t_{\rm int} = 1-2$~hours on-source per
configuration.  We used track-sharing of multiple targets per track to ensure
the best possible {\it uv} coverage.

Observations took place in a range of conditions from superb (atmospheric
opacities of $\tau_{\rm 225 \, GHz} = 0.04$, phase errors of $\Delta(\phi_{\rm
rms}) = 10\,^\circ$) to good (atmospheric opacities of $\tau_{\rm 225 \, GHz} =
0.1$, phase errors of $\Delta(\phi_{\rm rms}) = 40\,^\circ$).  Phase errors are
estimated from a fixed monitoring system on a variety of baselines.  For some
of the observations, 1 or more antennas were unavailable, so in those cases the
total number of antennas ($N_{\rm ant}$) used was less than 8.  Most notably,
in early 2011, the lower sideband of the SMA 345~GHz receiver in antenna 1 was
flagged due to significant instrumental noise.  In general, we optimized the
SMA single-polarization 345~GHz receivers for continuum detection by tuning to
a frequency of $\nu_{\rm LO} \approx 340$ GHz.  In some cases $\nu_{\rm LO}$
varied from this value by up to 10~GHz to avoid retuning in the middle of the
night when the SMA switched from another program.  Table~\ref{tab:observations}
presents details regarding the SMA observations, including the date, array
configuration, local oscillator frequency ($\nu_{\rm LO}$), $\tau_{\rm 225 \,
GHz}$, $\phi_{\rm rms}$, $N_{\rm ant}$, $t_{\rm int}$, and original reference
(some of the data used in this paper were originally presented elsewhere).

The SMA receivers make use of an intermediate frequency coverage of 4~GHz,
providing a total of 8~GHz bandwidth (considering both sidebands), with a
center-to-center sideband separation of 12$\,$GHz.  The primary goal of the
observations is to detect the continuum emission at the highest possible
significance.  When a spectroscopic redshift for a {\it Herschel} source was
available, we tuned the receivers to the closest CO rotational transition, as
long as the signal-to-noise (S/N) of the continuum data would not be
compromised by doing so.  This was possible for 8 of our targets.  Since the
background galaxies lie at $1.5 < z < 4.5$, our observations probe emission
from the $J = 10$ (or higher) levels in lines that are typically faint in SMGs
and therefore difficult to detect.  We defer a discussion of the molecular line
measurements based on the SMA data to a future publication.

Bandpass calibrators were chosen primarily based on their 880$\, \mu$m flux
densities (where possible, we used calibrators with $S_{880} > 5~$Jy) and their
observability at the beginning or end of each night.  For absolute flux density 
calibration, Titan was used whenever possible, followed by Callisto, Neptune
(in subcompact array only), and MWC349A (when no planets or moons were
available).  Amplitude and phase gain calibration was achieved by monitoring
nearby (angular separation from science target of $< 15\,^\circ$), bright
($S_{880} > 0.5\;$Jy) quasars.  Whenever possible, we used multiple quasars for
gain calibration, including a fainter quasar ($S_{880} > 0.1 \;$Jy) much closer
to the science target (angular separation $< 5\,^\circ$) to provide an
independent check of the reliability of the calibration, particularly phase
transfer.  We used the Interactive Data Language (IDL) {\sc MIR} package to
calibrate the {\it uv} visibilities.  

We used the Multichannel Image Reconstruction, Image Analysis, and Display
(MIRIAD) software package \citep{1995ASPC...77..433S} to reconstruct and
deconvolve the image from the visibilities.  We used natural weighting to
achieve maximum sensitivity for all targets.  We combine visibility data from
all available configurations for each target.  The beam size and shape in the
resulting images vary greatly from target to target.  This is primarily
because not every target was observed in all array configurations, but there is
also some dependence on the declination of the target, because the SMA {\it uv}
coverage is less complete at declinations near 0$\,^\circ$.  In general, we
achieve spatial resolutions of $\approx 0\farcs6$ full-width at half-maximum
(FWHM).

Figure~\ref{fig:imaging} shows the SMA image of each object in the SMA
subsample (red contours, beginning at $\pm3\sigma$ and increasing by factors of
$\sqrt{2}$) in comparison with the best available optical or near-IR image
(grayscale, see section~\ref{sec:opticalimaging}).  A detailed source-by-source
description is deferred to section~\ref{sec:objectbyobject}.  The position of
the 880${\rm \, \mu m}$ emission centroid (estimated by-eye) for each source is
presented in Table~\ref{tab:position}.  There is no absolute significance to
these centroid values, but they are necessary to undertake lens modeling. 

We use the SMA images in conjunction with knowledge of the redshifts of the
lenses and sources (see sections~\ref{sec:mmtobs},~\ref{sec:geminiobs}, 
\ref{sec:whtobs}, and \ref{sec:vltobs} for details) to characterize the nature of the lensing that is
occuring in the SMA subsample.  Those galaxies showing multiple images with a
morphology typical of strong lensing and that have known distinct lens and
source redshifts are given an A grade.  Galaxies with obvious strong lensing
morphology but with only a single known redshift measurement (either of the
lens or the source) receive a B grade.  We expect that all B grade systems are
strong lenses, but without distinct redshift measurements we cannot be certain.
Galaxies showing only a single image of the background source, but with known
distinct lens and source redshifts, are given a C grade.  Finally, an X grade
is given to those objects where the SMA imaging and the available spectroscopic
redshifts provide inconclusive evidence of lensing.  Additional data are needed
to determine whether lensing is occuring in these objects.  Our grades are
listed in Table~\ref{tab:position}.

\LongTables
\begin{deluxetable*}{lcccccccc}[!tbp]
\tablecolumns{9}
\tablewidth{7.5in}
\tablecaption{SMA Observations. \\
References key: N10 = \citet{Negrello:2010fk}; C11 = \citet{Conley:2011lr}; I11
= \citet{Ikarashi:2011qy}; F12 = \citet{Fu:2012uq}; B12 =
\citet{Bussmann:2012lr}; W13 = \citet{Wardlow:2013lr}; F13 = \citep{Fu:2013lr}; I13 = \citet{Ivison:2013fk}; W13a = Wardlow et al. (in prep.); F13a = Fu et al. (in prep.)}
\tablehead{
\colhead{} &
\colhead{} &
\colhead{} &
\colhead{Array} &
\colhead{$\nu_{\rm LO}$} &
\colhead{} &
\colhead{$\Delta(\phi_{\rm rms})$} &
\colhead{} &
\colhead{$t_{\rm int}$} 
\\
\colhead{IAU Name} &
\colhead{Reference} &
\colhead{UT Date} &
\colhead{Configuration\tnm{a}} &
\colhead{(GHz)} &
\colhead{$\tau_{\rm 225 GHz}$} &
\colhead{(deg)} &
\colhead{$N_{\rm ant}$} &
\colhead{(hr)}
}
\startdata
J021830.5$-$053124                & I11  &  2009-12-10  &  COM  &  339.925  &  0.10  &  20  &  7    &  3.7  \\
---                               & W13  &  2010-09-25  &  EXT  &  342.003  &  0.10  &  30  &  7    &  2.0  \\
---                               & New  &  2011-01-26  &  VEX  &  343.160  &  0.08  &  10  &  6.5\tnm{b} &  1.9  \\

J022016.5$-$060143                & W13  &  2010-08-14  &  SUB  &  342.017  &  0.07  &  10  &  8    &  0.9  \\
---                               & W13  &  2010-09-26  &  EXT  &  342.001  &  0.07  &  20  &  7    &  2.5  \\
---                               & F13  &  2011-01-04  &  VEX  &  340.226  &  0.06  &  25  &  6.5\tnm{b} &  3.2  \\

J083051.0$+$013224                & New  &  2011-12-08  &  COM  &  339.564  &  0.09  &  20  &  8    &  1.4  \\
---                               & New  &  2012-02-04  &  EXT  &  339.946  &  0.04  &  20  &  8    &  2.4  \\

J084933.4$+$021443                & I13  &  2011-12-08  &  COM  &  339.564  &  0.09  &  20  &  8    &  1.4  \\
---                               & I13  &  2012-01-31  &  EXT  &  339.537  &  0.05  &  30  &  7    &  4.9  \\
---                               & I13  &  2012-03-31  &  VEX  &  339.917  &  0.05  &  25  &  7    &  1.7  \\

J085358.9$+$015537                & New  &  2011-12-08  &  COM  &  339.564  &  0.09  &  20  &  8    &  1.4  \\
---                               & New  &  2012-02-04  &  EXT  &  339.946  &  0.04  &  20  &  8    &  2.4  \\
---                               & New  &  2012-03-31  &  VEX  &  339.917  &  0.05  &  25  &  7    &  1.7  \\
---                               & New  &  2012-04-09  &  VEX  &  339.915  &  0.05  &  10  &  7    &  1.3  \\

J090302.9$-$014127                & New  &  2010-12-16  &  COM  &  342.410  &  0.13  &  20  &  7.5\tnm{b} &  1.5  \\
---                               & New  &  2011-01-30  &  EXT  &  340.244  &  0.05  &  20  &  6.5\tnm{b} &  1.7  \\
---                               & New  &  2011-01-26  &  VEX  &  343.160  &  0.08  &  35  &  6.5\tnm{b} &  2.0  \\
---                               & New  &  2012-04-09  &  VEX  &  339.915  &  0.05  &  10  &  7    &  1.3  \\

J090311.6$+$003906                & N10  &  2010-03-16  &  SUB  &  340.725  &  0.05  &  10  &  5    &  2.5  \\
---                               & N10  &  2010-04-09  &  COM  &  341.609  &  0.07  &  25  &  6    &  2.6  \\
---                               & N10  &  2010-04-20  &  COM  &  340.714  &  0.06  &  30  &  7    &  2.4  \\
---                               & N10  &  2010-02-25  &  VEX  &  340.735  &  0.10  &  40  &  8    &  5.0  \\

J090740.0$-$004200                & New  &  2010-12-16  &  COM  &  342.410  &  0.13  &  20  &  7.5\tnm{b} &  1.5  \\
---                               & New  &  2011-01-30  &  EXT  &  340.244  &  0.05  &  20  &  6.5\tnm{b} &  1.7  \\
---                               & New  &  2012-02-06  &  EXT  &  339.989  &  0.05  &  30  &  8    &  0.9  \\

J091043.1$-$000321                & New  &  2010-12-16  &  COM  &  342.410  &  0.13  &  20  &  7.5\tnm{b} &  1.5  \\
---                               & New  &  2011-01-30  &  EXT  &  340.244  &  0.05  &  20  &  6.5\tnm{b} &  1.7  \\
---                               & New  &  2012-02-07  &  EXT  &  339.993  &  0.04  &  15  &  8    &  2.0  \\
---                               & New  &  2011-01-26  &  VEX  &  343.160  &  0.08  &  35  &  6.5\tnm{b} &  2.0  \\

J091305.0$-$005343                & N10  &  2010-03-16  &  SUB  &  340.725  &  0.05  &  10  &  5    &  2.5  \\
---                               & N10  &  2010-04-09  &  COM  &  341.609  &  0.07  &  25  &  6    &  2.6  \\
---                               & N10  &  2010-04-20  &  COM  &  340.714  &  0.06  &  30  &  7    &  2.4  \\
---                               & N10  &  2010-02-28  &  VEX  &  340.735  &  0.07  &   5  &  7    &  5.0  \\

J091840.8$+$023047                & New  &  2012-02-06  &  EXT  &  339.989  &  0.05  &  30  &  8    &  0.9  \\
---                               & New  &  2012-02-07  &  EXT  &  339.993  &  0.04  &  15  &  8    &  2.0  \\

J103826.6$+$581542                & W13  &  2010-05-16  &  COM  &  341.981  &  0.06  &  35  &  7    &  3.8  \\

J105712.2$+$565457                & W13  &  2010-12-06  &  COM  &  338.148  &  0.05  &  10  &  8    &  1.1  \\
---                               & New  &  2011-01-04  &  VEX  &  340.226  &  0.08  &  20  &  6.5\tnm{b} &  2.1  \\

J105750.9$+$573026                & C11  &  2010-05-14  &  COM  &  340.742  &  0.06  &  10  &  7    &  4.2  \\
---                               & New  &  2011-01-04  &  VEX  &  340.226  &  0.08  &  20  &  6.5\tnm{b} &  2.1  \\

J113526.3$-$014605                & New  &  2012-02-04  &  EXT  &  339.946  &  0.04  &  20  &  8    &  1.3  \\

J114637.9$-$001132                & F12  &  2012-01-14  &  SUB  &  336.929  &  0.15  &  20  &  7    &  2.0  \\
---                               & F12  &  2011-05-22  &  COM  &  339.579  &  0.08  &  25  &  7    &  1.0  \\
---                               & New  &  2012-02-04  &  EXT  &  339.946  &  0.04  &  20  &  8    &  1.3  \\

J125135.4$+$261457                & New  &  2012-05-22  &  COM  &  339.579  &  0.09  &  35  &  7    &  1.7  \\
---                               & New  &  2012-02-06  &  EXT  &  339.989  &  0.05  &  30  &  8    &  1.2  \\

J125632.7$+$233625                & New  &  2012-05-25  &  COM  &  340.045  &  0.08  &  20  &  7    &  0.8  \\
---                               & New  &  2012-02-06  &  EXT  &  339.989  &  0.05  &  30  &  8    &  1.2  \\

J132427.0$+$284452                & F13a &  2011-12-15  &  COM  &  339.561  &  0.05  &  10  &  8    &  3.0  \\
---                               & F13a &  2012-01-31  &  EXT  &  339.537  &  0.05  &  30  &  7    &  4.4  \\
---                               & F13a &  2012-04-09  &  VEX  &  339.915  &  0.05  &  10  &  7    &  2.3  \\

J132630.1$+$334410                & New  &  2012-02-07  &  EXT  &  339.993  &  0.04  &  15  &  8    &  1.4  \\

J132859.3$+$292317                & New  &  2013-05-03  &  SUB  &  340.757  &  0.12  &  25  &  6    &  0.8  \\
---                               & New  &  2012-02-07  &  EXT  &  339.993  &  0.04  &  15  &  8    &  1.4  \\
---                               & New  &  2012-04-24  &  VEX  &  339.960  &  0.06  &  20  &  7    &  2.5  \\

J133008.4$+$245900                & New  &  2012-05-25  &  COM  &  340.045  &  0.08  &  20  &  7    &  0.8  \\
---                               & New  &  2012-02-07  &  EXT  &  339.993  &  0.04  &  15  &  8    &  1.4  \\

J133649.9$+$291801                & New  &  2013-05-03  &  SUB  &  340.757  &  0.12  &  25  &  6    &  0.8  \\
---                               & New  &  2012-02-06  &  EXT  &  339.989  &  0.05  &  30  &  8    &  1.2  \\
---                               & New  &  2012-04-24  &  VEX  &  339.960  &  0.06  &  20  &  7    &  2.5  \\

J134429.4$+$303036                & New  &  2011-05-22  &  COM  &  339.579  &  0.09  &  35  &  7    &  1.7  \\
---                               & New  &  2011-07-26  &  EXT  &  341.037  &  0.06  &  15  &  8    &  1.7  \\
---                               & New  &  2012-03-17  &  EXT  &  339.949  &  0.04  &  35  &  7    &  3.2  \\
---                               & New  &  2012-04-09  &  VEX  &  339.915  &  0.05  &  10  &  7    &  2.3  \\

J141351.9$-$000026                & New  &  2011-05-23  &  COM  &  339.544  &  0.06  &  30  &  7    &  2.9  \\
---                               & New  &  2011-01-30  &  EXT  &  340.244  &  0.05  &  20  &  6.5\tnm{b} &  1.1  \\
---                               & New  &  2011-01-24  &  VEX  &  341.449  &  0.04  &  10  &  5.5\tnm{b} &  1.6  \\
---                               & New  &  2011-01-26  &  VEX  &  343.160  &  0.09  &  30  &  6.5\tnm{b} &  1.3  \\

J142413.9$+$022303                & B12  &  2010-06-16  &  COM  &  342.100  &  0.10  &  30  &  8    &  3.0  \\
---                               & B12  &  2011-01-28  &  EXT  &  340.711  &  0.08  &  30  &  6.5\tnm{b} &  1.3  \\
---                               & B12  &  2011-01-04  &  VEX  &  340.226  &  0.10  &  20  &  6.5\tnm{b} &  1.8  \\
---                               & B12  &  2011-01-06  &  VEX  &  340.225  &  0.04  &  10  &  6.5\tnm{b} &  1.7  \\

J142823.9$+$352619                & New  &  2011-01-28  &  EXT  &  340.711  &  0.08  &  10  &  6.5\tnm{b} &  1.0  \\
---                               & New  &  2011-02-04  &  EXT  &  350.086  &  0.12  &  35  &  7    &  1.3  \\

J142825.5$+$345547          & W13a       &  2012-05-25  &  COM  &  340.045  &  0.08  &  20  &  7    &  0.8  \\
---                              & W13a  &  2011-07-26  &  EXT  &  341.037  &  0.06  &  15  &  8    &  1.7  \\

J143330.8$+$345439                & W13  &  2010-12-16  &  COM  &  342.410  &  0.11  &  20  &  7.5\tnm{b} &  0.8  \\
---                               & W13  &  2011-01-28  &  EXT  &  340.711  &  0.08  &  10  &  6.5\tnm{b} &  1.0  \\
---                               & W13  &  2011-02-04  &  EXT  &  350.086  &  0.12  &  35  &  7    &  1.3  \\

J144556.1$-$004853                & New  &  2011-05-23  &  COM  &  339.544  &  0.06  &  30  &  7    &  2.9  \\
---                               & New  &  2011-01-30  &  EXT  &  340.244  &  0.05  &  20  &  6.5\tnm{b} &  1.1  \\
---                               & New  &  2011-01-26  &  VEX  &  343.160  &  0.09  &  30  &  6.5\tnm{b} &  1.3  \\
\enddata
\tnt{a}{SUB = subcompact (longest baseline length $\approx 25\,$m); COM =
compact (longest baseline length $\approx 75\,$m); EXT = extended (longest
baseline length $\approx 220\,$m); VEX = very extended (longest baseline length
$\approx 510\,$m)}
\tnt{b}{The lower sideband of one antenna was flagged for these observations.}
\label{tab:observations}
\end{deluxetable*}

We compute total flux densities at 880$\, \mu$m within rectangular apertures
customized to match the spatial extent of each object in the SMA images.
Uncertainties on these measurements are derived by placing apertures of the
same size and shape at random, non-overlapping locations within the SMA primary
beam field of view (excluding regions containing flux density from the source)
and computing the 1$\sigma$ root-mean-square (RMS) variation (which is
generally well-described by a Gaussian) in the distribution of aperture flux
densities.  The number of apertures varied from target to target, but was
typically $\approx 100$.

\begin{figure*}[!tbp] 
    \begin{centering}
\includegraphics[width=0.195\textwidth]{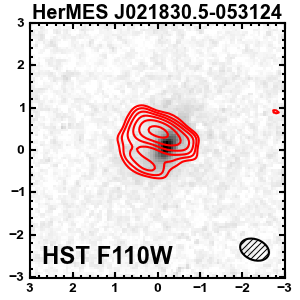}
\includegraphics[width=0.195\textwidth]{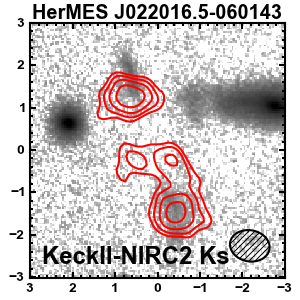}
\includegraphics[width=0.195\textwidth]{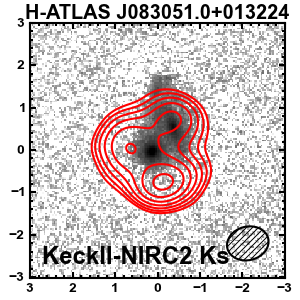}
\includegraphics[width=0.195\textwidth]{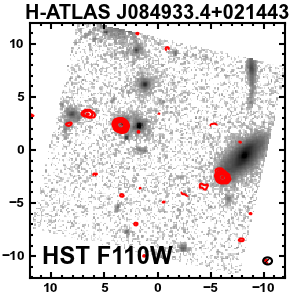}
\includegraphics[width=0.195\textwidth]{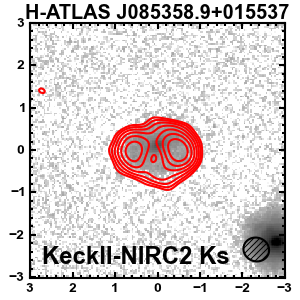}
\includegraphics[width=0.195\textwidth]{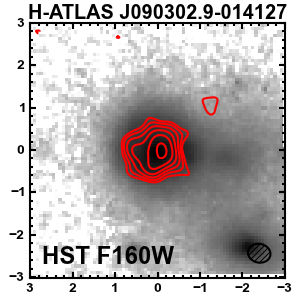}
\includegraphics[width=0.195\textwidth]{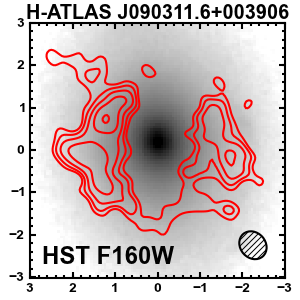}
\includegraphics[width=0.195\textwidth]{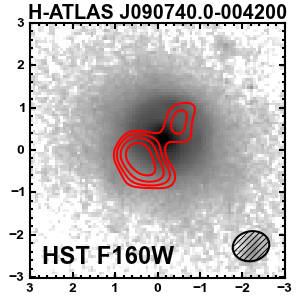}
\includegraphics[width=0.195\textwidth]{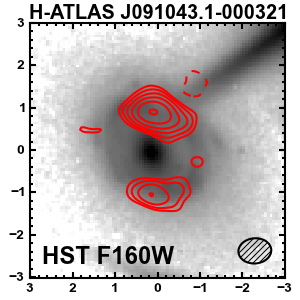}
\includegraphics[width=0.195\textwidth]{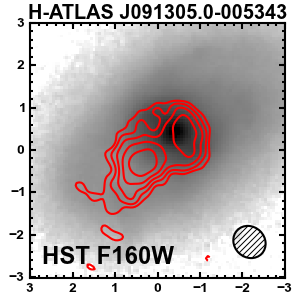}
\includegraphics[width=0.195\textwidth]{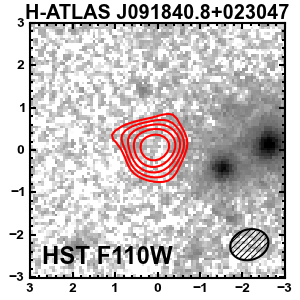}
\includegraphics[width=0.195\textwidth]{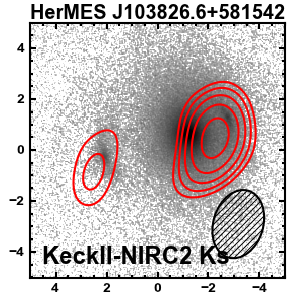}
\includegraphics[width=0.195\textwidth]{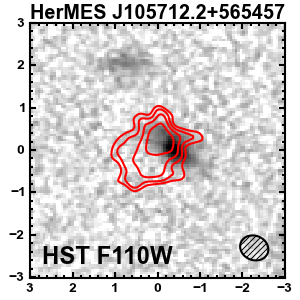}
\includegraphics[width=0.195\textwidth]{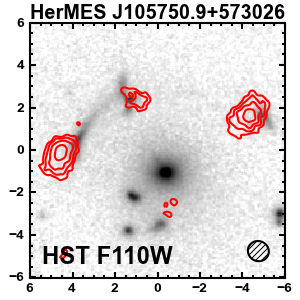}
\includegraphics[width=0.195\textwidth]{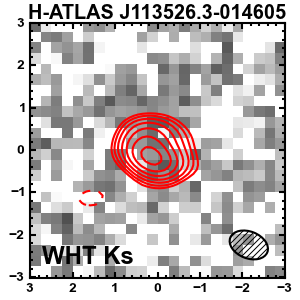}
\includegraphics[width=0.195\textwidth]{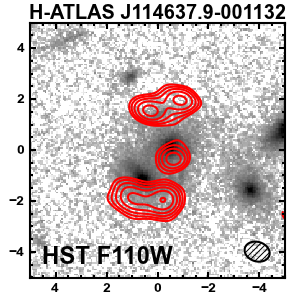}
\includegraphics[width=0.195\textwidth]{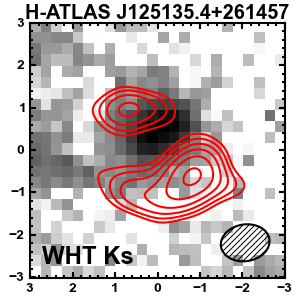}
\includegraphics[width=0.195\textwidth]{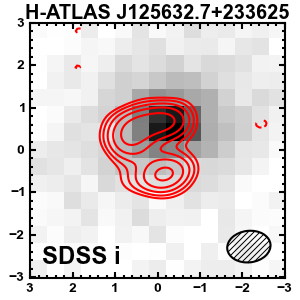}
\includegraphics[width=0.195\textwidth]{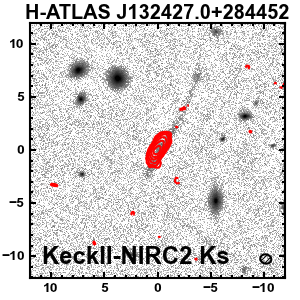}
\includegraphics[width=0.195\textwidth]{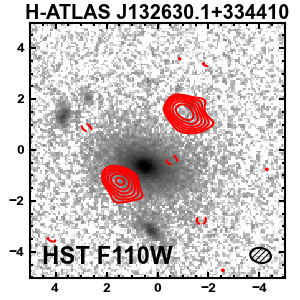}
\includegraphics[width=0.195\textwidth]{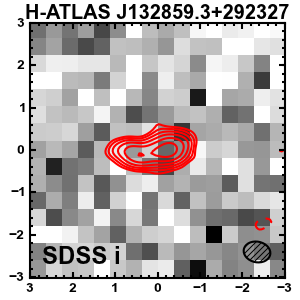}
\includegraphics[width=0.195\textwidth]{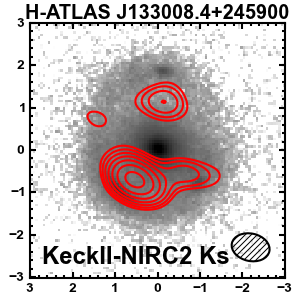}
\includegraphics[width=0.195\textwidth]{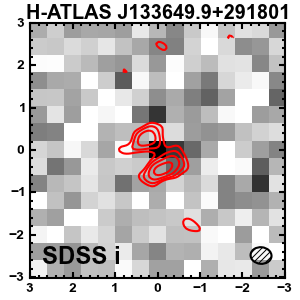}
\includegraphics[width=0.195\textwidth]{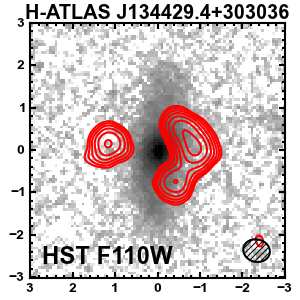}
\includegraphics[width=0.195\textwidth]{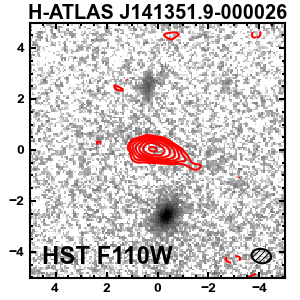}
\includegraphics[width=0.195\textwidth]{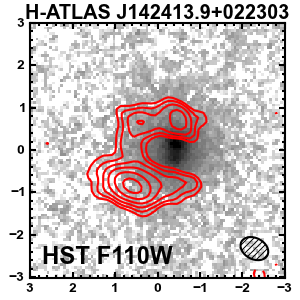}
\includegraphics[width=0.195\textwidth]{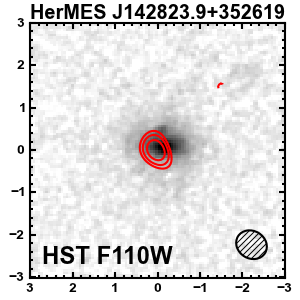}
\includegraphics[width=0.195\textwidth]{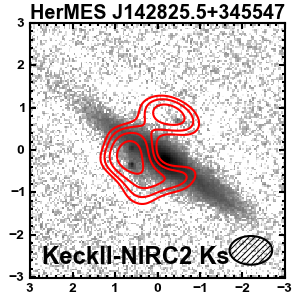}
\includegraphics[width=0.195\textwidth]{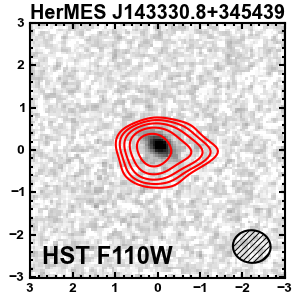}
\includegraphics[width=0.195\textwidth]{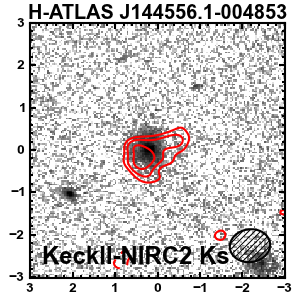}
\end{centering}

\caption{ SMA 880$\mu$m images (red contours, starting at $\pm3\sigma$ and
increasing by factors of $\sqrt{2}$) of candidate lensed SMGs from H-ATLAS and
HerMES, overlaid on best available optical or near-IR images (logarithmic
scaling; telescope and filter indicated in lower left corner of each panel).
North is up and east is left, with axes having units of arcseconds relative to
the 880$\,\mu$m centroid as given in Table~\ref{tab:position}.  The elliptical
FWHM of the SMA's synthesized beam is shown in the lower right corner of each panel.
The image separations are $\approx 1-2\arcsec$, suggesting gravitational
potential wells typical of isolated galaxies or small numbers of galaxies for
the lenses (only two lensed sources are associated with galaxy clusters:
J132427.0$+$284452 and J141351.9$-$000026).  \label{fig:imaging}}

\end{figure*}

\subsection{MMT Optical Spectroscopy}\label{sec:mmtobs}

Long-slit spectroscopic observations using the MMT Red Channel Spectrograph
\citep{Schmidt:1989fk} were conducted in the 2012A semester (PI:
R.~S.~Bussmann) and provided data on H-ATLAS J125632.7$+$233625, H-ATLAS
J132630.1$+$334410, H-ATLAS J133008.4$+$245900, and H-ATLAS J134429.4$+$303036.
The total on-source integration times for these targets were
70-120~minutes apiece.  The observations were obtained during dark time in
near-photometric conditions and seeing was typically $1\farcs0 - 1\farcs5$.
Details of the observations for each target are given in
Table~\ref{tab:opticalspectroscopy}.

We obtained lamp flats for all targets at the beginning of each night for flat
fields.  We obtained a sequence of up to six consecutive 10 minute exposures on
each target.  For wavelength calibration, before and after each of these
sequences we obtained comparison lamp observations using a He/Ar/Ne comparison
lamp.  We checked the focus periodically throughout the night.  

We used either the 270 or the 300 lines per mm grating with a central
wavelength ($\lambda_{\rm central}$) of $6996\,$\AA or $5504\,$\AA.  The slit
width was chosen to be as small as atmospheric seeing allowed and was
either $1\farcs5$ or $2\farcs0$.  The spectral resolution at $\lambda_{\rm
central}$ ($\Delta \lambda$) was 10-30$\,$\AA.

The long-slit data were reduced using standard {\sc IRAF} one-dimensional
spectroscopy routines.  The primary aim of these observations for the purpose
of this paper is to obtain a spectroscopic redshift for the putative lensing
system.  We accomplished this task with the {\sc xcsao} routine
in {\sc IRAF}, using as a template a 5~Gyr old simple stellar population from
\citet{2003MNRAS.344.1000B} with solar metallicity and a Chabrier~IMF.  This
template does not perfectly match the lensing galaxy spectra, but it is
sufficient to determine a robust redshift.  The lens redshifts are presented in
Table~\ref{tab:position}.  A more detailed exploration of the optical spectra
is deferred to a subsequent publication, so we do not show the spectra in this
paper.

\subsection{Gemini-South Optical Spectroscopy}\label{sec:geminiobs}

Long-slit spectroscopic observations using the Gemini Multi-Object
Spectrograph-South \citep[GMOS-S;][]{Hook:2004qy} were conducted in queue mode
during the 2012A semester as part of program GS-2012A-Q-52 (PI: R.~S.~Bussmann)
and provided data on H-ATLAS J090740.0$-$004200 
and H-ATLAS J141351.9$-$000026.  The total on-source integration times for each
of these targets were 1-4~hours.  The observations were obtained during
dark time in near-photometric conditions.  Observational details are given in
Table~\ref{tab:opticalspectroscopy}.

Some aspects of the observing strategy were common to both targets and
followed the guidelines given by the Gemini observatory
\footnote{http://www.gemini.edu/sciops/instruments/gmos/calibration}.  Flat
field observations were interspersed between the science exposures at each
wavelength setting.  CuAr arc lamp exposures were taken for the purpose of
wavelength calibration, using the same instrumental setup as for the science
exposures.  A 5$\,$\AA\ spectral dither between exposures was used to cover the
gap in the GMOS-S chip, and we binned the CCD pixels by a factor of 4 in both
the spatial and spectral directions, providing a spatial scale of
$0\farcs288\;$pixel$^{-1}$ and a spectral pixel scale of 2.69$\,$\AA.  We used
the GG455 blocking filter.  Aspects of the observing strategy that varied from
target to target are given in Table~\ref{tab:opticalspectroscopy}.  

The long-slit data were reduced using the {\sc IRAF} Gemini GMOS reduction
routines, following the standard GMOS-S reduction steps in the example taken
from the Gemini observatory webpage
\footnote{http://www.gemini.edu/sciops/data-and-results/processing-software/getting-started\#gmos}.
We used the same procedure as outlined in section~\ref{sec:mmtobs} to measure a
spectroscopic redshift using {\sc IRAF}'s xcsao task.  The lens redshifts are
presented in Table~\ref{tab:position}.  We plan a detailed exploration of the
optical spectra in future work, so we do not show the optical spectra in this
paper.

%
%

\subsection{WHT Optical Spectroscopy}\label{sec:whtobs}

Long-slit spectroscopic observations using the auxiliary port camera
\citep[ACAM;][]{Benn:2008lr} at the WHT were conducted in the 2011A semester
(PIs: I.~P\'erez-Fournon and A.~Verma) and provided three 1200~second exposures
of H-ATLAS~J083051.0$+$013224 on 2011 April 24.  ACAM provides fixed-format
spectroscopy, covering a spectral range of $3500 - 9400\, $\AA\ (spectral
resolution of $\approx3.3 \,$\AA\ $\, $pix$^{-1}$) using a 400~lines/mm
transmission Volume Phase Holographic grating.  The observing conditions were
photometric and the seeing was 0$\farcs$75.  We have included the observational
details for this target in Table~\ref{tab:opticalspectroscopy} for
completeness.

We obtained Tungsten lamp flats at the beginning of each night for flat fields.
For wavelength calibration, we obtained comparison lamp observations using a CuNe lamp.
We checked the focus periodically throughout the night using the ACAM imaging mode. 

A position angle of 114~deg east of north was chosen to include several objects
visible on the optical imaging close to the candidate lensing galaxy.  The slit
width was $1\farcs0$ and the corresponding spectral resolution provided by the
grating was $R \approx 450$, or 15$\,$\AA\ FWHM at 6750$\,$\AA.

The long-slit spectroscopic data were reduced using standard {\sc IRAF} two-dimensional
spectroscopy programs to correct for the distortions in the data and apply the
wavelength and flux density calibration.

The ACAM spectroscopy, based on G-band and Mg absorption features, shows that
the primary lens (i.e., the galaxy that is closest to the SMA emission
centroid) is located at $z_{\rm lens} = 0.626\pm0.001$. A nearby disk-like
galaxy was detected in the acquisition images and spatially resolved from the
primary lens (the existence of this additional galaxy is confirmed by ancillary
high-spatial resolution imaging from Keck-II adaptive optics imaging; see
section~\ref{sec:opticalimaging}). In the ACAM long-slit spectra an emission
line is detected at 7462$\,$\AA\ at the spatial location of the disk-like
galaxy. We cannot associate this line with emission lines at the same redshift
as the primary lens. The most plausible identification is [OII]3727 at $z = 1.002
\pm 0.001$, in which case the two galaxies are at different redshifts.

\subsection{VLT X-Shooter Spectroscopy}\label{sec:vltobs}

The optical and near-IR spectra of the southeastern lens in J114637$-$001132
\citep[i.e., the galaxy denoted as ``G2'' by][]{Fu:2012uq} were obtained with
the X-shooter spectrograph \citep{Vernet:2011lr} at the VLT. X-shooter provides
simultaneous spectral coverage from 300$\,$nm to 2.5$\,\mu$m.  The 1$\arcsec
\times 11\arcsec$ slit was used in the ``UVB'' arm ($300-560\,$nm), while the
$0.9\arcsec \times 11\arcsec$ slits were used in the ``VIS'' arm
($550-1020\,$nm) and ``NIR'' arm ($1020-2480\,$nm), yielding a resolving power
of $\lambda/\Delta \lambda$ of 4350, 7450 and 5300 in the three bands,
respectively.  The observations were obtained on 2012 September 18 and 19.  The
total integration time was 320 minutes and comprised individual exposures of 20
minutes each. The object was nodded along the slit by a few arcsec between one
exposure and the next.

Data reduction was performed by following the standard steps of the public
X-shooter pipeline \citep{Goldoni:2006lr}. Sky emission lines were subtracted
by exploiting temporally contiguous exposures in which the objects was nodded
in a different position of the slit. After flat-fielding, the pipeline extracts
the different orders of the echelle spectrum, which are then rectified,
wavelength calibrated and merged. Then the spectra were calibrated in flux by
using the observation of a spectrophotometric standard. The final
mono-dimensional spectrum was extracted from an aperture of 1".  

The lensing galaxy G2 is clearly detected with several emission lines
([OIII],Hb,Ha,[NII],[SII], [OII]) that imply a redshift of $z_{\rm
lens}=1.2247$.

\begin{deluxetable*}{lccccccc}[!bp]
\tablecolumns{8}
\tablewidth{0pt}
\tablecaption{Optical Spectroscopy Observations}
\tablehead{
\colhead{} &
\colhead{} &
\colhead{} &
\colhead{Grating} &
\colhead{$\lambda_{\rm central}$} &
\colhead{Slit Width} &
\colhead{$\Delta \lambda$} &
\colhead{$t_{\rm int}$} 
\\
\colhead{IAU Name} &
\colhead{Telescope} &
\colhead{UT Date} &
\colhead{(lines$\,$mm$^{-1}$)} &
\colhead{(\AA)} &
\colhead{($\arcsec$)} &
\colhead{(\AA)} &
\colhead{(min)}
}
\startdata
J083051.0$+$013224 & WHT  & 2011-04-24   & 400   &  6500  &  1.0  &  15    &  60  \\
J090740.0$-$004200 & Gemini-S & 2012-02-26 & 400 &  6710  &  1.5  &  10.5  & 120  \\
J114637.9$-$001132 & X-Shooter  & 2012-09-18/19 & --- &  16200  &  1.0  &  3.0 & 320  \\
J125632.7$+$233625 & MMT  &  2012-02-23  &  300  &  5504  &  1.5  &  17.9  &  80  \\
J132630.1$+$334410 & MMT  &  2012-02-23  &  270  &  6996  &  2.0  &  21.9  & 120  \\
J133008.4$+$245900 & MMT  &  2012-02-23  &  300  &  5504  &  1.5  &  16.4  &  70  \\
J134429.4$+$303036 & MMT  &  2012-02-22  &  270  &  6996  &  1.5  &  17.9  &  80  \\
J141351.9$-$000026 & Gemini-S & 2012-02-26 & 150 &  6720  &  1.5  &  22.3  & 240  \\
\enddata
\label{tab:opticalspectroscopy}
\end{deluxetable*}

\subsection{Ancillary Optical and Near-IR Imaging}\label{sec:opticalimaging}

In all cases where the SMA has clearly resolved multiple images of the
background source, there is no evidence for submm emission from the lens.
This means that detection of the foreground lens requires observations at
optical or near-IR wavelengths.  This paper makes use of the best available
optical or near-IR imaging to pinpoint the location of the lens and determine
whether it comprises multiple galaxies.  This imaging is shown in grayscale in
Figure~\ref{fig:imaging}, and the telescope and filter used are given in the
lower left corner of each panel.  Fifteen objects use {\it HST} Snapshot
imaging (marked as ``{\it HST} F110W'' in Figure~\ref{fig:imaging}), five use
Keck-II/NIRC2-LGSAO imaging (marked as ``Keck-II\_NIRC2 Ks'' in
Figure~\ref{fig:imaging}), five use full-orbit {\it HST} imaging (marked as
``{\it HST} F160W'' in Figure~\ref{fig:imaging}), four use SDSS $i$-band
imaging (marked as ``SDSS i'' in Figure~\ref{fig:imaging}), and two use WHT
$K_{\rm s}$-band imaging (marked as ``WHT Ks'' in Figure~\ref{fig:imaging}).

The focus of this paper is lens modeling of the SMA data.  A detailed analysis
of the optical and near-IR imaging will appear in a set of papers specific to
the {\it HST} Snapshot imaging (Amber et al., in prep.; Calanog et al. in
prep.), the full-orbit {\it HST} imaging (Negrello et al., in prep., Dye et
al., in prep.), and the Keck-II/NIRC2-LGSAO imaging (Calanog et al., in prep.)
and the WHT $K_{\rm s}$ imaging (Mart/'/nez-Navajas et al., in prep.).

To facilitate comparison with existing surveys for lenses based on SDSS
spectroscopy \citep[e.g.,][]{Bolton:2008wd, Brownstein:2012rt}, we compute
$i$-band photometry using SDSS Data Release 9 (DR9) for all of the objects in
the SMA subsample.
The imaging aspect of DR9 provides five optical bands: $u$, $g$, $r$, $i$, and
$z$.  The 95\% completeness levels for point sources are $u=22.0$, $g=22.2$,
$r=22.2$, $i=21.3$, and $z=20.5$ (AB mag), corresponding to flux densities of
$5.7 \, \mu$Jy, $4.8 \, \mu$Jy, $4.8 \, \mu$Jy, $11.0 \, \mu$Jy, and $23.0 \,
\mu$Jy, respectively.  The median seeing in the images at $r$-band is typically
1$\farcs$3.

We searched for counterparts in the DR9 catalog within a $2\arcsec$ radius of
each expected lens position based on the best available optical or near-IR
imaging.  If a counterpart was found, then it was assigned photometry directly
from the DR9 catalog.  If no counterpart was found, we used our own custom
aperture photometry code to measure the 2$\sigma$ limiting flux density at the
position of the target (note that at these wavelengths, the lens is typically
much brighter than the source).  We used a 4$\arcsec$ diameter circular aperture
and computed the sky background in an annulus with an inner radius of
2$\arcsec$ and an outer radius of 5$\arcsec$.  We measured the uncertainties by
placing $N$ random apertures (where $N \approx 300$) of the same size and shape
within 3$\arcmin$ of the lens candidate (taking care to avoid any objects found
in the DR9 catalog) and computing the 68\% confidence interval of the
dispersion in the measured flux densities.  The $i$-band AB magnitudes are
reported in Table~\ref{tab:fluxes} (limits indicate 2$\sigma$ values), along
with the {\it Herschel}/SPIRE and SMA 880$\, \mu$m measurements (the values
reported in the table do not include absolute flux density calibration
uncertainty of 7\%).


\begin{deluxetable*}{lcccccc}[b!]
\tabletypesize{\scriptsize} 
\tablecolumns{7}
\tablewidth{0pt}
\tablecaption{Spatially Integrated Flux Densities of Strong Lens Sample\tablenotemark{a}}
\tablehead{
\colhead{} & 
\colhead{$i$\tablenotemark{b}} &
\colhead{$S_{250}$} &
\colhead{$S_{350}$} &
\colhead{$S_{500}$} &
\colhead{$S_{880}$}
\\
\colhead{IAU name} & 
\colhead{(AB mag)} & 
\colhead{(mJy)} &
\colhead{(mJy)} &
\colhead{(mJy)} &
\colhead{(mJy)}
}
\startdata
J021830.5$-$053124   & $ > 22.6        $ & $92  \pm 7$ & $122\pm 8$ & $113\pm 7$ & $66.0\pm5.4$ \\
J022016.5$-$060143   & $ 20.32 \pm 0.06$ & $180 \pm 7$ & $192\pm 8$ & $132\pm 7$ & $28.3\pm3.4$ \\
J083051.0$+$013224   & $ 20.85 \pm 0.09$ & $260 \pm 7$ & $321\pm 8$ & $269\pm 9$ & $85.5\pm4.0$ \\
J084933.4$+$021443   & $ 19.01 \pm 0.02$ & $242 \pm 7$ & $293\pm 8$ & $231\pm 9$ & $50.0\pm3.5$ \\
J085358.9$+$015537   & $ > 22.3        $ & $389 \pm 7$ & $381\pm 8$ & $241\pm 9$ & $61.4\pm2.9$ \\
J090302.9$-$014127   & $ 20.92 \pm 0.11$ & $347 \pm 7$ & $339\pm 8$ & $219\pm 9$ & $54.7\pm3.1$ \\
J090311.6$+$003906   & $ 18.17 \pm 0.01$ & $138 \pm 7$ & $199\pm 8$ & $174\pm 9$ & $78.4\pm8.2$ \\
J090740.0$-$004200   & $ 20.94 \pm 0.07$ & $471 \pm 7$ & $343\pm 8$ & $181\pm 9$ & $24.8\pm3.3$ \\
J091043.1$-$000321   & $ 21.41 \pm 0.09$ & $417 \pm 6$ & $378\pm 7$ & $232\pm 8$ & $30.6\pm2.4$ \\
J091305.0$-$005343   & $ 18.74 \pm 0.02$ & $116 \pm 6$ & $140\pm 7$ & $108\pm 8$ & $36.7\pm3.9$ \\
J091840.8$+$023047   & $ > 22.4        $ & $142 \pm 7$ & $175\pm 8$ & $138\pm 9$ & $18.8\pm1.6$ \\
J103826.6$+$581542   & $ 18.71 \pm 0.02$ & $191 \pm 7$ & $157\pm10$ & $101\pm 7$ & $30.2\pm2.2$ \\
J105712.2$+$565457   & $ 22.0  \pm 0.4 $ & $114 \pm 7$ & $147\pm10$ & $114\pm 7$ & $50.3\pm5.9$ \\
J105750.9$+$573026   & $ 20.15 \pm 0.04$ & $403 \pm 7$ & $377\pm10$ & $249\pm 7$ & $55.7\pm5.8$ \\
J113526.3$-$014605   & $ > 22.5        $ & $290 \pm 7$ & $295\pm 8$ & $216\pm 9$ & $48.6\pm2.3$ \\
J114637.9$-$001132   & $ 21.44 \pm 0.10$ & $290 \pm 6$ & $356\pm 7$ & $295\pm 8$ & $86.0\pm4.9$ \\
J125135.4$+$261457   & $ > 22.2        $ & $145 \pm 7$ & $201\pm 8$ & $212\pm 9$ & $78.9\pm4.4$ \\
J125632.7$+$233625   & $ 18.70 \pm 0.02$ & $214 \pm 7$ & $291\pm 8$ & $261\pm 9$ & $97.2\pm6.5$ \\
J132427.0$+$284452   & $ > 22.6        $ & $347 \pm 7$ & $377\pm 8$ & $268\pm 9$ & $30.2\pm2.2$ \\
J132630.1$+$334410   & $ 20.25 \pm 0.07$ & $179 \pm 7$ & $279\pm 8$ & $265\pm 9$ & $65.2\pm2.3$ \\
J132859.3$+$292327   & $ > 22.6        $ & $264 \pm 9$ & $310\pm10$ & $261\pm10$ & $50.1\pm2.1$ \\
J133008.4$+$245900   & $ 20.00 \pm 0.03$ & $273 \pm 7$ & $282\pm 8$ & $214\pm 9$ & $59.2\pm4.3$ \\
J133649.9$+$291801   & $ > 22.7        $ & $295 \pm 8$ & $294\pm 9$ & $191\pm10$ & $36.8\pm2.9$ \\
J134429.4$+$303036   & $ 20.88 \pm 0.06$ & $481 \pm 9$ & $484\pm13$ & $344\pm11$ & $73.1\pm2.4$ \\
J141351.9$-$000026   & $ 22.0  \pm 0.2 $ & $190 \pm 7$ & $240\pm 8$ & $200\pm 9$ & $33.3\pm2.6$ \\
J142413.9$+$022303   & $ 21.62 \pm 0.12$ & $115 \pm 7$ & $192\pm 8$ & $203\pm 9$ & $90.0\pm5.0$ \\
J142823.9$+$352619   & $ 22.2  \pm 0.4 $ & $323 \pm 6$ & $244\pm 7$ & $140\pm33$ & $18.4\pm2.5$ \\
J142825.5$+$345547   & $ 19.89 \pm 0.04$ & $159 \pm 6$ & $196\pm 7$ & $157\pm33$ & $42.3\pm4.7$ \\
J143330.8$+$345439   & $ > 22.3        $ & $158 \pm 6$ & $191\pm 7$ & $160\pm33$ & $59.6\pm3.9$ \\
J144556.1$-$004853   & $ > 22.7        $ & $141 \pm 7$ & $157\pm 8$ & $130\pm 9$ & $ 9.0\pm2.1$ \\
\enddata
\tablenotetext{a}{Measurement uncertainties for {\it Herschel} photometry do not include absolute flux density calibration uncertainty of 7\%.}
\tablenotetext{b}{$i$-band magnitudes obtained from SDSS DR9.}
\label{tab:fluxes}
\end{deluxetable*}

\section{Lens Models}\label{sec:lensmodels}

The SMA data provide sufficient sensitivity and spatial resolution to permit
tight constraints on parameters of the lens models for a total of 25 lensed
SMGs out of the SMA subsample of 30 (those labeled with grade A, B, or C in
Table~\ref{tab:position}).  For some of these objects, deep {\it HST} or
Keck-II/NIRC2-LGSAO data exist that permit the assembly of lens models which
take into account simultaneously the optical, near-IR, and submm data.
However, because this sample of lensed SMGs are at $z > 1.5$ and are heavily
obscured by dust, the lensed emission is frequently detected only in the SMA
data and not in the optical or near-IR.  Therefore, for the current analysis we
focus our efforts on lens models based solely on SMA data (for the handful of
exceptions, see
section~\ref{sec:lensmethod} for details) and defer full SED lens modeling to
subsequent publications.  We describe the methodology behind the lens modeling
in section~\ref{sec:lensmethod} and give a detailed discussion of each object
in section~\ref{sec:objectbyobject}.  We defer an examination of the ensemble
properties of the lenses and lensed sources to
sections~\ref{sec:lensingresults} and \ref{sec:lensedresults}, respectively.  

\subsection{Methodology}\label{sec:lensmethod}

The SMA is an interferometer, so the surface brightness map of each lensed SMG
is obtained with incomplete sampling of the {\it uv} plane.  This means that
surface brightness is not necessarily conserved and that the pixel-to-pixel
errors in the surface brightness map are correlated.  
For these reasons, it is important to compare model and data visibilities
rather than surface brightness maps.  We follow the methodology used in
\citet{Bussmann:2012lr}, who presented the first lens model derived from a
visibility-plane analysis of interferometric imaging of a strongly lensed SMG
discovered in wide-field submm surveys.  We summarize important details here
and refer the interested reader to \citet{Bussmann:2012lr} for further
information.


We use the publicly available {\sc Gravlens} software
\citep{2001astro.ph..2340K} to map emission from the source plane to the image
plane for a given lensing mass distribution.  
To represent the lens mass profile, we use $N_{\rm lens}$ singular isothermal
ellipsoid (SIE) profiles, where $N_{\rm lens}$ is the number of lensing
galaxies found from the best available optical or near-IR imaging \citep[a
multitude of evidence supports the SIE as a reasonable choice; for a recent
review, see][]{Treu:2010fk}.  

The source(s) are assumed to have S\'ersic profile morphologies
\citep{1968adga.book.....S}.  
We always use a single S\'ersic profile in our fits, with the exception of
objects that are clearly only moderately lensed (i.e., singly imaged with $\mu
< 2$) and that show evidence of multiple source-plane components in the SMA
imaging.  This is true for J022016.5$-$060143 and J084933.4$+$021443.

Each SIE is fully described by the following five free parameters: the position
of the lens relative to the SMA emission centroid ($\Delta \alpha_{\rm lens} =
\alpha_{\rm lens} - \alpha_{\rm 880}$ and $\Delta \delta_{\rm lens} =
\delta_{\rm lens} - \delta_{\rm 880}$; these can be compared with the position
of the optical or near-IR counterpart relative to the SMA emission centroid:
$\Delta \alpha_{\rm NIR} = \alpha_{\rm NIR} - \alpha_{\rm 880}$ and $\Delta
\delta_{\rm NIR} = \delta_{\rm NIR} - \delta_{\rm 880}$), the mass of the lens
(parameterized in terms of the angular Einstein radius, $\theta_{\rm E}$), the
ellipticity of the lens ($\epsilon_{\rm lens}$; defined as $1 - b/a$), and the
position angle of the lens ($\phi_{\rm lens}$; degrees east of north).  When
there is evidence for additional lenses from optical or near-IR imaging (see
Figure~\ref{fig:imaging}), we estimate by-eye centroids for each lens (carrying
an uncertainty of order 1 pixel, or $0\farcs04$ and $0\farcs12$ in the
Keck-II/NIRC2-LGSAO and {\it HST} images, respectively) and fix the positions
of the additional lenses with respect to the primary lens.  Therefore, each
additional lens has only 3 free parameters: $\theta_{\rm E}$, $\epsilon_{\rm
lens}$, and $\phi_{\rm lens}$.  We assume secondary, tertiary, etc., lenses are
located at the same redshift as the primary lens, unless there is evidence
against that assumption (e.g., J083051.0$+$013224).  

Each S\'ersic profile is fully described by the following seven free
parameters: the position of the source relative to the primary lens ($\Delta
\alpha_{\rm s} = \alpha_{\rm s} - \alpha_{\rm lens}$ and $\Delta \delta_{\rm s}
= \delta_{\rm s} - \delta_{\rm lens}$), the intrinsic flux density ($S_{\rm
in}$), the S\'ersic index ($n_{\rm s}$), the half-light semi-major axis length
($a_{\rm s}$), the ellipticity ($\epsilon_{\rm s}$, defined as $1-b/a$), and
the position angle ($\phi_{\rm s}$, degrees east of north).  

The total number of free parameters for any given system is $N_{\rm free} = 5 +
3 \times (N_{\rm lens} - 1) + 7 * N_{\rm source}$, where $N_{\rm source}$ is
the number of S\'ersic profiles used.

We adopt loose, uniform priors for all model parameters.  The 1-$\sigma$
absolute astrometric solution between the SMA and optical/near-IR images is
generally $0\farcs2$, so in our modeling efforts the prior on the position of
the lens covers $\pm0\farcs6$ in both RA and Dec (i.e., 3$\sigma$ in each
direction).  In section~\ref{sec:objectbyobject}, we discuss the level of
agreement between the astrometry from the images and the astrometry from the
lens modeling.  For $\theta_{\rm E}$, the prior covers $0\farcs1 - 6\arcsec$.
The ellipticities of the lens and source are always restricted to be $<0.7$.
No prior is placed on the position angle of the lens or source.  The intrinsic
flux density is allowed to vary from 0.1$\,$mJy to the total flux density
observed by the SMA.  The source position is allowed to vary by $\pm 1\arcsec$
relative to the position of the primary lens.  The S\'ersic index varies from
0.1 to 4.0.  The half-light radius varies from $0\farcs05 - 1\farcs5$.

For a given set of model parameters, {\sc Gravlens} generates a surface
brightness map of the lensed emission (note that no model of the emission from
the lens is needed because the lenses are undetected in the SMA imaging).  This
surface brightness map can then be used as input to MIRIAD's {\sc uvmodel}
task, which produces a ``simulated visibility'' dataset ($V_{\rm model}$) by
computing the Fourier transform of the model lensed image and sampling the
resulting visibilities to match the sampling of the actual observed SMA
visibility dataset ($V_{\rm SMA}$).  The quality of fit for a given set of
model parameters is determined from the chi-squared value ($\chi^2$) according
to the following equations:

\begin{equation}
    \chi^2 = \chi^2_{\rm real} + \chi^2_{\rm imag},
\end{equation}
\begin{equation}
    \chi^2_{\rm real} = \sum_{u, v} \frac{[Re(V_{\rm SMA}(u,
    v)) - Re(V_{\rm model}(u, v))]^2}{\sigma_{\rm real}^2(u, v) + \sigma_{\rm
    imag}^2(u, v)}, 
\end{equation}
\begin{equation}
    \chi^2_{\rm imag} = \sum_{u, v} \frac{[Im(V_{\rm SMA}(u,
    v)) - Im(V_{\rm model}(u, v))]^2}{\sigma_{\rm real}^2(u, v) + \sigma_{\rm
    imag}^2(u, v)},
\end{equation}

\noindent where $\sigma_{\rm real}(u, v) = \sigma_{\rm imag}(u, v)$ is the
1$\sigma$ uncertainty level for each visibility and is determined from the
system temperatures (this corresponds to a natural weighting scheme).  

To sample the posterior probability density function (PDF) of our model
parameters, we use {\sc emcee} \citep{Foreman-Mackey:2013yq}, a Markov chain
Monte Carlo (MCMC) code that uses an affine-invariant ensemble sampler to
obtain significant performance advantages over standard MCMC sampling methods
\citep{goodmanweare}.  


We employ a ``burn-in'' phase with 250 walkers and 1000 iterations (i.e.,
250,000 samplings of the posterior PDF) to identify the best-fit model
parameters.  This position is then used to initialize the ``final'' phase with
250 walkers and 20 iterations (i.e., 5,000 samplings of the posterior PDF) to
determine uncertainties on the best-fit model parameters.  The autocorrelation
time for each parameter in a given ensemble of walkers and is of order unity
for each parameter, implying that we have 5,000 independent samplings of the
posterior PDF, more than enough to obtain a robust measurement of the mean and
uncertainty on each parameter of the model.

During each MCMC iteration, we also measure the magnification factor at
880$\,\mu$m, $\mu_{880}$ (we follow the nomenclature in the SMG literature here
and use $\mu$ to refer to the total magnification obtained by summing over all
individual lensed components).  Here, we describe how we measure $\mu_{880}$.

First, we take the unlensed, intrinsic source model and measure the total flux
density ($S_{\rm in}$) within an elliptical aperture ($A_{\rm in}$) centered on
the source with ellipticity and position angle equal to that of the source
model and with a semi-major axis length of $2 a_{\rm s}$.  Second, we take the
lensed image of the best-fit model and measure the total flux density ($S_{\rm
out}$) within the aperture $A_{\rm out}$, where $A_{\rm out}$ is determined by
using {\sc Gravlens} to map $A_{\rm in}$ in the source plane to $A_{\rm out}$
in the image plane (using the lens parameters which correspond to the best-fit
model).  The magnification is then computed simply as $\mu_{880} = S_{\rm out}
/ S_{\rm in}$.  The best-fit value and 1$\sigma$ uncertainty are drawn from the
posterior PDF, as with the other parameters of the model.

The choice of $A_{\rm in}$ has important implications for magnification
measurements.  For multiply imaged systems, apertures that are too large
include in the source plane too much flux density that is far away from the
caustic and relatively unmagnified.  This is a particularly important issue for
the models used here because the S\'ersic index of the background source is a
free parameter.  The S\'ersic index is partially degenerate with the half-light
radius of the source, in the sense that good fits to the data can be obtained
with a combination of small source size and small S\'ersic index or large
source size and large S\'ersic index.  In accordance with this, our model fits
sometimes include relatively large sources where significant fractions of the
unlensed flux density ($\approx 10-20\%$) arise from regions in the source
plane far away from the caustic and hence contribute nothing to the observed
lensed emission.  This situation biases the estimate of $\mu_{880}$ below the
true value.  Conversely, apertures that are too small will omit flux density in
the source plane that is detected at high significance in the SMA imaging, thus
biasing the estimate of $\mu_{880}$ above the true value.  Our choice---double
the semi-major axis length of the source---represents a compromise between
these two extremes.

\subsection{Descriptions of Individual Objects} \label{sec:objectbyobject}

In this section, we describe the basic characteristics of each object in the
SMA subsample, including the position of the lens relative to the SMA 880$\,
\mu$m emission centroid, the lensing configuration (where applicable), and any
unique notes for each object.  Figure~\ref{fig:modeling} shows the best-fit
model in comparison with the SMA data for every lensed SMG with a robust lens
model.  Tables~\ref{tab:lensesresults} and \ref{tab:sourcesresults} present,
for lenses and sources respectively, the model parameter mean values and
1$\sigma$ uncertainties as drawn from the posterior PDF for each parameter.
Note that in some cases the posterior PDFs are non-Gaussian and therefore the
best-fit model shown in Figure~\ref{fig:modeling} does not always correspond
perfectly to the model parameter mean values presented in
Tables~\ref{tab:lensesresults} and \ref{tab:sourcesresults}.

\begin{figure*}[!tbp] 
\begin{centering}
\includegraphics[width=0.5\textwidth]{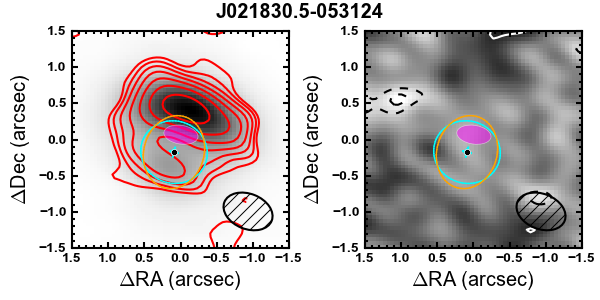}
\includegraphics[width=0.5\textwidth]{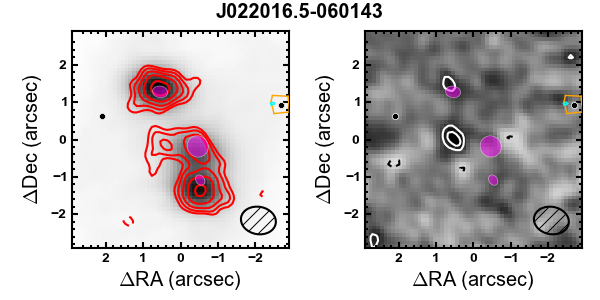}
\includegraphics[width=0.5\textwidth]{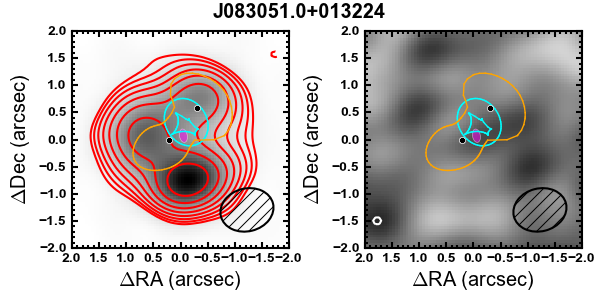}
\includegraphics[width=0.5\textwidth]{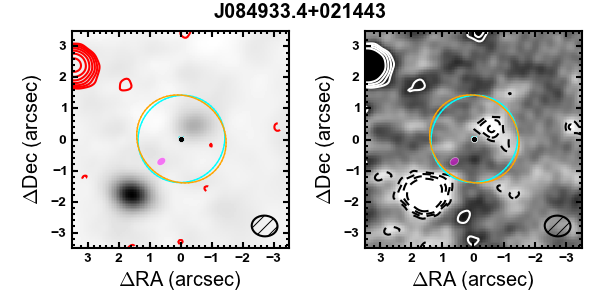}
\includegraphics[width=0.5\textwidth]{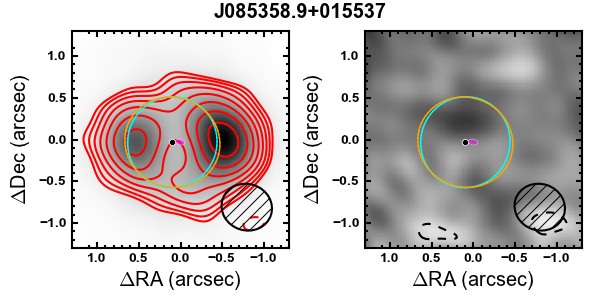}
\includegraphics[width=0.5\textwidth]{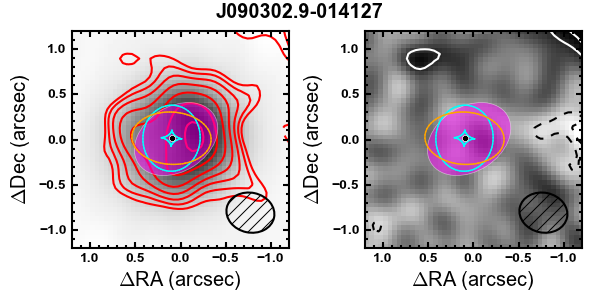}
\includegraphics[width=0.5\textwidth]{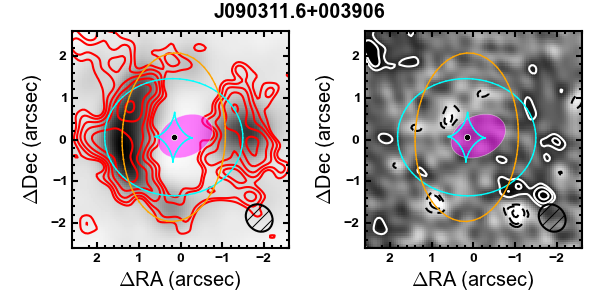}
\includegraphics[width=0.5\textwidth]{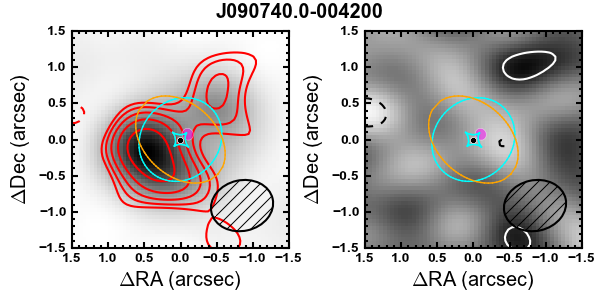}
\includegraphics[width=0.5\textwidth]{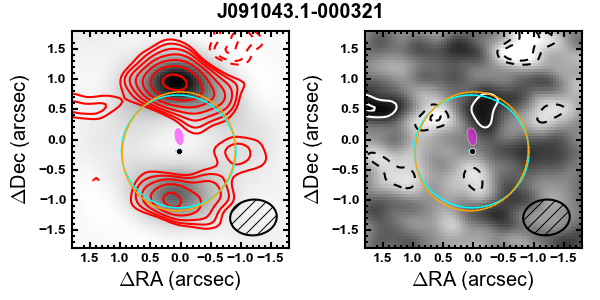}
\includegraphics[width=0.5\textwidth]{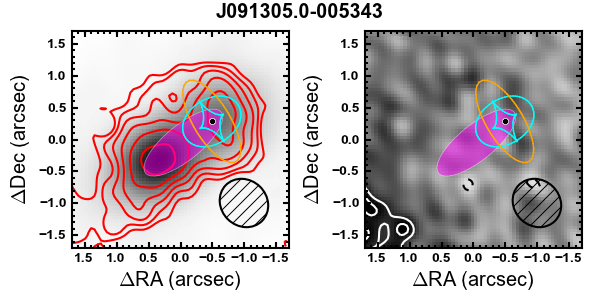}
\end{centering}

\caption{ Comparison of best-fit lens models with SMA data.  Odd columns show
the surface brightness maps of the best-fit model lensed emission (grayscale) in
comparison with the observed surface brightness maps from the SMA (red contours,
beginning at $\pm2\sigma$ and increasing by factors of $\sqrt{2}$).  Even
columns show the residual surface brightness maps obtained by subtracting the
best-fit lens model visibilities with the observed visibilities by the SMA
(contours drawn at same levels as odd panels).  For reference, all panels show
the critical curves (orange line), caustics (cyan lines), position of the
lens(es) (black circles), the half-light area of the background source(s)
(magenta filled ellipses), and the FWHM of the SMA synthesized beam (black
hatched ellipses).  \label{fig:modeling}} \addtocounter{figure}{-1}

\end{figure*}

\begin{figure*}[!tbp] 
\begin{centering}
\includegraphics[width=0.5\textwidth]{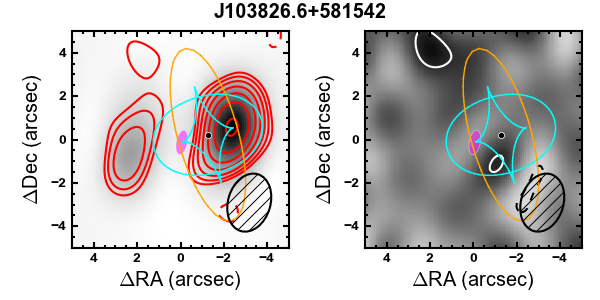}
\includegraphics[width=0.5\textwidth]{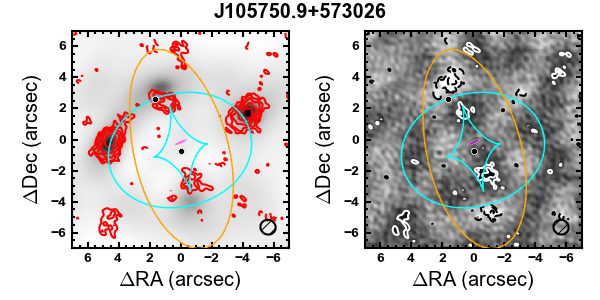}
\includegraphics[width=0.5\textwidth]{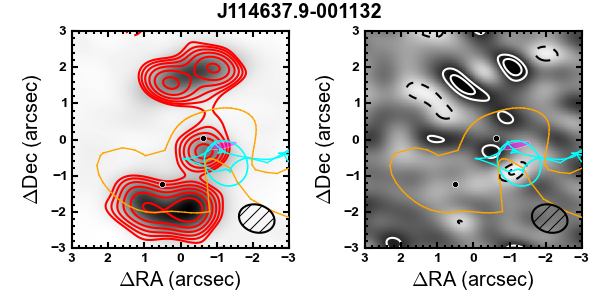}
\includegraphics[width=0.5\textwidth]{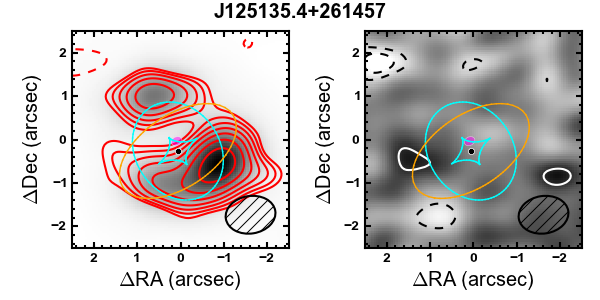}
\includegraphics[width=0.5\textwidth]{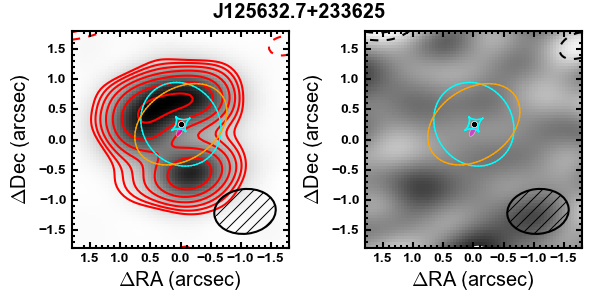}
\includegraphics[width=0.5\textwidth]{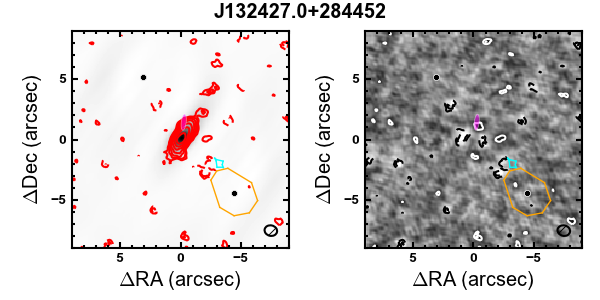}
\includegraphics[width=0.5\textwidth]{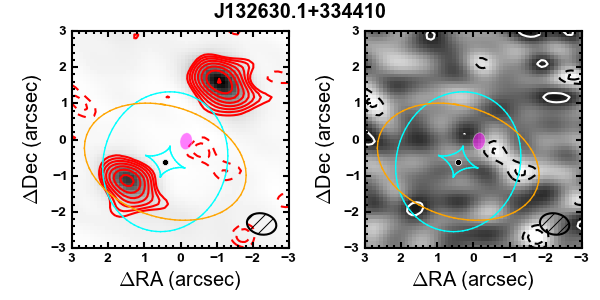}
\includegraphics[width=0.5\textwidth]{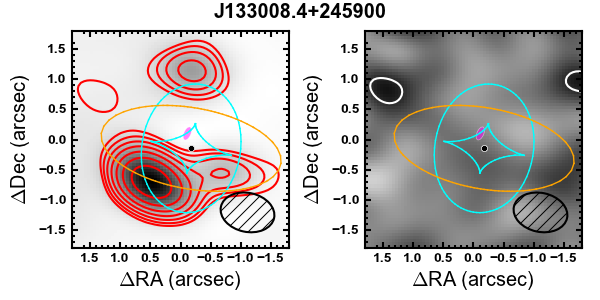}
\includegraphics[width=0.5\textwidth]{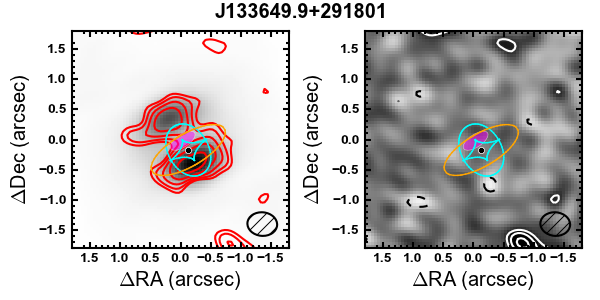}
\includegraphics[width=0.5\textwidth]{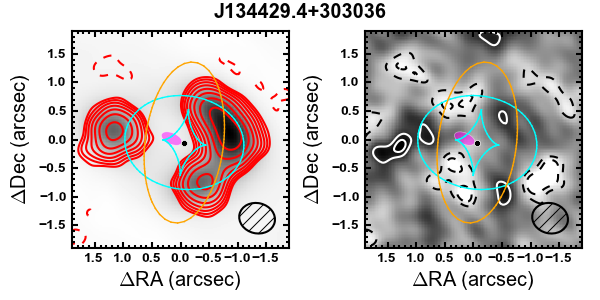}
\end{centering}

\caption{ Continued.  \label{fig:modeling1}}
\addtocounter{figure}{-1}

\end{figure*}

\begin{figure*}[!tbp] 
\begin{centering}
\includegraphics[width=0.5\textwidth]{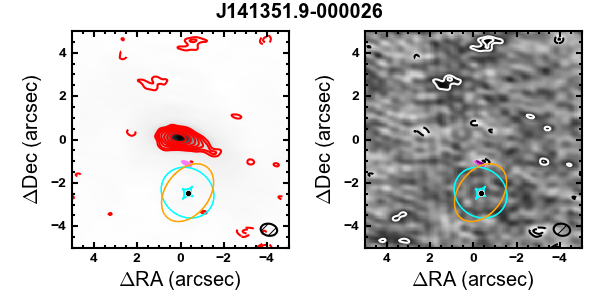}
\includegraphics[width=0.5\textwidth]{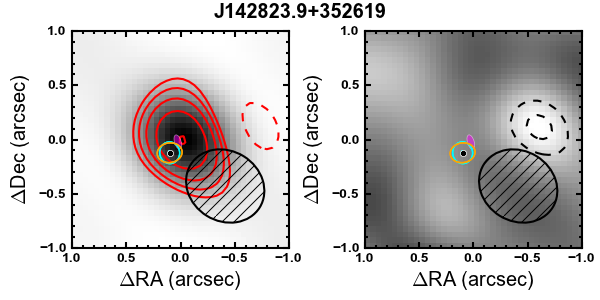}
\includegraphics[width=0.5\textwidth]{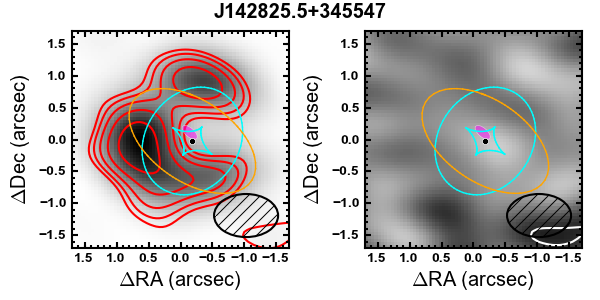}
\includegraphics[width=0.5\textwidth]{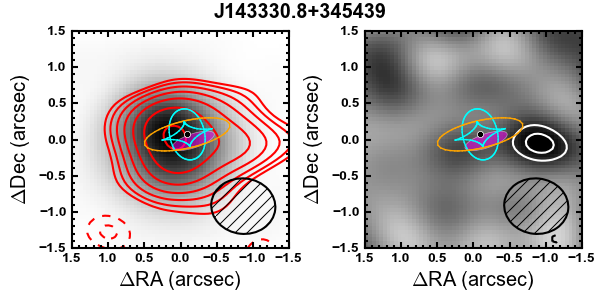}
\end{centering}

\caption{ Continued.  \label{fig:modeling2}}

\end{figure*}

{\bf J021830.5$-$053124:} SMA data from the compact and extended array
configurations were originally
presented in \citet{Ikarashi:2011qy} and \citet{Wardlow:2013lr}.  This paper
presents new data obtained the in the very extended array, permitting the first
resolved measurement of the 880$\, \mu$m emission from this object.  The
best-fit lens model finds a position for the lens that is offset relative to
that indicated in the {\it HST} image by $0\farcs26$ in RA and $-0\farcs11$ in
Dec.  These values are near the expected level of $0\farcs2-0\farcs3$ absolute
astrometric uncertainty between the SMA and {\it HST} reference frames.  The
lensed emission is barely resolved by the SMA due to the small Einstein radius
of the lens.  The bulk of the source-plane emission originates outside the
tangential caustic, favoring a two-image rather than four-image configuration
(excluding the central demagnified image, which is never detected at the
sensitivity levels probed in our data).  This is consistent with the best-fit
magnification factor of $\mu_{880} = 4.4 \pm 1.0$.

{\bf J022016.5$-$060143:} This object is the subject of a detailed study by
\citet{Fu:2013lr}.  We present it here mostly for completeness, but also to
test the validity of the model in \citet{Fu:2013lr} using the visibility-plane
lens modeling technique described in this paper.  We use a two-component lens
model and a three-component source model to reproduce the observed 880$\, \mu$m
emission.  We enforce a minimum Einstein radius of $\theta_{\rm E} > 0\farcs2$
for both lenses (corresponding to a minimum mass of $M_{\rm lens} > 1.6 \times
10^{10}\,$M$_\sun$).  This is intended to reproduce the prior on the lens
masses used by \citet{Fu:2013lr} based on the lens stellar masses and an
assumed relation between the stellar mass and dark matter halo mass.  The
best-fit parameters of this model are statistically consistent with those found
by \citet{Fu:2013lr}, despite a very different approach in modeling the data.
In particular, we find a modest magnification factor is appropriate for each of
the three components, with an average total magnification factor of
$\left\langle \mu_{\rm 880\, \mu m} \right\rangle = 1.3 \pm 0.1$.

{\bf J083051.0$+$013224:} The lensed emission has an unusual configuration that
reflects the complexity of the foreground mass distribution due to the presence
of a secondary lensing galaxy $<1\arcsec$ northwest of the primary lensing
galaxy.  WHT/ACAM spectroscopy shows that the primary lens is located at
$z_{\rm lens} = 0.626$ based on G-band and Mg absorption features, while the
secondary lens has very faint continuum and an emission line at observed frame
7462$\,$\AA. If this feature is [\ion{O}{2}]3727, then its redshift is $z_{\rm
lens2} = 1.002$.  The best-fit lens model for this system assigns each of the
foreground galaxies approximately equal Einstein radii ($\approx
0\farcs4$).  Given the stated lens and source redshifts, this implies lens
masses of $\approx0.7 \times 10^{11}\,$M$_\sun$ and $1.6 \times 10^{11}
\,$M$_\sun$, respectively.  

{\bf J084933.4$+$021443:} This object is the subject of a detailed study by
\citet{Ivison:2013fk}.  It is presented here for completeness and to ensure
that the use of a visibility-plane lens model provides the same results as
given in \citet{Ivison:2013fk}.  Note that the center and extent of the image
cutout shown in Figure~\ref{fig:modeling} has been adjusted from the SMA
emission centroid used in Figure~\ref{fig:imaging} to include only the lens and
the lensed source \citep[dubbed ``T'' in][]{Ivison:2013fk}.  This is done
simply to facilitate the comparison of model and data.  An additional two
sources used in the fitting process are not shown in this diagram.  These two
unlensed sources are modeled using the same visibility method but assuming no
magnification by foreground objects.  We find evidence for a larger
magnification factor for ``T'' than \citet{Ivison:2013fk}: $\mu_{880} = 2.8 \pm
0.2$ instead of $\mu_{880} = 1.5 \pm 0.2$.  A 2$\sigma$ emission peak located
just northwest of the lens can be seen in Figure~\ref{fig:modeling}.  If real,
this emission peak supports the notion of a higher magnification factor for
this object.  Overall, however, our results are in broad agreement with those
of \citet{Ivison:2013fk}.

{\bf J085358.9$+$015537:}  Although the image separations are small for this
object ($\theta_{\rm E} = 0.553 \pm 0.004 \arcsec$), the S/N is high. The
source appears to lie very close to the caustic (which is itself small due to
the low ellipticity of the lens), implying a high magnification factor
$\mu_{880} = 15.3 \pm 3.5$.

{\bf J090302.9$-$014127:} The image separations are close to the smallest values
found in the SMA subsample ($\theta_{\rm E} = 0.35 \pm 0.02 \arcsec$).  There
has been tentative evidence ($3\sigma$) of CO($J=5-4$) emission from the lens
at $z_{\rm lens} = 0.942 \pm 0.004$ \citep{Lupu:2012ly}, but this has not been
confirmed with subsequent, more sensitive observations that rule out lens
redshifts from 0.922 to 0.944 \citep{Omont:2011fk}.  In the lens model
presented here, the lens is assumed to be an insignificant submm emitter.

{\bf J090311.6$+$003906:} Because of the high S/N and well-separated images of
the background source ($\theta_{\rm E} = 1.52 \pm 0.03 \arcsec$), the lens
model is well-constrained for this object.  However, the image of the residual
visibilities shows emission at the $\pm 3\sigma$ level, possibly an indication
of complexity in the source structure that is not captured by a single S\'ersic
profile.

{\bf J090740.0$-$004200:} The counter image to the northwest of the lens is detected
at the $4\sigma$ level and provides good constraints on the lens model for this
object.

{\bf J091043.1$-$000321:} This is one of a handful of objects with emission at
the $>\pm 2\sigma$ level that can be seen in the surface brightness map made
from the residual visibilities.  An edge-on galaxy located $4\farcs4$ to the
northwest with a position angle of 135$\,^\circ$ east of north could be
responsible for an external shear that has not been included in the lens model.
Alternatively, the residual flux density may reflect a more complicated source
structure than can be represented by our choice of a single S\'ersic profile.

{\bf J091305.0$-$005343:} The best-fit lens models for this object are obtained when
the source is relatively large and most of it is located outside the region in
the source plane that produces multiple images.  For this reason, the best-fit
magnification factor is relatively low ($\mu_{880} = 2.1 \pm
0.3$).

{\bf J091840.8$+$023047:} This object is unique in having no counterpart
within $1\arcsec$ of the 880$\, \mu$m centroid in the {\it HST}/F110W Snapshot
imaging.  There is no obvious morphological signature of lensing based on the
SMA data, so this may be a rare unlensed SMG.  It is not included in
Figure~\ref{fig:modeling} since no lens model is available.  Further
investigation is needed to determine the nature of this object.


{\bf J103826.6$+$581542:} This object was originally presented by
\citet{Wardlow:2013lr}.  We present it here for completeness and to test the
validity of the model in \citet{Wardlow:2013lr} using the visibility-plane lens
modeling technique described in this paper.  Our results for the size ($0.45
\pm 0.18 \arcsec$ vs. $< 0.5 \arcsec$) and magnification factor of the
background source ($7.1 \pm 1.5$ vs.  $5.32^{+1.28}_{-1.06}$) are consistent
with those reported in \citet{Wardlow:2013lr}.  Since a redshift measurement
for the background source remains unavailable, no further analysis is possible
for this object.

{\bf J105712.2$+$565457:} This object was originally presented by
\citet{Wardlow:2013lr}.  We present a slightly modified reduction of this
object here.  Instead of MIRIAD's {\sc mossdi} task, which was used in
\citet{Wardlow:2013lr} to image this object, here we have shifted all of the
visibility datasets to have the same phase center and then used MIRIAD's {\sc
clean} task.  This resulted in slightly improved spatial resolution ($0\farcs68
\times 0\farcs57$ vs. $1\farcs18 \times 0\farcs97$), that is still not
sufficient to distinguish cleanly separated images from a single lensed source.
It is not included in Figure~\ref{fig:modeling} since no lens model is
available.  Further investigation is needed to determine the nature of this
object.

{\bf J105750.9$+$573026:} This object was originally analyzed in a series of
papers reporting its discovery, ISM properties, gas dynamics, and (on the basis
of $K_{\rm p}$-band imaging) lensing geometry \citep{Conley:2011lr,
2011ApJ...733...29S, Riechers:2011uq, Gavazzi:2011lr}.  We present new very
extended array data here and compare the SMA and Keck lens models.  The very
extended array data are not as sensitive as the compact array data, so natural
weighting provides a beamsize of $0\farcs99 \times 0\farcs94$.  This is
insufficient to resolve the individual images of the lensed source seen in the
image from the compact array only data and is further indication that the lens
model is primarily constrained by the compact array data.  Because the S/N and
spatial resolution in the Keck image are superior to those of the SMA image, we
fix the parameters of the lens to match those of the model found by
\citet{Gavazzi:2011lr}.  We find a similar magnification factor at $880 \,
\mu$m compared to $K_{\rm p}$ ($9.2 \pm 0.4$ vs.  $10.9 \pm 0.7$).  The model
has difficulty reproducing the locations of the images seen in the SMA data.
In addition, we find an offset in the position of the lens of $0\farcs54$ in RA
and $0\farcs4$ in Dec.  This represents a $\approx 2-3\sigma$ discrepancy in RA
and may be an indication that some of the assumptions in our model are
over-simplifications.

{\bf J113526.3$-$014605:} There is no counterpart detected in a 15-minute
$K_{\rm s}$ integration with the WHT.  The SMA 880$\, \mu$m image is clearly
resolved, but does not show individual, well-separated images of a lensed SMG.
It is not included in Figure~\ref{fig:modeling} since no lens model is
available.  Further investigation is required to determine whether strong
lensing is occuring in this object.  The main avenues for progress are
higher-spatial resolution submm imaging (e.g., SMA very extended array) and
deeper observations in the optical or near-IR to determine whether or not there
is a lensing galaxy.

{\bf J114637.9$-$001132:} This object was originally presented by
\citet{Fu:2012uq}.  Here, we present new SMA extended array data that resolve
the lensed emission into five striking images of the SMG.  The position and
morphology of the lensed SMG in the source plane reported here are
statistically consistent with those found by \citet{Fu:2012uq}.  The presence
of flux density at the $3\sigma$ level in the map of the residual visibilities
(see Figure~\ref{fig:modeling}) indicates our simple assumptions about the lens
mass model (singular isothermal ellipsoids at the locations of galaxies
identified in the Keck-II/NIRC2-LGSAO and {\it HST}/F110W imaging) and/or the
background source S\'ersic profile may be breaking down.

{\bf J125135.4$+$261457:} The offset between the position of the lens from the lens
model and from the WHT astrometry is $0\farcs01$ in RA and $0\farcs38$ in Dec
compared to the best-fit parameters from the lens model.  The offset in Dec is
larger than expected given the astrometric uncertainty in aligning the SMA and
WHT reference frames.  On the other hand, the lens model correctly predicts the
ellipticity and position angle of the lens potential (no priors were assumed
for the shape of the lens potential), a strong indication that the lens model
is robust.

{\bf J125632.7$+$233625:} The S/N is very high in this object and the images of the
lensed SMG are well-separated, making the parameters of the lens model very
robust.

{\bf J132427.0$+$284452:} This object is the subject of detailed studies by
\citep{George:2013qy} and Fu et al. (in prep.) to explore its dust, gas, and
stellar properties.  In this paper, we use the Keck-II/NIRC2-LGSAO image to
constrain the positions of the lenses and apply our visibility-plane lens
modeling technique to the SMA data.  We find that a two-lens mass model is
needed to reproduce the observed data.  These correspond directly to two
galaxies detected at high significance in the Keck imaging.  However, it should
be noted that there are two nearby clusters detected in the Red-Sequence
Cluster (RSC) survey \citep[RCS~J132427$+$2845.2 at $z=0.997 \pm 0.017$ and
RCS~J132419$+$2844.7 at $z=0.802 \pm 0.018$][]{Gladders:2005qy}.  The centers
of these clusters are uncertain due to a lack of X-ray data and no clear
brightest-cluster galaxy, but the RCS surface density map suggests that
RCS~J132427$+$2845.2 lies only 10$\arcsec$ away from lensed SMG.  Our lens
modeling here does not account for the presence of this cluster.  Furthermore,
the background source is not multiply imaged in the SMA data, so the
constraints on the lens parameters are weak.  The possibility of counter images falling
outside of the SMA primary beam (FWHM of 36$\arcsec$ at 880$\,\mu$m) is low due
to the lack of such a counterpart in the SPIRE maps.

{\bf J132630.1$+$334410:} Two well-separated images of the lensed SMG are
obvious in the SMA data, indicating the source is strongly lensed but is not
among the highest magnification sources.  The lens modeling result is
consistent with this ($\mu_{880} = 4.1 \pm 0.3$), suggesting a robust model
fit.

{\bf J132859.3$+$292317:} The nature of this object is unclear based on
existing data.  It is well-detected and clearly spatially resolved by the SMA
at 880$\, \mu$m.  The problem is the lack of optical or near-IR imaging at a
depth beyond what is achieved in SDSS, where the object is undetected in all
bands.  One plausible interpretation is that the object is lensed by a
relatively low-mass foreground galaxy at intermediate redshift so that it is
undetected in the SDSS images.  Alternatively, it is conceivable that the object
is not lensed and has an intrinsic submm flux density of $S_{880} = 51.8 \pm
2.0\,$mJy.  We consider this latter option unlikely since there are no known SMGs
with intrinsic submm flux densities that high.  However, without additional
data to confirm this intuition, we do not consider this object further in our
analysis.  It is not included in Figure~\ref{fig:modeling} since no lens model
is available.  

{\bf J133008.4$+$245900:} Multiple, well-separated images are detected in the SMA
data, consistent with the relatively large inferred magnification factor from
the lens model ($\mu_{880} = 13.0 \pm 1.5$), suggesting a robust
lens model has been obtained.

{\bf J133649.9$+$291801:} This object is similar to J132859.3$+$292317 in that
there is no significant detection in any of the SDSS optical bands, nor is
deeper optical or near-IR imaging available.  However, the SMA morphology
provides evidence typical of strong lensing.  Here, we assume that the object
is strongly lensed.

{\bf J134429.4$+$303036:} The lensed images are well-separated and
well-detected in the SMA data.  In fact, the S/N is so high that the map of the
residual visibilities reveals emission at the $\pm3\sigma$ level, likely
indicating that some of our model assumptions are over-simplifications.
Nevertheless, the model captures the vast majority of the SMA emission and
therefore provides a fair representation of intrinsic source size and
luminosity.  This object is similar to J125135.4$+$261457 in that the lens
model successfully predicts the ellipticity and position angle of the lens
potential without any non-standard priors placed on these parameters.

{\bf J141351.9$-$000026:} No counter image of this target is detected in the
SMA data, an immediate indication that this object is not strongly lensed.  The
possibility of counter images falling outside of the SMA primary beam (FWHM of
36$\arcsec$ at 880$\,\mu$m) is low due to the lack of such a counterpart in the
SPIRE maps.  The lens modeling confirms this, with $\mu_{880} = 1.8 \pm 0.3$.
The lens is located $7\farcs7$ northeast of a brightest cluster galaxy (BCG).
The lens redshift used here is from the BCG, and in order to derive the mass of
the lens from its $\theta_{\rm E}$ value, we have assumed that it is located at
the same redshift as the BCG.

{\bf J142413.9$+$022303:} This object is the subject of a detailed study by
\citet{Bussmann:2012lr}.  We do not reproduce the results of the previous work here
because it used the same visibility-plane lens modeling technique outlined in
this paper.

{\bf J142823.9$+$352619:} This object was originally discovered in {\it
Spitzer} mid-IR imaging of the Bo\"otes Field \citep{Borys:2006lr} and has
since been the subject of a great many follow-up observations, a thorough
summary of which may be found in \citet{Wardlow:2013lr}.  We present new SMA
extended array data which do not resolve the source, further corroborating the
idea that this object is very small and is not strongly lensed (i.e., $\mu <
2$).  Since we know that there is an intervening galaxy along the line of sight
to the {\it Herschel} source, we impose a minimum Einstein radius for the lens
of $\theta_{\rm E} > 0\farcs1$ (corresponding to a minimum lens mass of $M_{\rm
lens} > 10^{10}\,$M$_\sun$).  The constraints on the lens model are weak.  One
of the few robust claims we can make regarding this source is that it is very
small ($a_{\rm s} < 0\farcs2$).

{\bf J142825.5$+$345547:} This object is the subject of a detailed study by Wardlow
et al. (in prep.).  The lens model suggests the lens is located $0\farcs20$
west of the SMA emission centroid, but the peak of the edge-on spiral seen in
the astrometrically-aligned {\it HST} image occurs $0\farcs25$ east of the SMA
emission centroid.  This is a difference of $0\farcs45$ in RA (the difference
in Dec is insignificant) and is larger than the expected 1$\sigma$ astrometric
uncertainty of 0$\farcs2$.  However, estimating the peak position of the
lensing galaxy is difficult because it is nearly exactly edge-on with a
prominent dust lane and very little bulge.  Moreover, the lensed images are
well-resolved and well-detected in the SMA data, and the lens model correctly
predicts the position angle and approximate ellipticity of the lens without any
priors on these parameters.  For these reasons, we consider the lens model for
this object to be robust.

{\bf J143330.8$+$345439:} This object was originally presented in
\citet{Wardlow:2013lr}.  We present a slightly modified reduction of this
object here in which we shift all of the visibility datasets to have the same
phase center and then use MIRIAD's {\sc clean} task.  This provides slightly
improved spatial resolution that is still insufficient to identify clearly
separated images of a single lensed source.  Nevertheless, we use the
knowledge that spectroscopic redshifts are available for both the lens and
background source to infer that strong lensing is occuring.  We then model this
object using the same set of model parameters as applied to the other objects
in the SMA subsample.  Note that absolute astrometric calibration based on
alignment to existing ground-based imaging is not available in the {\it HST}
image for this object, but the lens model still finds a position for the lens
that is in reasonable agreement (within $0\farcs4$) with that in the {\it HST} image.  

{\bf J144556.1$-$004853:} No lens or source redshift is available for this
object.  The SMA data resolve the 880$\, \mu$m emission, but detect the source
at relatively low S/N (S/N$ < 6$).  Keck-II/NIRC2-LGSAO $K_{\rm s}$-band
imaging detects a counterpart that is located between the resolved components
identified in the SMA data.  This could be a detection of the lens, or it could
simply be an intrinsic component of an unlensed SMG.  Unlike
J132859.3$+$292317, the total 880$\, \mu$m flux density ($S_{880} = 9.0 \pm
2.1\,$mJy) is not unprecedented for unlensed SMGs.  It is not included in
Figure~\ref{fig:modeling} since no lens model is available.  Further
investigation is required to determine if strong lensing is occuring in this
object.

\begin{deluxetable*}{lcccccccccc}
\tabletypesize{\scriptsize} 
\tablecolumns{11}
\tablewidth{0pt}
\tablecaption{Gravitational Lens Model Results: Lens Properties}
\tablehead{
\colhead{} & 
\colhead{$\Delta \alpha_{\rm NIR}$} &
\colhead{$\Delta \delta_{\rm NIR}$} &
\colhead{$\Delta \alpha_{\rm lens}$} &
\colhead{$\Delta \delta_{\rm lens}$} &
\colhead{$\theta_{\rm E}$} &
\colhead{} &
\colhead{$\phi_{\rm lens}$} &
\colhead{} &
\colhead{} 
\\
\colhead{Object} & 
\colhead{($\arcsec$)} &
\colhead{($\arcsec$)} &
\colhead{($\arcsec$)} &
\colhead{($\arcsec$)} &
\colhead{($\arcsec$)} &
\colhead{$\epsilon_{\rm lens}$} &
\colhead{(deg)} &
\colhead{$\chi^2$} &
\colhead{$N_{\rm DOF}$} 
}
\startdata
J021830.5$-$053124        &  $-0.17$  &  $-0.02$ &   $0.09\pm0.04 $  &  $-0.13\pm0.05$ &  $0.44\pm0.02$  &  $0.35\pm0.10$  &  $156\pm18$  & 106027.4 & 114284 \\
J022016.5$-$060143        &    2.1    &    0.6   &   $2.07\pm0.01 $  &  $ 0.63\pm0.01$ &  $0.27\pm0.13$  &  $0.29\pm0.13$  &  $114\pm52$  & 131464.2 & 133819 \\
---                       &  $-4.76$  &  $ 0.3$  &     $-4.76$       &  $0.3$          &  $0.34\pm0.14$  &  $0.30\pm0.18$  &  $ 53\pm54$  &    ---   & ---    \\
J083051.0$+$013224        &  $ 0.18$  &  $ 0.10$ &   $ 0.20\pm0.01$  & $ 0.00\pm0.02$  &  $0.39\pm0.02$  &  $0.43\pm0.05$  &  $123\pm3 $  &  97761.3 & 100561 \\
---                       &    0.54   &   0.595  &       0.54        &   0.595         &  $0.43\pm0.02$  &  $0.25\pm0.07$  &  $ 47\pm 9$  &    ---   & ---    \\
J084933.4$+$021443        &  $-7.90$  &  $-0.60$ &   $ -7.90      $  &  $  -0.60     $ &  $1.41\pm0.04$  &  $0.11\pm0.06$  &  $ 62\pm22$  & 188965.5 & 168376 \\
J085358.9$+$015537        &  $ 0.04$  &  $ 0.02$ &   $ 0.09\pm0.03$  &  $-0.03\pm0.01$ & $0.553\pm0.004$ &  $0.06\pm0.02$  &  $ 70\pm12$  & 160445.4 & 161200 \\
J090302.9$-$014127        &  $ 0.20$  &  $ 0.16$ &   $0.092\pm0.006$ &  $ 0.03\pm0.02$ &  $0.33\pm0.02$  &  $0.39\pm0.07$  &  $ 83\pm 5$  & 124529.4 & 124378 \\
J090311.6$+$003906        &  $ 0.07$  &  $ 0.10$ &   $ 0.13\pm0.02$  &  $ 0.03\pm0.05$ &  $1.52\pm0.03$  &  $0.34\pm0.05$  &  $179\pm 4$  & 198305.2 & 188164 \\
J090740.0$-$004200        &  $-0.03$  &  $ 0.17$ &   $ 0.01\pm0.04$  &  $ 0.03\pm0.04$ &  $0.59\pm0.04$  &  $0.50\pm0.08$  &  $ 44\pm 5$  &  98244.2 &  91314 \\
J091043.1$-$000321        &  $ 0.22$  &  $-0.14$ &   $ 0.05\pm0.02$  &  $-0.18\pm0.01$ &  $0.95\pm0.02$  &  $0.08\pm0.03$  &  $152\pm24$  & 171966.9 & 159664 \\
J091305.0$-$005343        &  $-0.32$  &  $ 0.35$ &   $-0.45\pm0.05$  &  $ 0.38\pm0.08$ &  $0.43\pm0.07$  &  $0.51\pm0.13$  &  $ 43\pm14$  & 211928.5 & 200412 \\
J103826.6$+$581542        &  $-1.80$  &  $ 0.25$ &   $-1.1\pm0.2  $  &  $ 0.27\pm0.09$ &  $2.0 \pm0.2 $  &  $0.54\pm0.11$  &  $ 18\pm 3$  &  78239.2 &  76704 \\
J105750.9$+$573026        &  $-0.65$  &  $-1.19$ &   $-0.11\pm0.05$  &  $-0.80\pm0.08$ &  $3.86\pm0.01$  &  $0.52\pm0.01$  &  $ 14\pm 1$  & 142740.4 & 120764 \\
---                       &    1.7    &     3.4  &       1.7         &       3.4       &  $0.12\pm0.01$  &  $0.54\pm0.11$  &  $170\pm 9$  &    ---   & ---    \\
J114637.9$-$001132        &  $-0.40$  &  $ 0.02$ &   $-0.59\pm0.02$  &  $ 0.02\pm0.02$ &  $0.65\pm0.02$  &  $0.26\pm0.04$  &  $114\pm19$  &  90621.1 &  94863 \\
---                       &   1.154   &   1.270  &       1.154       &   1.270         &  $0.67\pm0.01$  &  $0.50\pm0.04$  &  $ 68\pm 2$  &    ---   & ---    \\
---                       & $-3.076$  & $-1.688$ &     $-3.076$      & $-1.688$        &  $0.61\pm0.03$  &  $0.33\pm0.07$  &  $ 95\pm12$  &    ---   & ---    \\
---                       & $-4.340$  &   0.632  &     $-4.340$      &   0.632         &  $0.51\pm0.04$  &  $0.50\pm0.10$  &  $ 77\pm 6$  &    ---   & ---    \\
J125135.4$+$261457        &  $ 0.04$  &  $ 0.13$ &   $ 0.05\pm0.04$  &  $-0.25\pm0.09$ &  $1.02\pm0.03$  &  $0.46\pm0.06$  &  $122\pm1 $  &  72453.5 &  62912 \\
J125632.7$+$233625        &  $ 0.11$  &  $ 0.25$ &   $ 0.00\pm0.01$  &  $ 0.25\pm0.01$ &  $0.68\pm0.01$  &  $0.31\pm0.03$  &  $123\pm1 $  &  62465.3 &  45880 \\
J132427.0$+$284452        &  3.09     &  5.22    &   3.09            &  5.22           &  $1.7 \pm0.4 $  &  $0.34\pm0.14$  &  $ 81\pm16$  & 228617.0 & 213957 \\
---                       & $-7.51$    & $-9.66$ &  $-7.51$          & $-9.66$         &  $2.2 \pm0.3 $  &  $0.14\pm0.09$  &  $ 88\pm15$  &    ---   & ---    \\
J132630.1$+$334410        &  $ 0.68$  &  $-0.78$ &   $ 0.53\pm0.05$  &  $-0.54\pm0.05$ &  $1.80\pm0.02$  &  $0.26\pm0.04$  &  $ 66\pm4 $  &  43605.2 &  38964 \\
J133008.4$+$245900        &  $-0.24$  &  $-0.02$ &   $-0.14\pm0.03$  &  $-0.13\pm0.02$ &  $0.88\pm0.02$  &  $0.52\pm0.03$  &  $ 81\pm1 $  &  71948.3 &  52744 \\
J133649.9$+$291801        &   ---     &   ---    &   $ 0.00\pm0.11$  &  $ 0.03\pm0.16$ &  $0.40\pm0.03$  &  $0.38\pm0.14$  &  $120\pm13$  & 100742.9 &  95844 \\
J134429.4$+$303036        &  $-0.03$  &  $-0.10$ &   $-0.03\pm0.02$  &  $-0.01\pm0.02$ &  $0.92\pm0.02$  &  $0.39\pm0.06$  &  $172\pm14$  & 182936.2 & 171864 \\
J141351.9$-$000026        &  $-0.32$  &  $-2.65$ &   $-0.32\pm0.03$  &  $-2.50\pm0.03$ &  $1.13\pm0.09$  &  $0.31\pm0.12$  &  $118\pm33$  & 132668.6 & 123420 \\
J142413.9$+$022303        &  $-0.11$  &  $ 0.79$ &   $-0.27\pm0.03$  &  $ 0.63\pm0.03$ &  $0.57\pm0.01$  &  $0.29\pm0.01$  &  $ 62\pm 1$  & 108899.5 & 144191 \\
---                       &  0.025    & $-0.327$ &   0.025           & $-0.327$        &  $0.40\pm0.01$  &  $0.06\pm0.02$  &  $133\pm14$  &    ---   &  ---   \\
J142823.9$+$352619        &  $-0.14$  &  $-0.02$ &   $-0.01\pm0.10$  &  $ 0.00\pm0.11$ &  $0.10\pm0.03$  &  $0.36\pm0.18$  &  $ 87\pm40$  &  17284.0 &  17036 \\
J142825.5$+$345547        &  $ 0.25$  &  $-0.05$ &   $-0.20\pm0.03$  &  $-0.04\pm0.03$ &  $0.77\pm0.03$  &  $0.46\pm0.06$  &  $ 56\pm 5$  &  93163.7 &  75448 \\
J143330.8$+$345439        &  $ 0.00$  &  $ 0.05$ &   $-0.07\pm0.03$  &  $ 0.07\pm0.02$ &  $0.28\pm0.02$  &  $0.59\pm0.08$  &  $104\pm 7$  &  91554.9 & 117732 \\
\enddata
\label{tab:lensesresults}
\end{deluxetable*}

\begin{deluxetable*}{lccccccc}
\tabletypesize{\scriptsize} 
\tablecolumns{8}
\tablecaption{Gravitational Lens Model Results: Source Properties}
\tablehead{
\colhead{} & 
\colhead{$\Delta \alpha_{\rm s}$} & 
\colhead{$\Delta \delta_{\rm s}$} &
\colhead{} & 
\colhead{$a_{\rm s}$} & 
\colhead{} &
\colhead{$\phi_{\rm s}$} &
\colhead{}
\\
\colhead{IAU Name} & 
\colhead{($\arcsec$)} & 
\colhead{($\arcsec$)} &
\colhead{$n_{\rm s}$} & 
\colhead{($\arcsec$)} & 
\colhead{$\epsilon_{\rm s}$} &
\colhead{(deg)} &
\colhead{$\mu_{\rm 880\, \mu m}$}
}
\startdata
J021830.5$-$053124      &  $-0.08\pm0.03$  & $ 0.22\pm0.04$  &  $2.1\pm0.9$  &  $0.33\pm0.12$  & $0.29\pm0.13$ & $ 82\pm22$ & $ 4.4\pm1.0$ \\
J022016.5$-$060143      &  $-1.54\pm0.02$  & $ 0.65\pm0.02$  &  $2.1\pm0.8$  &  $0.16\pm0.05$  & $0.25\pm0.10$ & $ 68\pm34$ & $ 1.5\pm0.3$ \\
---                     &  $-2.56\pm0.02$  & $-0.82\pm0.07$  &  $1.7\pm0.8$  &  $0.30\pm0.07$  & $0.40\pm0.13$ & $ 45\pm53$ & $ 1.2\pm0.1$ \\
---                     &  $-2.61\pm0.02$  & $-2.05\pm0.02$  &  $1.4\pm0.7$  &  $0.28\pm0.06$  & $0.19\pm0.09$ & $ 72\pm44$ & $ 1.2\pm0.0$ \\
J083051.0$+$013224      &  $-0.25\pm0.01$  & $ 0.08\pm0.02$  &  $1.8\pm0.3$  &  $0.14\pm0.01$  & $0.32\pm0.06$ & $ 20\pm 7$ & $ 6.9\pm0.6$ \\
J084933.4$+$021443      &  $ 0.67\pm0.03$  & $-0.73\pm0.03$  &  $2.1\pm0.7$  &  $0.15\pm0.03$  & $0.20\pm0.12$ & $103\pm35$ & $ 2.8\pm0.2$ \\ 
---                     &  $11.3$          &  $ 3.0$         &  $0.4\pm0.3$  &  $0.25\pm0.02$  & $0.33\pm0.10$ & $ 24\pm18$ & $1.0$        \\ 
---                     &  $14.3$          &  $ 4.0$         &  $2.0\pm0.9$  &  $0.17\pm0.07$  & $0.43\pm0.17$ & $ 93\pm34$ & $ 1.0   $    \\ 
J085358.9$+$015537      &  $-0.09\pm0.03$  & $0.001\pm0.002$ &  $2.0\pm0.7$  &  $0.06\pm0.01$  & $0.33\pm0.14$ & $ 83\pm17$ & $15.3\pm3.5$ \\
J090302.9$-$014127      &  $-0.06\pm0.01$  & $-0.01\pm0.01$  &  $2.5\pm0.6$  &  $0.42\pm0.12$  & $0.22\pm0.12$ & $116\pm23$ & $ 4.9\pm0.7$ \\
J090311.6$+$003906      &  $-0.26\pm0.03$  & $ 0.04\pm0.05$  &  $2.2\pm0.4$  &  $0.52\pm0.10$  & $0.35\pm0.06$ & $ 94\pm11$ & $11.1\pm1.1$ \\
J090740.0$-$004200      &  $-0.15\pm0.03$  & $ 0.14\pm0.03$  &  $2.0\pm1.1$  &  $0.16\pm0.10$  & $0.31\pm0.14$ & $ 73\pm57$ & $ 8.8\pm2.2$ \\
J091043.1$-$000321      & $-0.017\pm0.008$ & $ 0.23\pm0.02$  &  $1.5\pm0.5$  &  $0.13\pm0.02$  & $0.35\pm0.12$ & $  5\pm20$ & $10.9\pm1.3$ \\
J091305.0$-$005343      &   $0.42\pm0.05$  & $-0.37\pm0.06$  &  $3.0\pm0.6$  &  $0.76\pm0.12$  & $0.56\pm0.13$ & $129\pm10$ & $ 2.1\pm0.3$ \\
J103826.6$+$581542      &    $1.1\pm0.2$   & $-0.33\pm0.06$  &  $3.0\pm0.7$  &  $0.45\pm0.18$  & $0.44\pm0.19$ & $129\pm32$ & $ 7.1\pm1.5$ \\
J105750.9$+$573026      & $ -0.07\pm0.04$  &  $0.76\pm0.06$  &  $2.4\pm0.8$  &  $0.57\pm0.08$  & $0.58\pm0.11$ & $122\pm6$  & $ 9.2\pm0.4$ \\
J114637.9$-$001132      &  $-0.50\pm0.06$  & $-0.28\pm0.04$  &  0.5\tnm{a}   &  $0.38\pm0.03$  & $0.77\pm0.03$ & $107\pm4$  & $ 9.5\pm0.6$ \\
J125135.4$+$261457      &  $ 0.02\pm0.04$  & $ 0.22\pm0.09$  &  $1.9\pm0.6$  &  $0.15\pm0.03$  & $0.22\pm0.10$ & $109\pm32$ & $11.0\pm1.0$ \\
J125632.7$+$233625      &  $0.014\pm0.006$ & $-0.12\pm0.01$  &  $1.1\pm0.6$  &  $0.07\pm0.01$  & $0.37\pm0.14$ & $140\pm21$ & $11.3\pm1.7$ \\
J132427.0$+$284452      &  $-3.8 \pm0.4$   & $-5.1 \pm0.6 $  &  $2.4\pm0.4$  &  $0.72\pm0.09$  & $0.69\pm0.01$ & $169\pm17$ & $ 2.8\pm0.4$ \\
J132630.1$+$334410      &  $-0.60\pm0.03$  & $ 0.60\pm0.03$  &  $1.1\pm0.3$  &  $0.22\pm0.02$  & $0.18\pm0.09$ & $150\pm15$ & $ 4.1\pm0.3$ \\
J133008.4$+$245900      &   $0.05\pm0.01$  & $ 0.23\pm0.02$  &  $1.6\pm0.7$  &  $0.09\pm0.03$  & $0.37\pm0.10$ & $129\pm29$ & $13.0\pm1.5$ \\
J133649.9$+$291801      &  $-0.04\pm0.09$  & $-0.05\pm0.15$  &  $1.2\pm0.5$  &  $0.19\pm0.03$  & $0.43\pm0.12$ & $125\pm13$ & $ 4.4\pm0.8$ \\
J134429.4$+$303036      &   $0.22\pm0.02$  & $ 0.04\pm0.02$  &  $1.9\pm0.5$  &  $0.24\pm0.06$  & $0.39\pm0.07$ & $100\pm15$ & $11.7\pm0.9$ \\
J141351.9$-$000026      &  $ 0.13\pm0.13$  & $ 1.40\pm0.09$  &  $1.5\pm0.5$  &  $0.30\pm0.04$  & $0.48\pm0.12$ & $ 62\pm 7$ & $ 1.8\pm0.3$ \\
J142413.9$+$022303      &  $-0.24\pm0.03$  & $-0.53\pm0.03$  &  $2.9\pm0.3$  &  $0.64\pm0.07$  & $0.27\pm0.09$ & $ 77\pm12$ & $ 4.6\pm0.5$ \\
J142823.9$+$352619      &  $-0.00\pm0.07$  & $ 0.03\pm0.08$  &  $1.9\pm1.2$  &  $0.10\pm0.07$  & $0.33\pm0.18$ & $ 64\pm62$ & $ 3.0\pm1.5$ \\
J142825.5$+$345547      &  $ 0.05\pm0.01$  & $ 0.15\pm0.03$  &  $0.7\pm0.4$  &  $0.16\pm0.04$  & $0.50\pm0.09$ & $ 49\pm10$ & $10.3\pm1.7$ \\
J143330.8$+$345439      &  $-0.08\pm0.02$  & $-0.07\pm0.02$  &  $1.9\pm0.7$  &  $0.31\pm0.06$  & $0.55\pm0.14$ & $116\pm13$ & $ 4.5\pm0.4$ \\
\tnt{a}{$n_{\rm s} = 0.5$ was assumed for this source.}
\enddata
\label{tab:sourcesresults}
\end{deluxetable*}

\section{Properties of Lenses Discovered by {\it
Herschel}}\label{sec:lensingresults}

\begin{figure*}[!tbp] 
\centering
\includegraphics[trim = 0mm 0mm -4mm 0mm, width=0.48\linewidth]{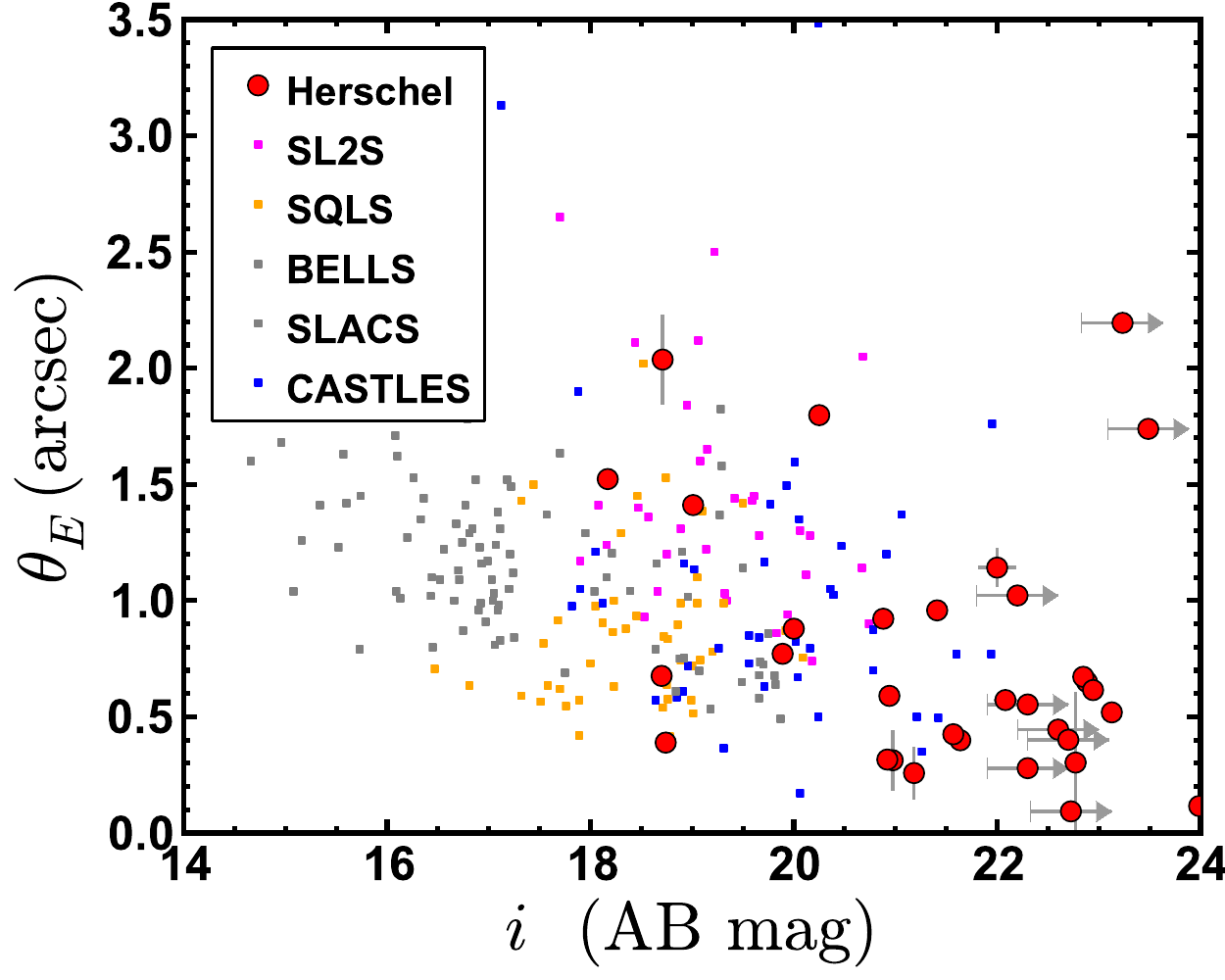}
\includegraphics[trim = -4mm 0mm 0mm 0mm, width=0.48\linewidth]{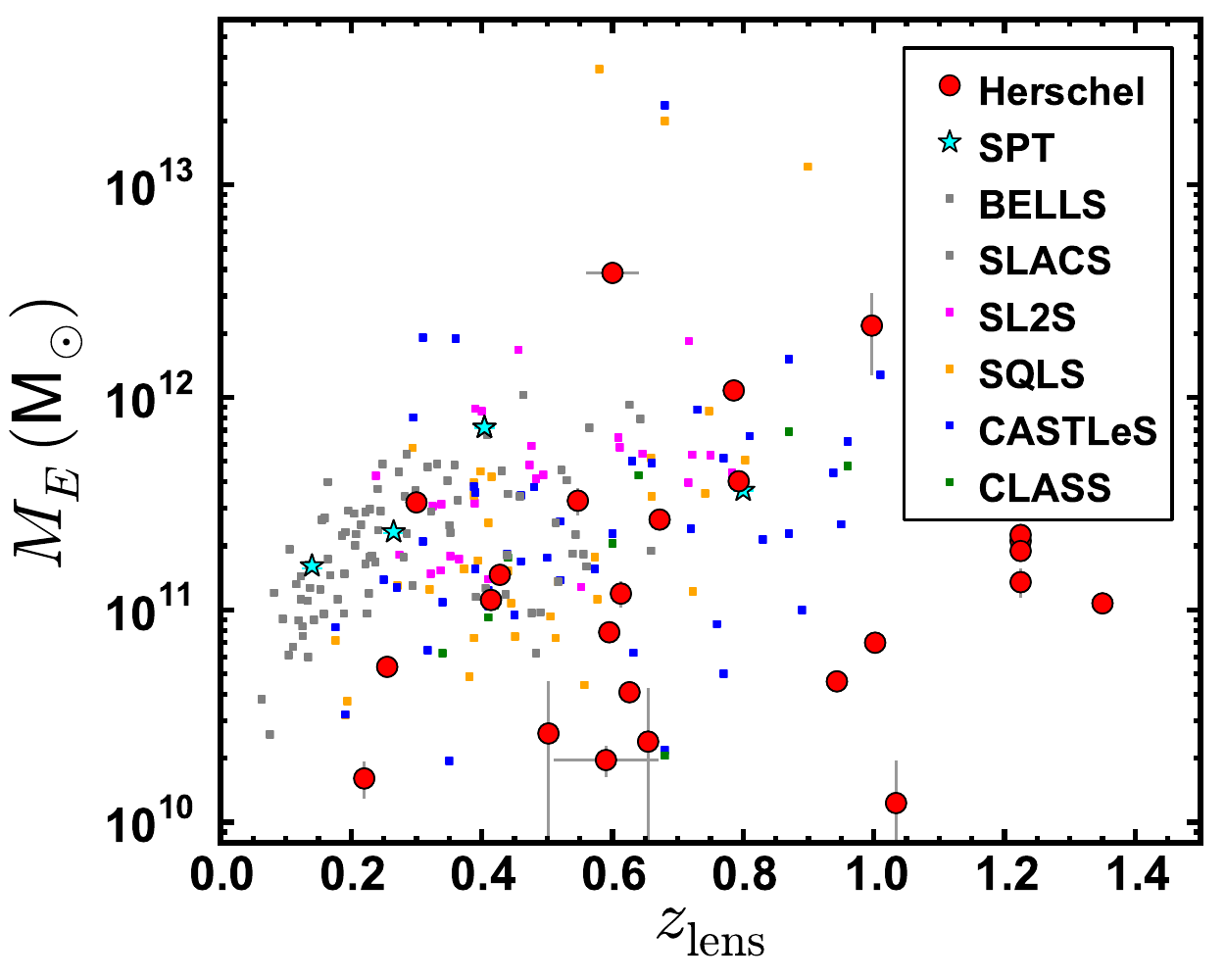}

\caption{ Properties of lenses discovered by {\it Herschel} (red circles) and
SPT (cyan stars), compared with a complilation of lenses from BELLS and SLACS
(gray squares), SL2S (magenta squares), CASTLES (blue squares), CLASS (green
squares), and SQLS (yellow squares).  {\it Left}: Einstein radius ($\theta_{\rm
E}$) as a function of $i$-band AB magnitude.  The {\it Herschel} sample is
fainter and shows a wider range in $\theta_{\rm E}$ values than any of the
previous samples of lenses.  {\it Right}: Mass enclosed within $\theta_E$ as a
function of lens redshift.  {\it Herschel} has identified lenses that are lower
in mass or higher in redshift than any of the optically-based searches (SLACS,
BELLS, and SL2S).  The range in parameter space occupied by the {\it Herschel}
data points is comparable to that of CASTLES, CLASS, and SQLS, but {\it
Herschel} promises to provide a sample size that is over an order of magnitude
larger \citep{Gonzalez-Nuevo:2012lr}.  \label{fig:lensingresults}}

\end{figure*}

Two of the most basic properties of strong gravitational lensing galaxies are
their $i$-band magnitudes (tracing the stellar light emitted by the lens, since
the background sources detected by {\it Herschel} are dust-obscured and
high-redshift, and therefore faint in the optical) and Einstein radii.  The
left panel of Figure~\ref{fig:lensingresults} shows these values for the
objects in the SMA subsample with robust lens models and compares them to other
strong lenses found in SLACS \citep{Bolton:2008wd}, BELLS
\citep{Brownstein:2012rt}, CASTLeS \citep{Munoz:1998mz}, and CLASS
\citep{Myers:2003lr, Browne:2003lr}.  

In this figure, the $i$-band magnitudes for the SMA subsample come from SDSS
DR9 (see section~\ref{sec:opticalimaging} for details).  We account for
multiple lens systems by assigning a fraction of the total SDSS $i$-band flux
density to each lens.  The appropriate fraction is determined from the ratio of
the $\theta_{\rm E}$ for that lens to the sum of the $\theta_{\rm E}$ values
for that system.  This explains why a few objects appear in the left panel of
Figure~\ref{fig:lensingresults} to have $i$-band magnitudes below the SDSS
limit of $i \approx 22.5$ (AB mag).

Figure~\ref{fig:lensingresults} helps clarify the observational distinctions
between lenses discovered via wide-field (sub)mm surveys (i.e., {\it Herschel}
and SPT) and optically-selected surveys like SLACS, BELLS, SL2S, and SQLS.
Note that the CASTLeS and CLASS samples of lenses have similar properties to
the {\it Herschel} and SPT lenses, but will not grow further in size.  Although
BELLS and SL2S go much deeper in the $i$-band than SLACS, they are still biased
towards brighter lenses than the source-selected surveys by the need for
detections in SDSS optical spectroscopy.  Even SQLS, which does not require
optical spectroscopic detections from SDSS, has lenses with brighter $i$-band
magnitudes than the source-selected samples.  Indeed, 10 out of 30 objects in
the SMA subsample are undetected in SDSS imaging, likely indicating that the
SMA subsample probes higher redshifts or lower lens masses than any of the
previous surveys.  There is very little overlap in the observational properties
of {\it Herschel}-selected lenses and SLACS or BELLS lenses, indicating that
the two techniques are highly complementary.

With only a modest investment of observing time ($\approx 1$~hour on-source per
target) with 4-6$\,$m class optical telescopes, it is possible to measure
spectroscopic redshifts for most of the lenses in the SMA subsample (see
sections~\ref{sec:mmtobs} and \ref{sec:geminiobs} for details).  We use the
standard equations from \citet{Schneider:1992fk} to compute the mass inside
$\theta_{\rm E}$ for the lensing galaxies, $M_{\rm E}$.  These values are shown
for the SMA subsample as a function of lens redshift, $z_{\rm lens}$, in the
right panel of Figure~\ref{fig:lensingresults}.  This plot emphasizes the new
range in parameter space that is probed by {\it Herschel}-selected lenses
compared to SDSS-based lens searches: high redshift ($z_{\rm lens} > 0.6$) and
low mass ($M_{\rm E} < 10^{11} \; $M$_\sun$).  The distinction in redshift
confirms the evidence based on photometric redshifts presented by
\citet{Gonzalez-Nuevo:2012lr}.  The objects from the literature with the most
overlap with the SMA subsample are those from CASTLeS, CLASS, and SQLS---as
expected, since these are source-selected samples of lenses.  It should be
noted that this comparison is not entirely fair to the {\it Herschel}-selected
lenses because the SMA subsample is missing lens redshifts for optically-faint
targets that are likely to have lower mass or lie at higher redshift ($z_{\rm
lens} \gtrsim 1$).  A handful of lenses identified by the SPT
\citep{Hezaveh:2013fk} are also shown in this diagram, and are likely to have
ensemble properties similar to those of the {\it Herschel}-selected lenses once
statistically significant sample sizes are available.

Lenses discovered by {\it Herschel} have relatively high ellipticities compared
to lenses selected from optical surveys.  We measure a median ellipticity of
$\epsilon_{\rm lens} = 0.35 \pm 0.15$.  In comparison, \citet{Brewer:2012lr}
study a subset of the SLACS sample where the foreground deflector has an
inclined disk and measure a median ellipticity of $\epsilon_{\rm lens} = 0.39
\pm 0.07$.  There is a theoretical basis for why such an effect could occur:
optical surveys for lenses might miss a large segment of highly inclined lenses
due to dust obscuration by the foreground deflector
\citep[e.g.,][]{Bartelmann:1998qy, Blain:1999fr}.  A submm survey for lenses
(like ours) is not affected by this limitation.  However, our models do not
include the effect of shear, which has a well-known degeneracy with lens
ellipticity \citep[e.g.,][]{Keeton:1997ys}.  We therefore view this as an
interesting line of research for further study and urge caution when readers
consider this result.  

As a final note on the properties of the lensing galaxies discovered by {\it
Herschel}, it is worth emphasizing the sample size at present and how large
it might grow in the future.  The SMA subsample consists of a subset of 30
candidate lensed SMGs selected from 104 objects with $S_{500} >
100\;$mJy within $\approx 400\;$deg$^2$ of wide-field {\it Herschel} surveys.
When the {\it Herschel} catalogs are complete, a total of $\approx
1000\;$deg$^2$ of sky will be surveyed and should provide a sample of $\approx
250$ lens candidates.  This is comparable to the expected number of lensed SMGs
found by the full SPT survey \citep{Vieira:2013fk}, but is already a factor of
$\approx 5$ larger than other source-selected or heterogenous surveys such as
CLASS or CASTLeS.  It is also comparable in size to SLACS and the initial release of
strong lenses from BELLS.  


\section{Intrinsic Properties of Lensed SMGs Discovered by {\it Herschel}}\label{sec:lensedresults}

In this section, we focus on the intrinsic properties of the SMA subsample of
lensed SMGs.  We begin by discussing the size bias inherent to samples of
strongly lensed galaxies.  We then compare our magnification measurements with
statistical predictions.  Next, we describe our methodology for measuring dust
temperatures, intrinsic (i.e., unlensed) FIR luminosities and FIR luminosity
surface densities for the SMA subsample by combining modified blackbody fitting
of the SPIRE and SMA photometry with the magnification factors and intrinsic
source sizes predicted by our lens models.  Throughout, we define the $L_{\rm
FIR}$ as integrated over $40 - 120\,\mu$m in the rest-frame \citep[studies
indicate a bolometric correction factor from $L_{\rm FIR}$ to $L_{\rm IR}$ of
1.91 is typical; e.g.][]{Dale:2001fj}.

\subsection{Size Bias in the SMA Subsample of Lensed SMGs}\label{sec:sizebias} 

Strong gravitational lensing permits the study of SMGs at higher spatial
resolutions than would otherwise be possible \citep[e.g., a highly magnified
SMG at $z=2$ has been studied at 100$\,$pc resolution with the
SMA;][]{Swinbank:2010lr}.  While this is a highly attractive feature of strong
lensing, it does necessitate certain unique considerations when transferring
conclusions regarding lensed SMGs to the unlensed population.  Chief amongst
these considerations is the size bias inherent in any flux-limited sample of lensed
galaxies.  This bias has been investigated in a quantitative manner by a number
of authors \citep{Serjeant:2012lr, Hezaveh:2012fk, Wardlow:2013lr}, who find
that those objects with the brightest apparent flux densities should also have
higher magnification factors and smaller sizes, on average.

Objects with high magnification factors are preferentially selected to have
small sizes by flux-limited surveys like those of {\it Herschel} and SPT.  This
is because the degree of magnification depends primarily on the fraction of the
source that is close to the caustic.  A source that is extended relative to the
size of the caustic will inevitably have a significant fraction of its flux
density emitted in a region that is not highly magnified, so the total
magnification factor summed over the entire source is not critically dependent
on the exact location of the source relative to the caustic.  Conversely, a
population of lensed sources which are intrinsically compact will have a
bimodal distribution of magnification factors that depends primarily on how far
from the caustic the source is located.  If a source is not highly magnified,
it will likely not be bright enough to appear in our sample.

\begin{figure}[!tbp] 
\begin{centering}
\includegraphics[width=\linewidth]{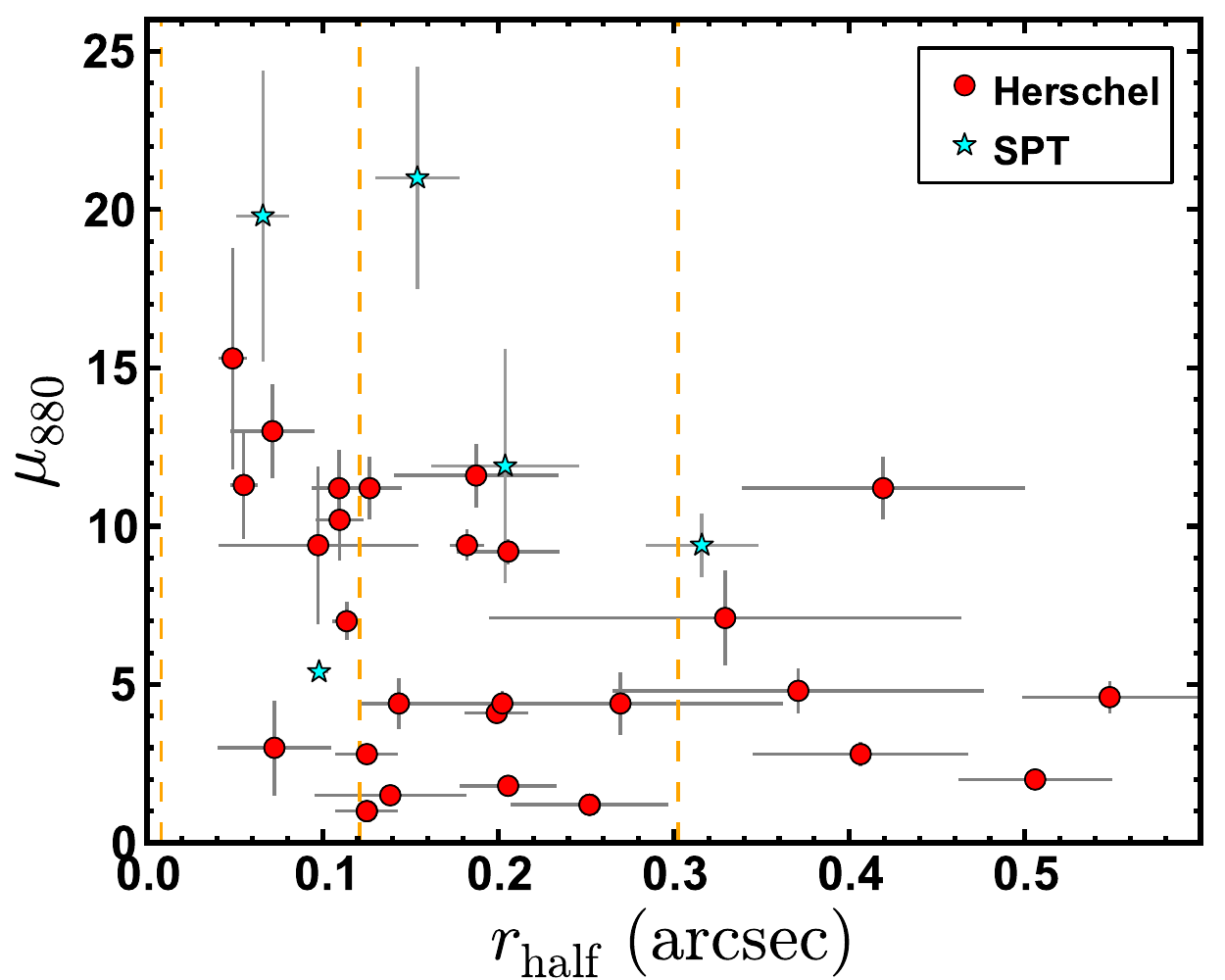}
\end{centering}

\caption{ Magnification factor at 880$\,\mu$m as a function of half-light
radius for lensed SMGs discovered by {\it Herschel} (red circles) and SPT (cyan
stars).  The most highly magnified sources are also the smallest, consistent
with expectations from theoretical models \citep{Serjeant:2012lr,
Hezaveh:2012fk}.  Vertical dashed lines represent maximum achievable spatial
resolution at 880$\, \mu$m for unlensed sources with (from right to left) SMA,
ALMA Cycle~1, and full ALMA.  In some cases, strong lensing permits the SMA to
resolve sources that would otherwise require baseline lengths of $>
10\,$km (i.e., full ALMA).  \label{fig:mu_rhalf}}

\end{figure}

Figure~\ref{fig:mu_rhalf} demonstrates this degeneracy between size (shown as
half-light radius or $r_{\rm half}$, where $r_{\rm half} = a_{\rm s}
\sqrt{(1-e_{\rm s})}$) and magnification factor ($\mu_{880}$) for the SMA
subsample and a handful of objects from SPT \citep{Hezaveh:2013fk}.  Our
determinations of these values for the SMA subsample are presented in
Table~\ref{tab:sourcesresults}.  Nearly every lensed SMG with $r_{\rm half} <
0\farcs1$ is associated with $\mu_{880} > 10$.  However, there are a surprising
number of sources with $0\farcs1 < r_{\rm half} < 0\farcs2$ and relatively
modest magnification factors of $3 < \mu_{880} < 10$.  We will return to the
implications of the large number of low-magnification objects in
Section~\ref{sec:statistics}.  

For a source at $z=2$, these sizes correspond to a physical scale of $\approx
1$kpc, which is at the low end of sizes measured for unlensed SMGs
\citep{Tacconi:2006lr}.  It must be noted, however, that there is a subtle bias
when comparing the sizes of lensed and unlensed SMGs: if intrinsically low
surface brightness regions are preferentially located farther from the caustic
than high surface brightness regions (as expected for the reasons outlined
above), then we expect the lensed SMGs to show smaller sizes than unlensed
SMGs, because differential lensing makes detecting the low surface brightness
regions more difficult than in the unlensed scenario.

Definitive measurements of the relative bias in the size measurements of lensed
and unlensed SMGs will require spatially resolved observations of unlensed
SMGs.  The dashed vertical lines in Figure~\ref{fig:mu_rhalf} illustrate both
the advantages offered by lensing and the difficulties that must be overcome to
assemble a statistically significant sample of unlensed SMGs with spatially
resolved imaging.  From right to left, the three lines indicate the maximum
spatial resolutions available with the SMA, Cycle~1 ALMA, and
full ALMA.  It is only with the full ALMA and baselines $> 10\;$km that
spatial resolution better than $0\farcs1$ can be achieved at these
wavelengths---i.e., matching the best the SMA can do today for lensed
SMGs discovered by {\it Herschel}.


\subsection{Testing Predictions Derived from Lens
Statistics}\label{sec:statistics}

The number counts of unlensed SMGs fall off dramatically at the bright end of
the luminosity function \citep[e.g.,][]{Barger:1999rt, Coppin:2006lr,
2010A&A...518L..21O, Clements:2010fk}.  This is the central reason why
wide-field surveys at (sub-)mm wavelengths are useful tools for discovering
strongly lensed galaxies \citep[e.g.][]{1996MNRAS.283.1340B}.  There are
several key elements of astrophysical interest in models which predict the
magnification factor as a function of (sub-)mm flux density for strongly lensed
galaxies found in wide-field (sub-)mm surveys.  These are discussed in detail
elsewhere \citep{2002MNRAS.329..445P, 2007MNRAS.377.1557N, Paciga:2009uq,
Hezaveh:2011kx, Wardlow:2013lr}, so we provide only the briefest of summaries
here.  In short, they are the lens mass profile (typically assumed to match the
analytical form of Navarro-Frenk-White \citep[NFW;][]{Navarro:1997ys} or a
singular isothermal sphere) and the number densities of lenses and (unlensed)
sources as functions of mass and redshift.

\begin{figure}[!tbp] 
\includegraphics[width=\linewidth]{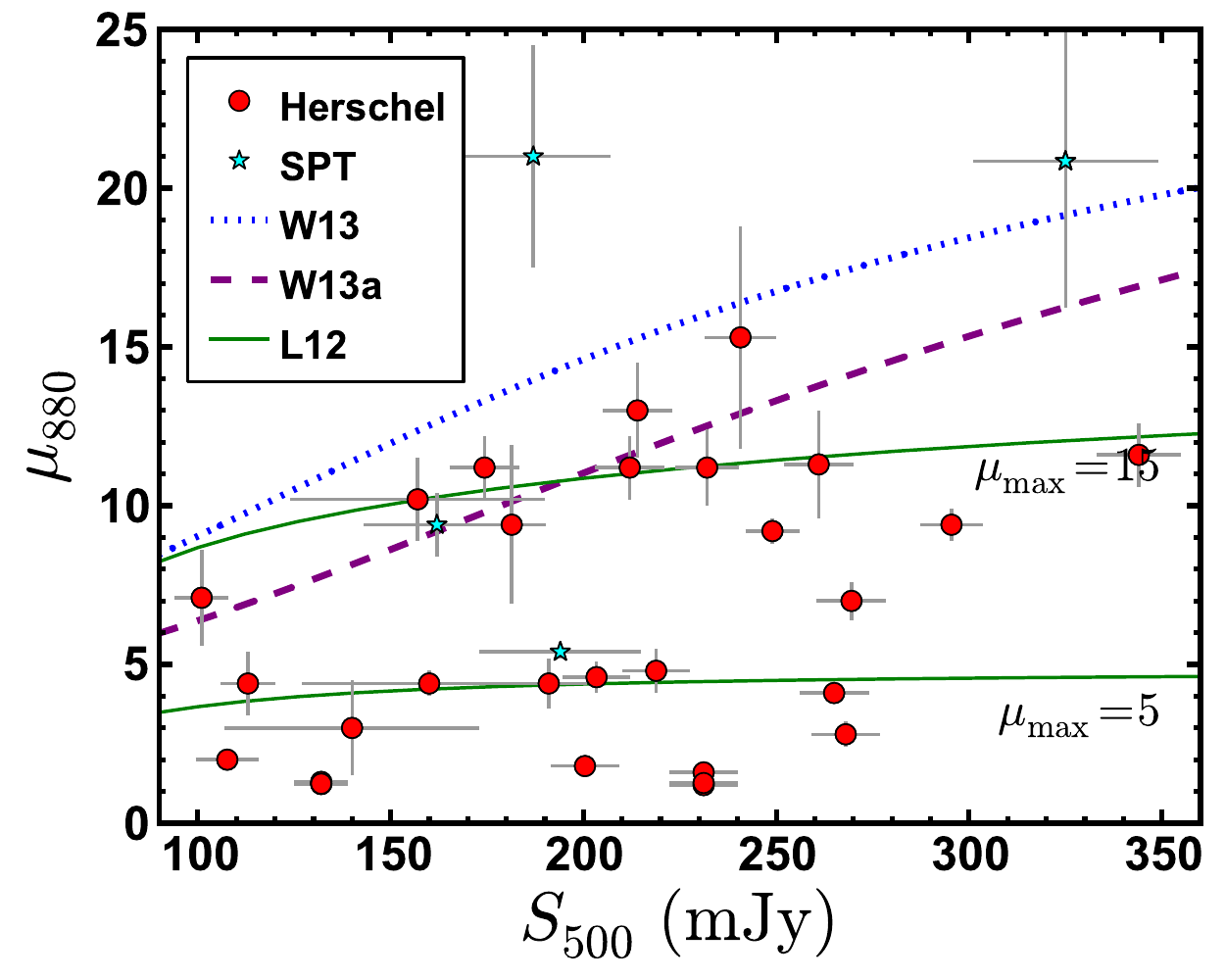}

\caption{Magnification factor from the SMA lens model as a function of 500$\,
\mu$m flux density.  The prediction for these values from
\citet{Wardlow:2013lr} is shown by the dotted blue line, whereas the dashed
purple line traces the same model, but with parameters tuned to jointly match
the observed number counts and magnification factors shown in this diagram.
The solid green lines show the effects of different maximum magnifications and
are taken from \citet{Lapi:2012kx}.} \label{fig:mu_f500}

\end{figure}

Figure~\ref{fig:mu_f500} shows the magnification factor as a function of the
$500\, \mu$m flux density for each strongly lensed SMG in the SMA subsample.
Recall that we are complete for $S_{500} > 170\,$mJy (see
Section~\ref{sec:select}), except for one object which is the subject of a
paper by Messias et al., (in prep.).  The blue line is taken from
\citet{Wardlow:2013lr} and represents the predicted mean $\mu$ as a function of
$S_{500}$ for a complete sample of strongly lensed SMGs ($\mu > 2$).  There are
far more low-$\mu_{880}$ objects than expected based on the model.  In fact,
only two objects (J085358.9$+$015537 and J142825.5$+$345547) have $\mu_{880}$
values that are consistent at the 1$\sigma$ level with the model predictions.

It is not presently clear why the model over-predicts the magnification factors
at a given $S_{500}$ value.  One possibility is that our assumption of a
single, smooth S\'ersic profile for the background source leads to
underestimates in some cases of the magnification factor.  For example, one
might imagine that a multi-component, clumpy model for the source morphology
could reproduce the observed data while yielding larger magnification factors
on average.  Testing such models is beyond the scope of this paper, but we 
acknowledge that this possibility exists.

If the discrepancy between model and data is not simply a product of limited data
quality, there still exist several possible explanations.  Investigation into
the dependence of the predicted mean magnification on the various model
parameters shows that this prediction is sensitive to a number of factors.  The
first of these is the shape of the intrinsic SMG number counts.  The intrinsic
counts are not well constrained at the bright end, primarily because of the
contribution from lensed SMGs \citep{Wardlow:2013lr}.  Modest changes in the
parameters of the Schechter function used to characterize the counts have a
significant effect on the predicted mean magnification factor.  This can be
seen by a comparison of the dotted blue line in Figure~\ref{fig:mu_f500}, which
traces the predicted $\mu_{880} (S_{500})$ curve from \citet{Wardlow:2013lr}
based on the observed number counts of SMGs, and the dashed purple line, which
shows $\mu_{880} (S_{500})$ found from a satisfactory joint fit to the SMG number counts and
the magnification measurements reported in Table~\ref{tab:sourcesresults}.  The
Schechter function from the joint fit has a flatter slope (by 25\%), a brighter
characteristic flux density (by 15\%), and a lower normalization (by 30\%).  It is also
possible that a Schechter function provides an inadequate description of the
bright end shape of the number counts \citep[e.g., if blending is not properly
taken into account; see][]{Fu:2013lr, Ivison:2013fk, Hodge:2013qy,
Karim:2013lr}.   

Another important parameter for predicting magnification factors is a fixed
maximum magnification ($\mu_{\rm max}$), that is intended to reflect both the
intrinsic sizes of the background sources as well as the typical angular
separation between the centers of the sources and the centers of the lenses.
The green lines shown in Figure~\ref{fig:mu_f500} are taken from
\citet{Lapi:2012kx} and show the effect of this parameter on $\mu_{880}
(S_{500})$.  These curves help to highlight the apparent bimodality in
$\mu_{880}$ values in the SMA subsample.  One speculative explanation for this
is a bimodal distribution of sizes in SMGs (this is supported by the sizes from
the lens modeling, see Table~\ref{tab:sourcesresults} and
Figure~\ref{fig:mu_rhalf}) and hence a bimodal distribution in $\mu_{\rm max}$.
Larger samples and higher-spatial resolution are needed to confirm this
intriguing possibility, which could plausibly result from SMGs comprising a
population of rotating disks as well as major mergers.  Such a study is beyond
the scope of this paper and is therefore deferred to future publications.

\subsection{SED Fitting Methodology}

All but two sources in this sample have short baseline SMA data (i.e., $D < 50
\,$m) that provide a robust total flux density at 880$\, \mu$m.  The exceptions
are J132630.1$+$334410, which is resolved into two distinct images of the
background source, and J142823.9$+$352619, which appears unresolved in extended
array only data.  We therefore expect that the SMA total flux density measurements are
reliable for the entire sample.  In conjunction with the {\it Herschel}/SPIRE
photometry, these data can be used to measure the apparent (i.e., lensed)
bolometric luminosity as well as the shape of the far-IR SED.  In this section,
we describe the methodology we use to undertake this task.

For galaxies at redshifts of $1.5 < z < 4.5$, {\it Herschel}/SPIRE and
SMA~880$\, \mu$m photometry probe rest-frame 45$\, \mu$m to 350$\, \mu$m.  At
these wavelengths, the dominant contribution is thermal emission from dust
heated by star-formation or an active galactic nucleus (AGN).  We fit
single-temperature, optically-thin, modified blackbody curves to the data,
assuming an emissivity parameter of $\beta = 1.5$ \citep{1983QJRAS..24..267H}.
Studies based on {\it Infrared Astronomical Satellite} data have shown that
this simple model gives a useful measure of the heating conditions of the
interstellar medium in galaxies \citep{Desert:1990lr}.   For the highest
redshift sources ($z_{\rm source} > 3.5$), the 250$\mu$m channel of SPIRE
probes the Wien side of the modified blackbody curve.  The observed
250$\,\mu$m flux density is under-predicted by this model if there are alternative
powering sources that drive mid-IR luminosity (e.g., hot dust from AGN or
intense SF).  If such powering sources exist, the best-fit dust temperatures
will be artificially inflated to compensate for the stronger 250$\,\mu$m
emission.  However, for the four objects in the SMA subsample at $z > 3.5$), we
do not find evidence for significantly higher dust temperatures or poor fits to
the data (see Table~\ref{tab:intrinsic}, indicating that a simple modified
blackbody is a reasonable choice even for the high-redshift objects.  

Another consideration that is important at high redshift is the influence of
the cosmic microwave background (CMB) radiation.  The CMB acts as an additional
source of heating of the dust that shifts the SED to warmer temperatures and
boosts the observed flux densities.  However, for the dust temperatures and redshifts
of the SMA subsample, this effect is insignificant \citep{da-Cunha:2013lr} and
we therefore do not incorporate it into our model fitting process.

The modified blackbody curve used here has the following form for the flux
density, $S_{\nu}$, given a dust temperature, $T_{\rm dust}$:

\begin{equation}
S_\nu \propto \frac{\nu^{3 + \beta}}{ {\rm exp}(h \nu / k T_{\rm dust}) - 1 }
\end{equation}

An analysis of more complicated models that incorporate additional temperature
components \citep[e.g.,][]{2001MNRAS.327..697D, Planck-Collaboration:2011qy} or
allow the frequency at which the thermal emission becomes optically thick to
vary \citep[e.g.,][]{Hayward:2012lr} is beyond the scope of this paper.  This
is because we are chiefly concerned with the apparent far-IR luminosities ($\mu
L_{\rm FIR}$) of the sources, which are relatively insensitive to the
particular details of the chosen modified blackbody model.  The dust
temperature is also of interest here, but mainly for the purpose of comparison
to existing samples of SMGs.  Our choices here are well-matched to those that
have been made previously, thereby facilitating direct comparisons to the
literature \citep[e.g.,][]{Magnelli:2012lr}.  

For a given $\mu L_{\rm FIR}$ and $T_{\rm dust}$, we compute model flux
densities at the {\it Herschel}/SPIRE and SMA bands by multiplying the modified
blackbody curve with the appropriate filter function and integrating (for the
SMA, we assume a top-hat filter shape with 8~GHz of bandwidth centered on
$\nu_{\rm LO}$).  Calibration uncertainties are accounted for by adding 7\%
uncertainty in quadrature to each photometric measurement.  Confusion noise is
also included, though it is largely insignificant at the flux densities of the
lensed SMGs.  We use the {\sc emcee} software package to iterate over 
plausible ranges in $\mu L_{\rm FIR}$ ($10^{10} - 10^{15}~$L$_\sun$) and $T_{\rm
dust}$ ($20 - 100$~K) values for each lensed SMG (see
section~\ref{sec:lensmethod} for a description of the behavior of {\sc emcee}).
We use 100 walkers and 200 iterations in the ``burn-in'' stage to converge on
the best-fit model, keeping in mind the additional 7\% absolute flux density
calibration uncertainty in the {\it Herschel}/SPIRE and SMA photometry.  In the
final stage, we use 100 walkers and 10 iterations for a total of 1000 sets of
model parameters.  These provide the shape of the posterior probability density
functions for $\mu L_{\rm FIR}$ and $T_{\rm dust}$, which are then used to
compute the best-fit values and the 1$\sigma$ confidence intervals.  We use our
measurements of $\mu_{880}$ from section~\ref{sec:lensedresults} to
obtain the intrinsic, unlensed FIR luminosity, $L_{\rm FIR}$.  Finally, we
compute dust masses using the standard equation \citep{1983QJRAS..24..267H},

\begin{equation}
    M_{\rm dust} = \frac{1}{1+z_{\rm s}} \frac{S_{880} d_{\rm
    L}^2}{\kappa_d^{\rm rest} B(\nu^{\rm rest}, T_{\rm dust})},
\end{equation}

\noindent where $\kappa_{\rm d}^{\rm rest}$ is the mass absorption coefficient and
$B(\nu^{\rm rest}, T_{\rm dust}$) is the value of the blackbody function for
the given $T_{\rm dust}$ and computed at the rest-frame frequency $\nu^{\rm
rest}$.  We obtain $\kappa_{\rm d}^{\rm rest}$ by interpolation of the values in
\citet{2003ARA&A..41..241D} over the appropriate range in rest-frame
wavelength.  The uncertainty in $\kappa_{\rm d}^{\rm rest}$ is a factor of a
few and dominates the total error budget for our dust mass estimates.

We follow the procedure outlined here for all SMGs with {\it Herschel}/SPIRE
and submm photometry.  Besides the objects in this paper, this also includes
SMGs from \citet{Hezaveh:2013fk} \citep[with photometry and redshifts coming
from][]{Weis:2013qy, Bothwell:2013fk} and \citet{Magnelli:2012lr}.

\subsection{The SEDs of {\it Herschel}-selected Lensed SMGs}

The results of our SED fitting procedure are given for the SMA subsample in
Table~\ref{tab:intrinsic}, including the best-fit reduced chi-squared value
($\chi^2_{\rm min}$), the dust temperature ($T_{\rm dust}$), the dust mass
($M_{\rm dust}$; error bars do not include the factor of a few uncertainty in
the mass opacity coefficient), the FIR luminosity ($L_{\rm FIR}$), the
half-light radius ($r_{\rm half}$), and the FIR luminosity surface density
($\Sigma_{\rm FIR}$).  The error bars are pertinent only to our chosen model of
a single optically thin modified blackbody.  In this table, the magnification
factor and uncertainty inferred from the lens model have been used to compute
intrinsic values and their uncertainties.  For 13 objects, $\chi^2_{\rm min} >
2$, suggesting that the single-component modified blackbody model is an
over-simplification in nearly half of the SMA subsample.  

\begin{deluxetable*}{lcccccccc}
\tabletypesize{\scriptsize} 
\tablecolumns{8}
\tablewidth{0pt}
\tablecaption{Intrinsic (i.e., unlensed) properties of SMA Sample (assuming
$\beta=1.5$, optically thin modified blackbody, and $L_{\rm FIR}$ integrated over $40 -
120\,\mu$m.  Error bars are pertinent only to our chosen model and therefore
underestimate the true uncertainties.)}
\tablehead{
\colhead{} & 
\colhead{} & 
\colhead{$T_{\rm dust}$} &
\colhead{log$(M_{\rm dust})$} &
\colhead{log$(L_{\rm FIR})$} &
\colhead{$r_{\rm half}$} &
\colhead{log$(\Sigma_{\rm FIR})$}
\\
\colhead{IAU name} & 
\colhead{$\chi^2_{\rm min}$} &
\colhead{(K)} & 
\colhead{($M_\sun$)} & 
\colhead{($L_\sun$)} & 
\colhead{(kpc)} & 
\colhead{($L_\sun \; {\rm kpc}^{-2}$)}
}
\startdata
J021830.5$-$053124 &   7.11 & $36\pm1$ & $ 9.49\pm0.11$ & $12.79\pm0.09$  & $2.03\pm0.71 $  & $11.45\pm0.41$ \\
J022016.5$-$060143 &   4.46 & $37\pm1$ & $ 9.10\pm0.10$ & $12.79\pm0.05$  & $1.14\pm0.37 $  & $11.93\pm0.34$ \\
---                &   ---  &  ---     & $ 9.07\pm0.10$ & $12.76\pm0.05$  & $2.09\pm0.35 $  & $11.34\pm0.15$ \\
---                &   ---  &  ---     & $ 9.00\pm0.10$ & $12.67\pm0.05$  & $2.09\pm0.39 $  & $11.24\pm0.17$ \\
J083051.0$+$013224 &   0.86 & $44\pm1$ & $ 9.16\pm0.06$ & $13.09\pm0.05$  & $0.85\pm0.07 $  & $12.44\pm0.07$ \\
J084933.4$+$021443 &   9.62 & $36\pm1$ & $ 8.92\pm0.05$ & $12.72\pm0.04$  & $1.10\pm0.22 $  & $11.86\pm0.18$ \\
---                &   ---  &  ---     & $ 9.31\pm0.05$ & $13.11\pm0.04$  & $1.69\pm0.19 $  & $11.87\pm0.10$ \\
---                &   ---  &  ---     & $ 8.76\pm0.05$ & $12.56\pm0.04$  & $1.04\pm0.42 $  & $11.85\pm0.51$ \\
J085358.9$+$015537 &   0.98 & $36\pm1$ & $ 8.91\pm0.09$ & $12.37\pm0.09$  & $0.41\pm0.08 $  & $12.37\pm0.17$ \\
J090302.9$-$014127 &   1.13 & $38\pm1$ & $ 9.29\pm0.07$ & $12.92\pm0.05$  & $3.05\pm0.92 $  & $11.20\pm0.30$ \\
J090311.6$+$003906 &   0.87 & $34\pm1$ & $ 9.18\pm0.06$ & $12.45\pm0.04$  & $3.30\pm0.65 $  & $10.63\pm0.18$ \\
J090740.0$-$004200 &  10.30 & $43\pm2$ & $ 8.73\pm0.10$ & $12.58\pm0.11$  & $1.09\pm0.62 $  & $11.92\pm0.74$ \\
J091043.1$-$000321 &  25.66 & $41\pm1$ & $ 8.69\pm0.06$ & $12.50\pm0.06$  & $0.89\pm0.16 $  & $11.82\pm0.16$ \\
J091305.0$-$005343 &   0.04 & $35\pm1$ & $ 9.56\pm0.08$ & $12.94\pm0.07$  & $4.14\pm0.72 $  & $10.92\pm0.15$ \\
J105750.9$+$573026 &   2.64 & $47\pm1$ & $ 8.83\pm0.05$ & $12.94\pm0.03$  & $1.83\pm0.26 $  & $11.62\pm0.13$ \\
J114637.9$-$001132 &   1.86 & $41\pm1$ & $ 9.12\pm0.06$ & $12.90\pm0.04$  & $1.59\pm0.13 $  & $11.70\pm0.07$ \\
J125135.4$+$261457 &   2.24 & $39\pm1$ & $ 9.02\pm0.06$ & $12.68\pm0.05$  & $0.93\pm0.21 $  & $11.97\pm0.20$ \\
J125632.7$+$233625 &   0.41 & $40\pm1$ & $ 9.10\pm0.07$ & $12.80\pm0.06$  & $0.40\pm0.07 $  & $12.80\pm0.16$ \\
J132427.0$+$284452 &  47.59 & $37\pm1$ & $ 9.39\pm0.07$ & $12.92\pm0.07$  & $3.44\pm0.44 $  & $11.05\pm0.11$ \\
J132630.1$+$334410 &  13.33 & $36\pm1$ & $ 9.48\pm0.04$ & $13.00\pm0.04$  & $1.57\pm0.17 $  & $11.81\pm0.09$ \\
J133008.4$+$245900 &   1.14 & $43\pm1$ & $ 8.78\pm0.06$ & $12.66\pm0.05$  & $0.55\pm0.24 $  & $12.52\pm0.56$ \\
J133649.9$+$291801 &   7.62 & $39\pm1$ & $ 9.15\pm0.09$ & $12.87\pm0.08$  & $0.87\pm0.34 $  & $12.27\pm0.41$ \\
J134429.4$+$303036 &   6.90 & $38\pm1$ & $ 9.06\pm0.04$ & $12.73\pm0.05$  & $1.50\pm0.40 $  & $11.62\pm0.26$ \\
J141351.9$-$000026 &  23.28 & $38\pm1$ & $ 9.50\pm0.09$ & $13.18\pm0.08$  & $1.73\pm0.31 $  & $11.92\pm0.16$ \\      
J142413.9$+$022303 &   0.49 & $41\pm1$ & $ 9.36\pm0.05$ & $13.17\pm0.03$  & $3.79\pm0.38 $  & $11.22\pm0.09$ \\
J142823.9$+$352619 &   6.14 & $39\pm2$ & $ 9.19\pm0.20$ & $12.77\pm0.20$  & $0.71\pm0.43 $  & $12.52\pm0.84$ \\
J142825.5$+$345547 &   0.88 & $38\pm1$ & $ 8.85\pm0.10$ & $12.45\pm0.09$  & $0.89\pm0.18 $  & $11.77\pm0.19$ \\
J143330.8$+$345439 &   0.84 & $39\pm1$ & $ 9.34\pm0.06$ & $12.96\pm0.05$  & $1.59\pm0.34 $  & $11.78\pm0.19$ \\
\enddata
\label{tab:intrinsic}
\end{deluxetable*}

\begin{figure}[!tbp] 
\includegraphics[width=\linewidth]{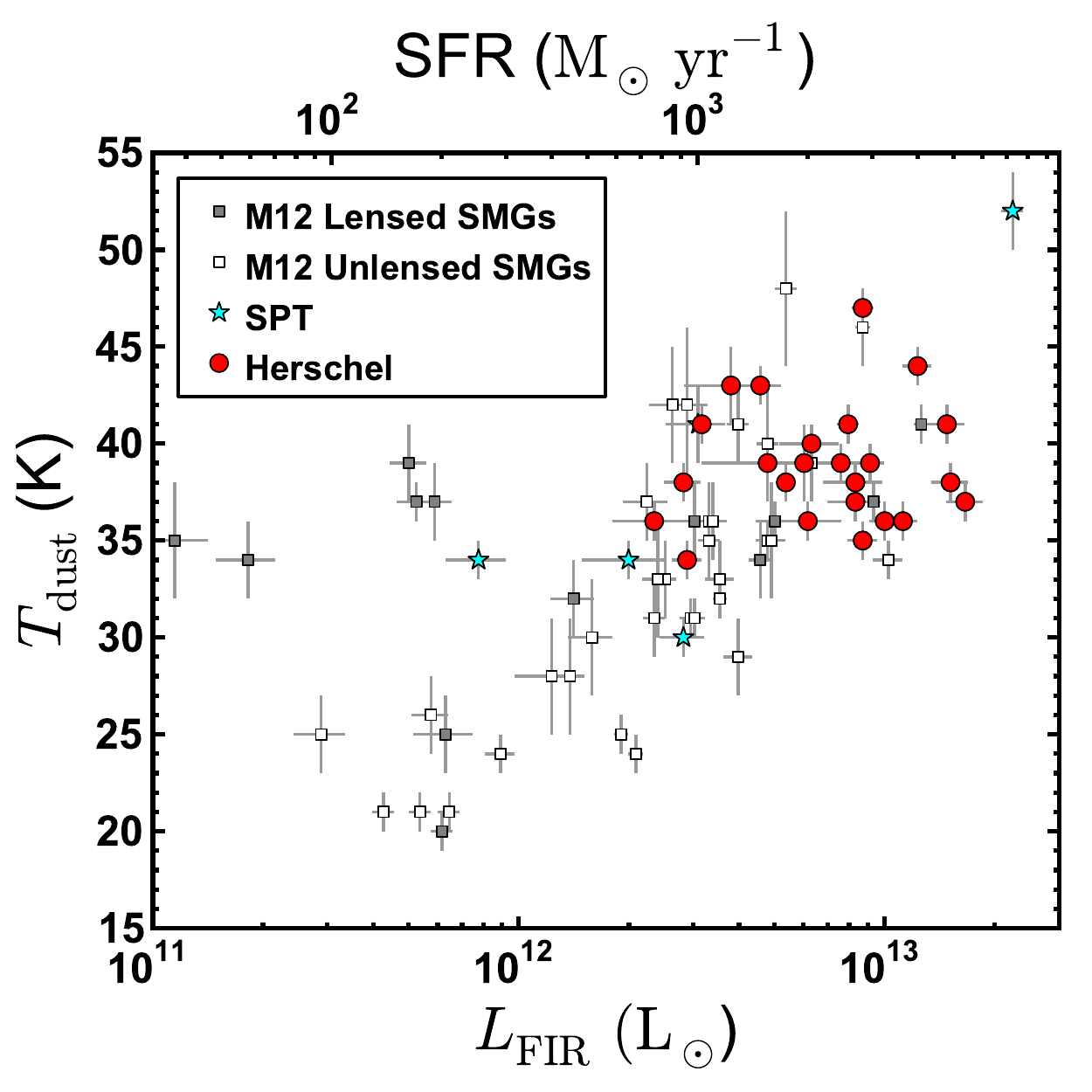}

\caption{ Dust temperature as a function of intrinsic (i.e., unlensed) FIR
luminosity for lensed SMGs discovered by {\it Herschel} (red circles) and SPT
(cyan stars), as well as unlensed and lensed SMGs from \citet{Magnelli:2012lr}
(grey squares).  Error bars include uncertainty in magnification factors.} \label{fig:lircomparison}

\end{figure}

The $T_{\rm dust}-L_{\rm FIR}$ diagram is useful for characterizing the shape
and amplitude of the rest-frame far-IR SED of dusty galaxies.
Figure~\ref{fig:lircomparison} shows these parameters for the objects in the
SMA subsample, a handful of SPT lensed SMGs \citep{Hezaveh:2013fk}, and a
sample of primarily unlensed
SMGs \citep{Magnelli:2012lr}.  Some of the objects in this diagram are known to
have multiple components in the source plane (e.g., J022016.5$-$060143).  In
these cases, we show the $L_{\rm FIR}$ value integrated over all components in
the source plane.  This is primarily done because our measurements of $T_{\rm
dust}$ are based on {\it Herschel} photometry in large part, which does not
resolve the individual components in the source plane.  

Objects in the SMA subsample populate the high$-L_{\rm FIR}$, high$-T_{\rm
dust}$ regime (median $L_{\rm FIR} = 7.9 \times 10^{12} \; $L$_\sun$, median
$T_{\rm dust} = 39 \;$K) compared to unlensed SMGs (median $L_{\rm FIR} = 2.8
\times 10^{12} \; $L$_\sun$, median $T_{\rm dust} = 32 \;$K).  The selection of
the brightest objects found in wide-field surveys is the dominant reason for
the high $L_{\rm FIR}$ values reported here.  However, thanks to lensing, we
probe a relatively wide range in $L_{\rm FIR}$: $2.7 - 17.0 \times 10^{12} \;
$L$_\sun$, despite selecting the brightest objects at 500$\, \mu$m.  The top
axis in Figure~\ref{fig:lircomparison} shows the SFRs that are inferred based
on the \citet{1998ARA&A..36..189K} prescription for converting $L_{\rm IR}$
\citep[computed from $L_{\rm FIR}$ assuming a bolometric correction factor of
1.91;][]{Dale:2001fj} to SFR assuming a Salpeter IMF.  Finally, the high dust
temperatures in the SMA subsample reflect the fact that a greater portion of
the IR luminosity is emitted at short wavelengths in the highest luminosity
sources.  This result is consistent with what has been found previously for
lensed SMGs \citep{Harris:2012fr} as well as for unlensed SMGs
\citep{Magnelli:2012lr} and is one of the first indications that {\it
Herschel}-selected lensed SMGs are not greatly dissimilar in their physical
properties from unlensed SMGs \citep[the same conclusion is reached
by][]{Harris:2012fr}.

\subsection{Size Scale of Star-formation in {\it Herschel}-selected Lensed SMGs}

One of the central themes in understanding the evolution of the most luminous
galaxies during the epoch of peak star-formation rate density in the Universe
\citep[i.e., $1 < z < 4$; e.g.,][]{Burgarella:2013lr} is the role played by
major mergers.  It has been clear for decades that the most luminous galaxies
at $z \sim 0$---commonly known as ultra-luminous infrared galaxies (ULIRGs) and
defined to have $L_{\rm IR} > 10^{12}\,$L$_\sun$--- are powered by major
mergers \citep[e.g.,][]{1987AJ.....94..831A, 1996MNRAS.279..477C,
1996AJ....111.1025M}.  However, it is also clear that such systems contribute
only trivially to the star-formation rate density in the Universe today because
they are so rare \citep{1986ApJ...303L..41S}.  Since ULIRGs contribute an
increasing fraction of the total star-formation rate density as a function of
redshift \citep[e.g.,][]{2005ApJ...632..169L, Magnelli:2011ul}, some
theoretical efforts have suggested that an increase in the merger rate in
conjunction with strong inflows of gas from the intergalactic medium at high
redshift could allow the merger paradigm to explain a large fraction of the
luminous galaxies at these epochs \citep[e.g.,][]{Hopkins:2010bh}.  Providing
support for this theoretical paradigm are sub-arcsecond observations of CO
emission in a handful of SMGs which show that a significant fraction of SMGs
that are spatially resolved have multiple, similar mass components---i.e., they
are major mergers \citep{2008ApJ...680..246T, 2010ApJ...724..233E,
2011MNRAS.412.1913I, 2011ApJ...733L..11R}.

On the other hand, some recent theoretical attempts to simulate the formation
of galaxies on cosmological scales (i.e., cubes that are $\approx 200\,$Mpc on
a side) have found evidence that favors a model of galaxy formation in which
smooth flows of gas from the IGM feed large, extended reservoirs of gas in disk
galaxies \citep[e.g.,][]{Keres:2009kx, 2010MNRAS.404.1355D}.  There is even
observational evidence based on dynamical models that disk-like systems exist
at high redshift \citep{Hodge:2012fk}.  Ultimately, detailed dynamical models
of statistically significant samples can resolve the dispute between the merger
and disk paradigms.  However, assembling the requisite datasets is extremely expensive
in terms of telescope time.  A far more feasible goal is spatially resolved
observations of the dust emission in SMGs at high redshift.

The dust emission in SMGs is critically important because it represents
reprocessed emission from young massive stars which provide a reliable measure
of the instantaneous SFR (within the past $\approx10\,$Myr).  This assumes no
significant contribution from a cold, diffuse ISM component (supported by our
measurements of $T_{\rm dust}$) and no significant contribution from an AGN
(our use of the FIR luminosity integrated over $40-120\, \mu$m rest-frame is
intended to mitigate this possibility).  Measuring the size-scale and the
luminosity of this dust can therefore in principle be used to contrast extended
galaxy morphologies expected from accretion-fuelled disks
\citep[e.g.,][]{Dekel:2009yq} and compact morphologies expected from
dissipational mergers of gas-rich disks \citep[e.g.,][]{1996ApJ...464..641M}.

\begin{figure*}[!tbp] 
\includegraphics[width=0.5\linewidth]{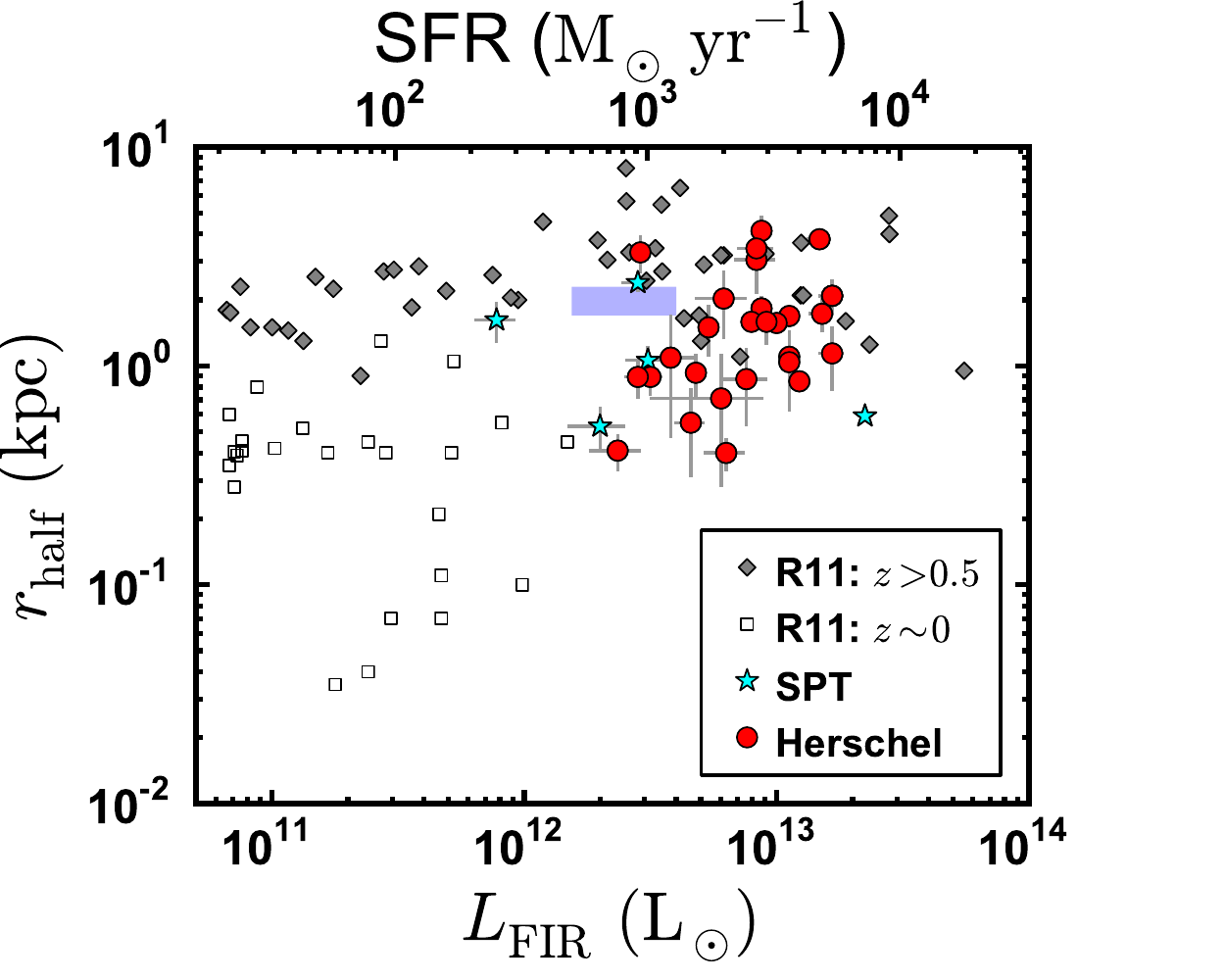}
\includegraphics[width=0.5\linewidth]{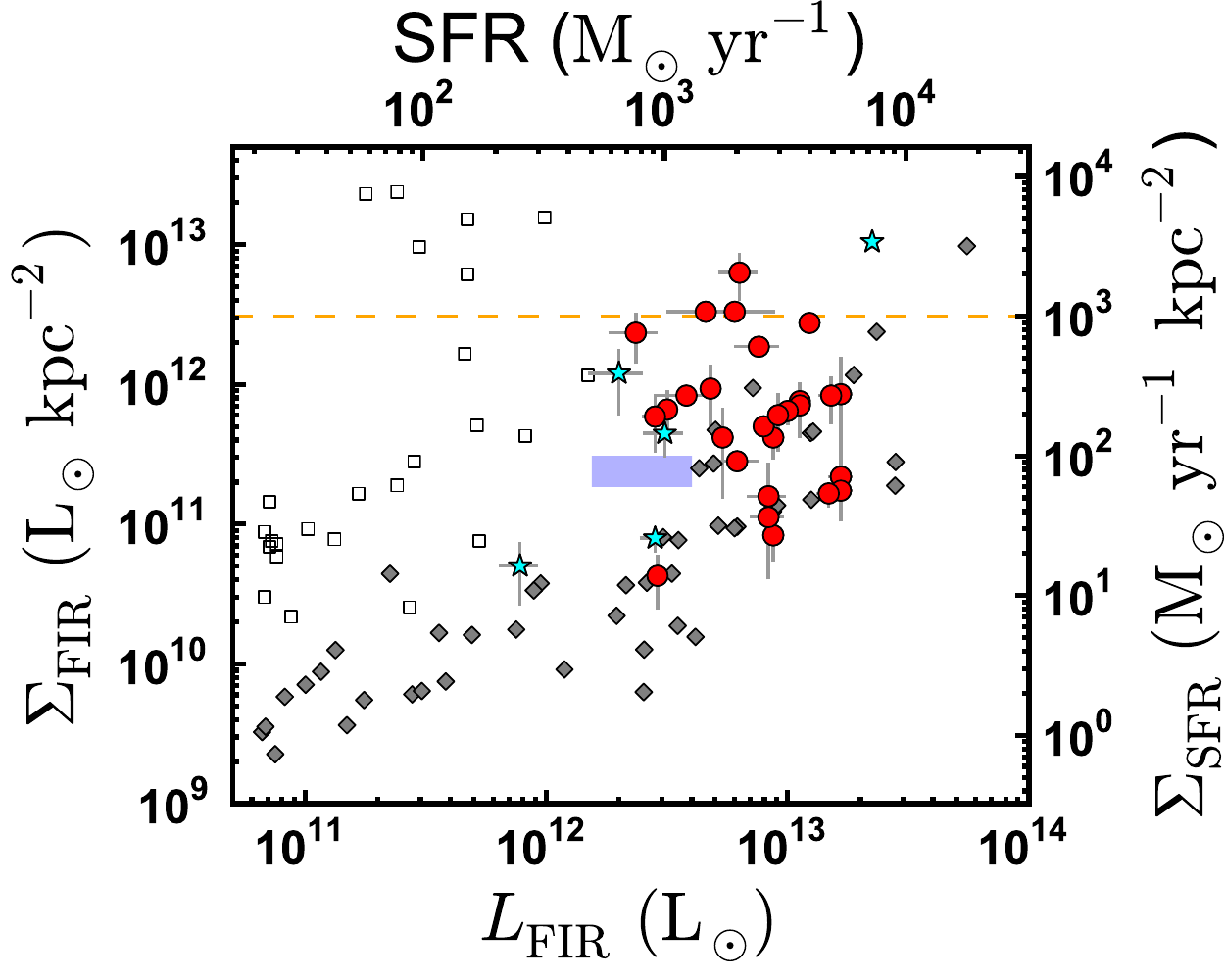}

\caption{ {\it Left}: Half-light radius as a function of FIR luminosity for
lensed SMGs discovered by {\it Herschel} (red circles) and SPT (cyan stars), as
well as galaxies from a compilation in \citet{Rujopakarn:2011dq} at $z > 0.5$
(filled grey diamonds) and at $z \sim 0$ (open squares).  The blue shaded
region represents the median and 1$\sigma$ range found for unlensed SMGs by
\citet{Tacconi:2006lr}.  {\it Right}: Far-IR luminosity surface density as a
function of FIR luminosity.  The orange dashed line traces the theoretical
limit of $\Sigma_{\rm SFR}$ for an optically thick disk
\citep{Thompson:2005vn}.  The SMA subsample spans nearly one decade in $L_{\rm
FIR}$ and two decades in $\Sigma_{\rm FIR}$.  A handful of sources approach or
exceed the highest values observed in local LIRGs and ULIRGs ($\Sigma_{\rm FIR}
= 10^{13} \, $L$_\sun \; {\rm kpc}^{-2}$).  \label{fig:lfirsigfir} }

\end{figure*}

The left panel of Figure~\ref{fig:lfirsigfir} shows the physical sizes
(circularized radii, $r_{\rm half}$) as a function of $L_{\rm FIR}$ for the SMA
subsample and the SPT sample \citep{Hezaveh:2013fk}.  In this figure, objects
with multiple components in the source plane are plotted individually, unlike
in Figure~\ref{fig:lircomparison}.  In doing this, we have assumed that the FIR
luminosity in each component is proportional to its intrinsic flux density at
880$\,\mu$m from the lens model (i.e., $\mu_{\rm FIR} \approx \mu_{880}$.  This
assumption is consistent with our modeling efforts because the only time we use
multiple components in the source plane is when it is clear that only moderate
lensing is occurring, in which case $\mu_{\rm FIR} \approx \mu_{880}$ is a
valid approximation.  Two comparison samples of unlensed LIRGs and ULIRGs from
\citet{Rujopakarn:2011dq} are shown: one at $0.5 < z < 2.5$ and one at $z \sim
0$.  The sample of intermediate and high redshift LIRGs and ULIRGs relies on
size measurements based primarily on radio observations \citep{Chapman:2004uq,
Muxlow:2005fj, Biggs:2008kx, Casey:2009yq}, but also includes a handful of with
mm size measurements \citep{Tacconi:2006lr, Tacconi:2010vn, Daddi:2010rt} and
the assumption that the radio and far-IR sizes are correlated.  Finally,
Figure~\ref{fig:lfirsigfir} also shows the median and 1$\sigma$ range in these
values for a handful of unlensed SMGs with high-spatial resolution imaging by
\citet[blue rectangle;][]{Tacconi:2006lr}.  

We find a wide range in sizes for the SMA subsample: the minimum, median, and
maximum $r_{\rm half}$ values are 0.41$\,$kpc, 1.53$\,$kpc, and 4.16$\,$kpc.
In comparison, \citet{Rujopakarn:2011dq} find values of 0.9$\,$kpc, 2.6$\,$kpc,
and 8.0$\,$kpc for the intermediate and high redshift sample.  Part of this
difference is a result of lensing, which lets us access spatial resolutions that are
otherwise inaccessible (the radio observations used to measure sizes in the
comparison sample have a typical angular resolution of $0\farcs15$,
corresponding to a physical scale of $\approx 1\,$kpc at the redshifts of
interest).  Local LIRGs and ULIRGs have smaller sizes than their higher
redshift counterparts \citep{Rujopakarn:2011dq}, but the sizes of objects in
the SMA subsample begin to overlap with those of the local LIRGs and ULIRGs.  Very strong
lensing (e.g., $\mu \gtrsim 30$) can reach $\approx 100\,$pc scale spatial
resolution \citep{Swinbank:2010lr}, but these occurrences must be rare because
no such object is found in the SMA subsample.

The right panel of Figure~\ref{fig:lfirsigfir} shows the FIR luminosity surface
density ($\Sigma_{\rm FIR}$) as a function of $L_{\rm FIR}$ for the same set of
objects as in the left panel.  A wide range in $\Sigma_{\rm FIR}$ is evident
for the SMA subsample: the minimum, median, and maximum $\Sigma_{\rm FIR}$ values
are 0.05, 0.6, and $6.0 \times 10^{12} \,$L$_\sun\,$kpc$^{-2}$, respectively.  A
horizontal line drawn at 1000$\,$M$_\sun\,$yr$^{-1}\,$kpc$^{-2}$ represents the
theoretical limit for star-formation rate surface density in a sustainable
radiation-pressure supported disk \citep{Thompson:2005vn}.  A handful of
objects in the SMA subsample and one object in the SPT sample lie near to or in
excess of this limit, possibly indicating that they are in a very short-lived
phase of evolution (e.g., the coalescence stage of a major merger).  There are
also, however, a number of objects with $\Sigma_{\rm FIR}$ values more than an order
of magnitude below the limit, suggesting that multiple modes of star-formation
are viable in high redshift SMGs.

\begin{figure}[!tbp] 
\includegraphics[width=\linewidth]{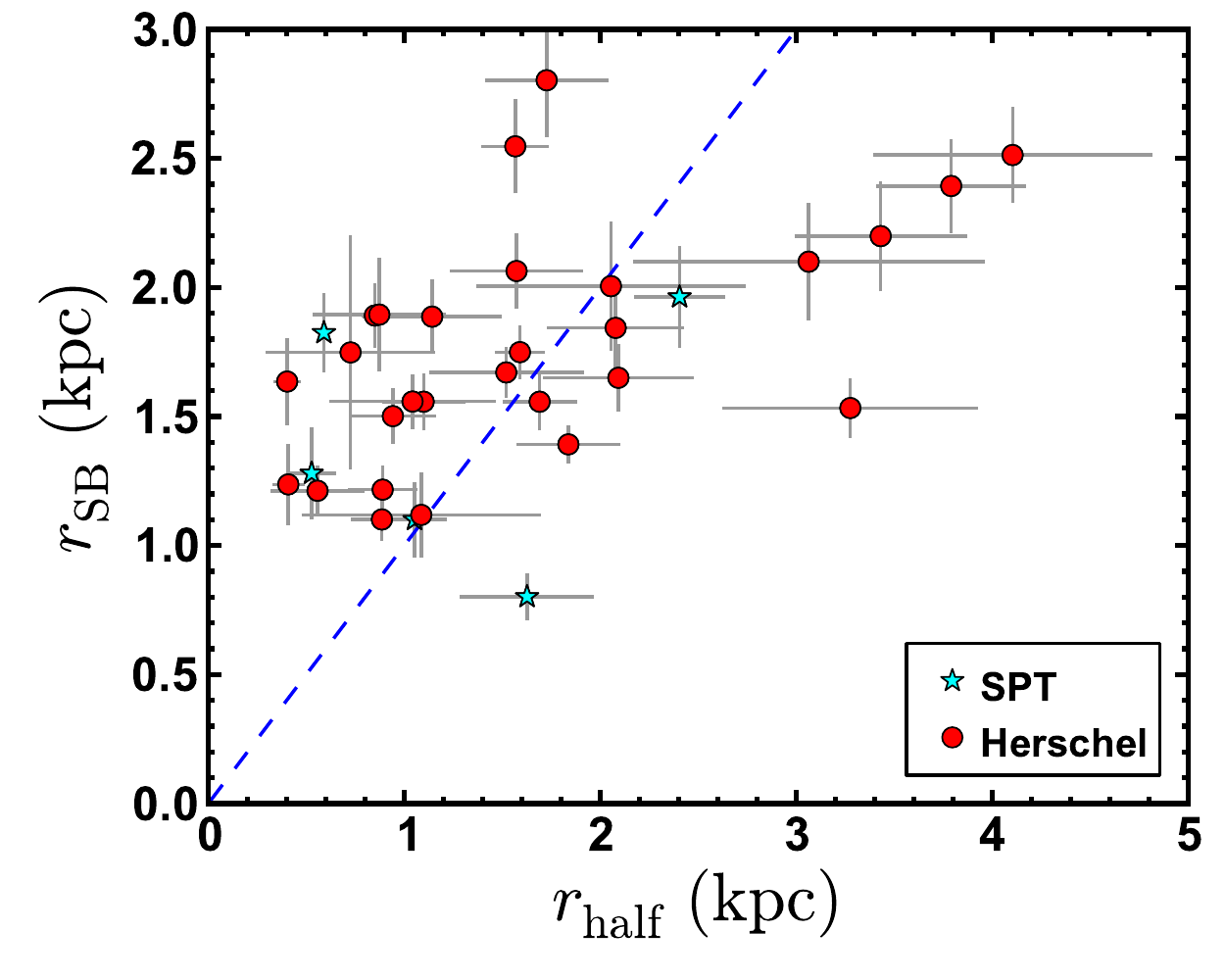}

\caption{ Comparison of size measurements ($r_{\rm SB}$) computed from
application of Stefan-Boltzmann law to intrinsic IR luminosities and dust
temperatures with those obtained from direct measurements from lens models
(circularized half-light radius, $r_{\rm half}$).  A dashed blue line traces
$r_{\rm SB} = r_{\rm half}$.  Approximately 75\% of the sample has $r_{\rm
half} \lesssim r_{\rm SB}$.  For these sources, the FIR emission is likely to
be optically thick. \label{fig:rsb_rhalf} }

\end{figure}

A spherically symmetric dust source radiating as a blackbody obeys the
Stefan-Boltzmann law relating emitted flux density and the temperature of the dust.  We
use this fact to infer an alternative measure of the size of the lensed SMG, which we
denote as $r_{\rm SB}$:

\begin{equation}
    r_{\rm SB} = \sqrt{\frac{L_{\rm IR}}{4 \pi \sigma_{\rm SB} T_{\rm dust}^4}}
\end{equation}

\noindent where $\sigma_{\rm SB}$ is the Stefan-Boltzmann constant.  This
quantity is similar in scope to the ``effective radius'' described in
\citet{Greve:2012uq} for lensed SMGs discovered by the SPT.
Figure~\ref{fig:rsb_rhalf} shows the results of this analysis for the SMA and
SPT samples.  In this diagram, the error bars reflect the formal values
obtained for our given set of model assumptions---e.g., the uncertainty
introduced by the assumption of a single-temperature modified blackbody is not
included.  A dashed line traces the 1:1 relation between these two measures of
the size (i.e., $r_{\rm SB} = r_{\rm half}$).  

A total of 15 out of 23 sources satisfy the unphysical criterion of $r_{\rm SB}
> r_{\rm half}$.  The best explanation for the large number of sources that
violate the blackbody limit is that our $T_{\rm dust}$ values are
underestimated by the assumption of optically thin emission.  In fact, $r_{\rm
SB} \approx r_{\rm half}$ suggests optically thick FIR emission.  Adopting an
optically thick model for the SED fitting would lead to larger dust
temperatures by $25-50$\%, an increase that is nearly sufficient for all of the
sources in our sample to satisfy $r_{\rm SB} < r_{\rm half}$.  The exact
geometry of the source is unknown and is therefore an additional complicating
factor.  Nevertheless, this crude line of analysis is one indication that the
FIR emission is optically thick in lensed SMGs discovered by {\it Herschel},
similar to local ULIRGs, which have $r_{\rm half} \approx 2-3 \times r_{\rm
SB}$ \citep{Solomon:1997qy}.  Only a handful of sources have large $r_{\rm
half}$ values and low $\Sigma_{\rm FIR}$ values typical of optically thin disks
far from the Eddington limit (e.g., J091305.0$-$005343).

\section{Conclusions} \label{sec:conclusions}

We present sub-arcsecond submm imaging from the SMA for 30 strong lens
candidates discovered by {\it Herschel} in the H-ATLAS and HerMES wide-field
surveys.  The candidates are selected to have $S_{500} > 100
\,$mJy and the SMA subsample contains nearly all targets with $S_{500} > 170 \,$mJy.
We also present optical spectroscopy from the MMT, Gemini-S, and WHT that
provide new redshift measurements for 8 of the putative lenses.
Nearly all candidates in the SMA subsample have existing spectroscopic redshifts
for the putative lensed SMGs from blind CO searches with the GBT
\citep{Harris:2012fr}, CSO \citep{Lupu:2012ly}, CARMA (Riechers et al., in
prep.), and PdBI \citep[][Krips et al., in prep.]{Cox:2011fk}.  

Out of the SMA subsample of 30, there are 16 that have distinct lens and source
redshifts and obvious lensed morphology in the submm (``grade A'' lenses).
Four objects have convincing morphological signatures of lensing, but only one
spectroscopic redshift measurement---we consider these to be highly likely to
be strongly lensed (``grade B'' lenses).  Another five objects have distinct
redshift measurements for lens and source, but the SMA imaging reveals only one
image of the background source, suggesting modest magnification factors:
$\mu_{880} < 2$ (``grade C'' lenses).  Finally, there are five
objects that lack distinct redshift measurements and do not show obvious
morphological signatures of lensing---additional data are needed to determine
whether or not these ``grade X'' systems are strongly lensed.  In
total, the strong lensing rate is $70-87\%$ ($83-100\%$ if moderately lensed
systems are included as well).

We use the SMA data to develop lens models in the visibility plane, as is
appropriate for interferometers like the SMA.  We derive lens models for the 25
objects with obvious signatures of lensing (either strong or moderate) in the
submm.  In conjunction with redshifts from optical and mm-wave
spectroscopy, the lens models provide measurements of the mass of the
lenses inside the Einstein radius, as well as the size and far-IR
luminosity of the lensed SMGs.  

We find that the lenses are at higher redshifts and have lower masses than
lenses found in surveys based on SDSS optical spectroscopy, in agreement with
expectations for a source-selected (rather than lens-selected) survey for
lenses.  The number of lenses that will be found from wide-field (sub-)mm
surveys \citep{Gonzalez-Nuevo:2012lr} promises to be comparable to that from
SDSS-based searches, but the former provide access to a population of lenses with
fundamentally different properties.  For this reason, lenses found by {\it
Herschel} and SPT are highly complementary to those found by SDSS and will
remain so for the forseeable future.

The lensed SMGs probe over a decade in sizes (median circularized
half-light radii of 1.6$\,$kpc) and intrinsic (i.e., unlensed) FIR luminosity
(median $L_{\rm FIR}$ of $7.9 \times 10^{12} \, $L$_\sun$).  Applying the
\citet{1998ARA&A..36..189K} prescription to convert $L_{\rm IR}$ to SFR, we use
the sizes and $L_{\rm FIR}$ values to infer a nearly two-decade range in SFR
surface density (median $\Sigma_{\rm SFR} = 200 \, $M$_\sun \, {\rm yr}^{-1}
\, {\rm kpc}^{-2}$).  A handful of lensed SMGs lie near or above the
theoretical limit of $\Sigma_{\rm SFR} = 1000 \, $M$_\sun \, {\rm yr}^{-1} \,
{\rm kpc}^{-2}$ for an optically thick disk, but there are also several objects
with $\Sigma_{\rm SFR}$ values over an order of magnitude below this limit,
implying that multiple modes of star-formation may be required to explain SMGs
at $z \gtrsim 1.5$.

The magnification factors we measure for the lensed SMGs are
significantly lower than predicted from models based on number counts of
unlensed SMGs.  This may be an indication that the bright end of the SMG
luminosity function or the intrinsic sizes of SMGs are currently poorly
understood.

Finally, it is worth emphasizing that the advent of ALMA makes the future in
this field looks very promising.  Many of the unsolved questions from this work
can be addressed in a direct manner by obtaining more sensitive submm
observations at higher spatial resolution.  For a given amount of integration
time, ALMA (when fully operational) will provide factors of $10-100$
improvement in these quantities compared to the SMA.

\begin{acknowledgments}

The results described in this paper are based on observations obtained with
{\it Herschel}, an ESA space observatory with science instruments provided by
European-led Principal Investigator consortia and with important participation
from NASA.  The {\it Herschel}-ATLAS is a project with {\it Herschel}.  The
H-ATLAS website is http://www.h-atlas.org/.  US participants in H-ATLAS
acknowledge support from NASA through a contract from JPL.  

This research has made use of data from the HerMES project
(http://hermes.sussex.ac.uk/). HerMES is a Herschel Key Programme utilizing
Guaranteed Time from the SPIRE instrument team, ESAC scientists, and a mission
scientist. HerMES is described in \citet{Oliver:2012lr}.  The HerMES data presented in
this paper will be released through the {\em Herschel} Database in Marseille
HeDaM (\footnote{http://hedam.oamp.fr/HerMES}).

SPIRE has been developed by a consortium of institutes led by Cardiff Univ.
(UK) and including: Univ. Lethbridge (Canada); NAOC (China); CEA, LAM (France);
IFSI, Univ. Padua (Italy); IAC (Spain); Stockholm Observatory (Sweden);
Imperial College London, RAL, UCL-MSSL, UKATC, Univ. Sussex (UK); and Caltech,
JPL, NHSC, Univ. Colorado (USA). This development has been supported by
national funding agencies: CSA (Canada); NAOC (China); CEA, CNES, CNRS
(France); ASI (Italy); MCINN (Spain); SNSB (Sweden); STFC, UKSA (UK); and NASA
(USA).

R. S. B.  acknowledges support from the SMA Fellowship program. H. F., A. C.,
and J. L.  W. acknowledge support from NSF CAREER AST-0645427.  A portion of
this work was completed at the Aspen Center for Physics during a 2013 summer
workshop on dusty galaxies at high redshift.  R. S. B. acknowledges the
hospitality of the Aspen Center for Physics, which is supported by the National
Science Foundation Grant No.  PHY-1066293.  S. J. O., L. W., and A. S.
acknowledge support from the Science and Technology Facilities Council (grant
number ST/I000976/1).  M. N. acknowledges financial support from PRIN INAF 2012
project ``Looking into the dust-oscured phase of galaxy formation through
cosmic zoom lenses in the Herschel Astrophysical Large Area Survey.''  AO and
RG acknowledge support from the Programme National Cosmologie et Galaxies
(PNCG).  J.G.N. acknowledges ﬁnancial support from Spanish CSIC for a JAE-DOC
fellowship and partial ﬁnancial support from the Spanish Ministerio de Ciencia
e Innovacion project AYA2010-21766-C03-01.  I. P.-F.,  P. M.-N., N. L. and A.
S.  acknowledge support from the Spanish grant AYA2010-21697-C05-04.  We thank
K. Rosenfeld for assistance in implementing the visibility-plane aspect of the
lens modeling software used in this paper.  We thank the anonymous referee for
a timely review that provided useful comments and helped improved the clarity
of the manuscript.

The SMA is a joint project between the Smithsonian Astrophysical Observatory
and the Academia Sinica Institute of Astronomy and Astrophysics and is funded
by the Smithsonian Institution and the Academia Sinica. The authors wish to
recognize and acknowledge the very significant cultural role and reverence that
the summit of Mauna Kea has always had within the indigenous Hawaiian
community.  We are most fortunate to have the opportunity to conduct
observations from this mountain.  

Observations reported here were obtained at the MMT Observatory, a joint
facility of the Smithsonian Institution and the University of Arizona.

Based on observations obtained at the Gemini Observatory, which is operated by
the Association of Universities for Research in Astronomy, Inc., under a
cooperative agreement with the NSF on behalf of the Gemini partnership: the
National Science Foundation (United States), the National Research Council
(Canada), CONICYT (Chile), the Australian Research Council (Australia),
Minist\'erio da Ci\^encia, Tecnologia e Inova\c{c}\~ao (Brazil) and Ministerio
de Ciencia, Tecnolog\'ia e Innovaci\'on Productiva (Argentina).  

The William Herschel Telescope is operated on the island of La Palma by the
Isaac Newton Group in the Spanish Observatorio del Roque de los Muchachos of
the Instituto de Astrof\'isica de Canarias

Facilities: SMA, MMT, Gemini-S, WHT, VLT.

\end{acknowledgments}



\begin{thebibliography}{126}
\expandafter\ifx\csname natexlab\endcsname\relax\def\natexlab#1{#1}\fi

\bibitem[{{Armus} {et~al.}(1987){Armus}, {Heckman}, \&
  {Miley}}]{1987AJ.....94..831A}
{Armus}, L., {Heckman}, T., \& {Miley}, G. 1987, \aj, 94, 831

\bibitem[{{Barger} {et~al.}(1999){Barger}, {Cowie}, {Smail}, {Ivison}, {Blain},
  \& {Kneib}}]{Barger:1999rt}
{Barger}, A.~J., {Cowie}, L.~L., {Smail}, I., {Ivison}, R.~J., {Blain}, A.~W.,
  \& {Kneib}, J.-P. 1999, \aj, 117, 2656

\bibitem[{{Bartelmann} \& {Loeb}(1998)}]{Bartelmann:1998qy}
{Bartelmann}, M. \& {Loeb}, A. 1998, \apj, 503, 48

\bibitem[{{Becker} {et~al.}(1995){Becker}, {White}, \&
  {Helfand}}]{Becker:1995fj}
{Becker}, R.~H., {White}, R.~L., \& {Helfand}, D.~J. 1995, \apj, 450, 559

\bibitem[{{Benn} {et~al.}(2008){Benn}, {Dee}, \& {Ag{\'o}cs}}]{Benn:2008lr}
{Benn}, C., {Dee}, K., \& {Ag{\'o}cs}, T. 2008, in Society of Photo-Optical
  Instrumentation Engineers (SPIE) Conference Series, Vol. 7014, Society of
  Photo-Optical Instrumentation Engineers (SPIE) Conference Series

\bibitem[{{Biggs} \& {Ivison}(2008)}]{Biggs:2008kx}
{Biggs}, A.~D. \& {Ivison}, R.~J. 2008, \mnras, 385, 893

\bibitem[{{Blain}(1996)}]{1996MNRAS.283.1340B}
{Blain}, A.~W. 1996, \mnras, 283, 1340

\bibitem[{{Blain} \& {Longair}(1993)}]{Blain:1993lr}
{Blain}, A.~W. \& {Longair}, M.~S. 1993, \mnras, 264, 509

\bibitem[{{Blain} {et~al.}(1999){Blain}, {Moller}, \& {Maller}}]{Blain:1999fr}
{Blain}, A.~W., {Moller}, O., \& {Maller}, A.~H. 1999, \mnras, 303, 423

\bibitem[{{Bolton} {et~al.}(2008){Bolton}, {Burles}, {Koopmans}, {Treu},
  {Gavazzi}, {Moustakas}, {Wayth}, \& {Schlegel}}]{Bolton:2008wd}
{Bolton}, A.~S., {Burles}, S., {Koopmans}, L.~V.~E., {Treu}, T., {Gavazzi}, R.,
  {Moustakas}, L.~A., {Wayth}, R., \& {Schlegel}, D.~J. 2008, \apj, 682, 964

\bibitem[{{Borys} {et~al.}(2006)}]{Borys:2006lr}
{Borys}, C. {et~al.} 2006, \apj, 636, 134

\bibitem[{{Bothwell} {et~al.}(2013)}]{Bothwell:2013fk}
{Bothwell}, M.~S. {et~al.} 2013, ArXiv e-prints

\bibitem[{{Brewer} {et~al.}(2012)}]{Brewer:2012lr}
{Brewer}, B.~J. {et~al.} 2012, \mnras, 422, 3574

\bibitem[{{Browne} {et~al.}(2003)}]{Browne:2003lr}
{Browne}, I.~W.~A. {et~al.} 2003, \mnras, 341, 13

\bibitem[{{Brownstein} {et~al.}(2012)}]{Brownstein:2012rt}
{Brownstein}, J.~R. {et~al.} 2012, \apj, 744, 41

\bibitem[{{Bruzual} \& {Charlot}(2003)}]{2003MNRAS.344.1000B}
{Bruzual}, G. \& {Charlot}, S. 2003, \mnras, 344, 1000

\bibitem[{{Burgarella} {et~al.}(2013)}]{Burgarella:2013lr}
{Burgarella}, D. {et~al.} 2013, \aap, 554, A70

\bibitem[{{Bussmann} {et~al.}(2012)}]{Bussmann:2012lr}
{Bussmann}, R.~S. {et~al.} 2012, \apj, 756, 134

\bibitem[{{Carlstrom} {et~al.}(2011)}]{Carlstrom:2011qy}
{Carlstrom}, J.~E. {et~al.} 2011, \pasp, 123, 568

\bibitem[{{Casey} {et~al.}(2009)}]{Casey:2009yq}
{Casey}, C.~M. {et~al.} 2009, \mnras, 399, 121

\bibitem[{{Chapman} {et~al.}(2004){Chapman}, {Smail}, {Windhorst}, {Muxlow}, \&
  {Ivison}}]{Chapman:2004uq}
{Chapman}, S.~C., {Smail}, I., {Windhorst}, R., {Muxlow}, T., \& {Ivison},
  R.~J. 2004, \apj, 611, 732

\bibitem[{{Clements} {et~al.}(1996){Clements}, {Sutherland}, {McMahon}, \&
  {Saunders}}]{1996MNRAS.279..477C}
{Clements}, D.~L., {Sutherland}, W.~J., {McMahon}, R.~G., \& {Saunders}, W.
  1996, \mnras, 279, 477

\bibitem[{{Clements} {et~al.}(2010)}]{Clements:2010fk}
{Clements}, D.~L. {et~al.} 2010, \aap, 518, L8

\bibitem[{{Condon} {et~al.}(1998){Condon}, {Cotton}, {Greisen}, {Yin},
  {Perley}, {Taylor}, \& {Broderick}}]{Condon:1998uq}
{Condon}, J.~J., {Cotton}, W.~D., {Greisen}, E.~W., {Yin}, Q.~F., {Perley},
  R.~A., {Taylor}, G.~B., \& {Broderick}, J.~J. 1998, \aj, 115, 1693

\bibitem[{{Conley} {et~al.}(2011)}]{Conley:2011lr}
{Conley}, A. {et~al.} 2011, \apjl, 732, L35

\bibitem[{{Coppin} {et~al.}(2006)}]{Coppin:2006lr}
{Coppin}, K. E.~K. {et~al.} 2006, \mnras, 372, 1621

\bibitem[{{Cox} {et~al.}(2011)}]{Cox:2011fk}
{Cox}, P. {et~al.} 2011, \apj, 740, 63

\bibitem[{{da Cunha} {et~al.}(2013)}]{da-Cunha:2013lr}
{da Cunha}, E. {et~al.} 2013, \apj, 766, 13

\bibitem[{{Daddi} {et~al.}(2010)}]{Daddi:2010rt}
{Daddi}, E. {et~al.} 2010, \apj, 713, 686

\bibitem[{{Dale} {et~al.}(2001){Dale}, {Helou}, {Contursi}, {Silbermann}, \&
  {Kolhatkar}}]{Dale:2001fj}
{Dale}, D.~A., {Helou}, G., {Contursi}, A., {Silbermann}, N.~A., \&
  {Kolhatkar}, S. 2001, \apj, 549, 215

\bibitem[{{Dav{\'e}} {et~al.}(2010){Dav{\'e}}, {Finlator}, {Oppenheimer},
  {Fardal}, {Katz}, {Kere{\v s}}, \& {Weinberg}}]{2010MNRAS.404.1355D}
{Dav{\'e}}, R., {Finlator}, K., {Oppenheimer}, B.~D., {Fardal}, M., {Katz}, N.,
  {Kere{\v s}}, D., \& {Weinberg}, D.~H. 2010, \mnras, 404, 1355

\bibitem[{{de Zotti} {et~al.}(2005){de Zotti}, {Ricci}, {Mesa}, {Silva},
  {Mazzotta}, {Toffolatti}, \& {Gonz{\'a}lez-Nuevo}}]{2005A&A...431..893D}
{de Zotti}, G., {Ricci}, R., {Mesa}, D., {Silva}, L., {Mazzotta}, P.,
  {Toffolatti}, L., \& {Gonz{\'a}lez-Nuevo}, J. 2005, \aap, 431, 893

\bibitem[{{Dekel} {et~al.}(2009){Dekel}, {Sari}, \& {Ceverino}}]{Dekel:2009yq}
{Dekel}, A., {Sari}, R., \& {Ceverino}, D. 2009, \apj, 703, 785

\bibitem[{{Desert} {et~al.}(1990){Desert}, {Boulanger}, \&
  {Puget}}]{Desert:1990lr}
{Desert}, F.-X., {Boulanger}, F., \& {Puget}, J.~L. 1990, \aap, 237, 215

\bibitem[{{Diolaiti} {et~al.}(2000){Diolaiti}, {Bendinelli}, {Bonaccini},
  {Close}, {Currie}, \& {Parmeggiani}}]{Diolaiti:2000qy}
{Diolaiti}, E., {Bendinelli}, O., {Bonaccini}, D., {Close}, L.~M., {Currie},
  D.~G., \& {Parmeggiani}, G. 2000, in Society of Photo-Optical Instrumentation
  Engineers (SPIE) Conference Series, Vol. 4007, Society of Photo-Optical
  Instrumentation Engineers (SPIE) Conference Series, ed. P.~L. {Wizinowich},
  879--888

\bibitem[{{Draine}(2003)}]{2003ARA&A..41..241D}
{Draine}, B.~T. 2003, \araa, 41, 241

\bibitem[{{Dunne} \& {Eales}(2001)}]{2001MNRAS.327..697D}
{Dunne}, L. \& {Eales}, S.~A. 2001, \mnras, 327, 697

\bibitem[{{Eales} {et~al.}(2010)}]{2010PASP..122..499E}
{Eales}, S. {et~al.} 2010, \pasp, 122, 499

\bibitem[{{Eisenstein} {et~al.}(2011){Eisenstein}, {Weinberg}, {Agol},
  {Aihara}, {Allende Prieto}, {Anderson}, {Arns}, {Aubourg}, {Bailey},
  {Balbinot}, \& et~al.}]{Eisenstein:2011fr}
{Eisenstein}, D.~J., {Weinberg}, D.~H., {Agol}, E., {Aihara}, H., {Allende
  Prieto}, C., {Anderson}, S.~F., {Arns}, J.~A., {Aubourg}, {\'E}., {Bailey},
  S., {Balbinot}, E., \& et~al. 2011, \aj, 142, 72

\bibitem[{{Engel} {et~al.}(2010){Engel}, {Tacconi}, {Davies}, {Neri}, {Smail},
  {Chapman}, {Genzel}, {Cox}, {Greve}, {Ivison}, {Blain}, {Bertoldi}, \&
  {Omont}}]{2010ApJ...724..233E}
{Engel}, H., {Tacconi}, L.~J., {Davies}, R.~I., {Neri}, R., {Smail}, I.,
  {Chapman}, S.~C., {Genzel}, R., {Cox}, P., {Greve}, T.~R., {Ivison}, R.~J.,
  {Blain}, A., {Bertoldi}, F., \& {Omont}, A. 2010, \apj, 724, 233

\bibitem[{{Foreman-Mackey} {et~al.}(2013){Foreman-Mackey}, {Hogg}, {Lang}, \&
  {Goodman}}]{Foreman-Mackey:2013yq}
{Foreman-Mackey}, D., {Hogg}, D.~W., {Lang}, D., \& {Goodman}, J. 2013, \pasp,
  125, 306

\bibitem[{{Frayer} {et~al.}(2011)}]{2011ApJ...726L..22F}
{Frayer}, D.~T. {et~al.} 2011, \apjl, 726, L22+

\bibitem[{{Fu} {et~al.}(2012)}]{Fu:2012uq}
{Fu}, H. {et~al.} 2012, \apj, 753, 134

\bibitem[{{Fu} {et~al.}(2013)}]{Fu:2013lr}
---. 2013, \nat, 498, 338

\bibitem[{{Gavazzi} {et~al.}(2011)}]{Gavazzi:2011lr}
{Gavazzi}, R. {et~al.} 2011, \apj, 738, 125

\bibitem[{{George} {et~al.}(2013)}]{George:2013qy}
{George}, R.~D. {et~al.} 2013, ArXiv e-prints

\bibitem[{{Gladders} \& {Yee}(2005)}]{Gladders:2005qy}
{Gladders}, M.~D. \& {Yee}, H.~K.~C. 2005, \apjs, 157, 1

\bibitem[{{Goldoni} {et~al.}(2006){Goldoni}, {Royer}, {Fran{\c c}ois},
  {Horrobin}, {Blanc}, {Vernet}, {Modigliani}, \& {Larsen}}]{Goldoni:2006lr}
{Goldoni}, P., {Royer}, F., {Fran{\c c}ois}, P., {Horrobin}, M., {Blanc}, G.,
  {Vernet}, J., {Modigliani}, A., \& {Larsen}, J. 2006, in Society of
  Photo-Optical Instrumentation Engineers (SPIE) Conference Series, Vol. 6269,
  Society of Photo-Optical Instrumentation Engineers (SPIE) Conference Series

\bibitem[{{Gonz{\'a}lez-Nuevo} {et~al.}(2012)}]{Gonzalez-Nuevo:2012lr}
{Gonz{\'a}lez-Nuevo}, J. {et~al.} 2012, \apj, 749, 65

\bibitem[{{Goodman} \& {Weare}(2010)}]{goodmanweare}
{Goodman}, J. \& {Weare}, J. 2010, Communications in Applied Mathematics and
  Computational Science, 5, 65

\bibitem[{{Greve} {et~al.}(2012)}]{Greve:2012uq}
{Greve}, T.~R. {et~al.} 2012, \apj, 756, 101

\bibitem[{{Griffin} {et~al.}(2010)}]{2010A&A...518L...3G}
{Griffin}, M.~J. {et~al.} 2010, \aap, 518, L3+

\bibitem[{{Harris} {et~al.}(2012)}]{Harris:2012fr}
{Harris}, A.~I. {et~al.} 2012, \apj, 752, 152

\bibitem[{{Hayward} {et~al.}(2012){Hayward}, {Jonsson}, {Kere{\v s}},
  {Magnelli}, {Hernquist}, \& {Cox}}]{Hayward:2012lr}
{Hayward}, C.~C., {Jonsson}, P., {Kere{\v s}}, D., {Magnelli}, B., {Hernquist},
  L., \& {Cox}, T.~J. 2012, \mnras, 424, 951

\bibitem[{{Hezaveh} \& {Holder}(2011)}]{Hezaveh:2011kx}
{Hezaveh}, Y.~D. \& {Holder}, G.~P. 2011, \apj, 734, 52

\bibitem[{{Hezaveh} {et~al.}(2012){Hezaveh}, {Marrone}, \&
  {Holder}}]{Hezaveh:2012fk}
{Hezaveh}, Y.~D., {Marrone}, D.~P., \& {Holder}, G.~P. 2012, \apj, 761, 20

\bibitem[{{Hezaveh} {et~al.}(2013)}]{Hezaveh:2013fk}
{Hezaveh}, Y.~D. {et~al.} 2013, \apj, 767, 132

\bibitem[{{Hildebrand}(1983)}]{1983QJRAS..24..267H}
{Hildebrand}, R.~H. 1983, \qjras, 24, 267

\bibitem[{{Ho} {et~al.}(2004){Ho}, {Moran}, \& {Lo}}]{Ho:2004lr}
{Ho}, P.~T.~P., {Moran}, J.~M., \& {Lo}, K.~Y. 2004, \apjl, 616, L1

\bibitem[{{Hodge} {et~al.}(2012){Hodge}, {Carilli}, {Walter}, {de Blok},
  {Riechers}, {Daddi}, \& {Lentati}}]{Hodge:2012fk}
{Hodge}, J.~A., {Carilli}, C.~L., {Walter}, F., {de Blok}, W.~J.~G.,
  {Riechers}, D., {Daddi}, E., \& {Lentati}, L. 2012, \apj, 760, 11

\bibitem[{{Hodge} {et~al.}(2013)}]{Hodge:2013qy}
{Hodge}, J.~A. {et~al.} 2013, \apj, 768, 91

\bibitem[{{Hook} {et~al.}(2004){Hook}, {J{\o}rgensen}, {Allington-Smith},
  {Davies}, {Metcalfe}, {Murowinski}, \& {Crampton}}]{Hook:2004qy}
{Hook}, I.~M., {J{\o}rgensen}, I., {Allington-Smith}, J.~R., {Davies}, R.~L.,
  {Metcalfe}, N., {Murowinski}, R.~G., \& {Crampton}, D. 2004, \pasp, 116, 425

\bibitem[{{Hopkins} {et~al.}(2010){Hopkins}, {Younger}, {Hayward}, {Narayanan},
  \& {Hernquist}}]{Hopkins:2010bh}
{Hopkins}, P.~F., {Younger}, J.~D., {Hayward}, C.~C., {Narayanan}, D., \&
  {Hernquist}, L. 2010, \mnras, 402, 1693

\bibitem[{{Ikarashi} {et~al.}(2011)}]{Ikarashi:2011qy}
{Ikarashi}, S. {et~al.} 2011, \mnras, 415, 3081

\bibitem[{{Inada} {et~al.}(2012)}]{Inada:2012lr}
{Inada}, N. {et~al.} 2012, \aj, 143, 119

\bibitem[{{Ivison} {et~al.}(2011){Ivison}, {Papadopoulos}, {Smail}, {Greve},
  {Thomson}, {Xilouris}, \& {Chapman}}]{2011MNRAS.412.1913I}
{Ivison}, R.~J., {Papadopoulos}, P.~P., {Smail}, I., {Greve}, T.~R., {Thomson},
  A.~P., {Xilouris}, E.~M., \& {Chapman}, S.~C. 2011, \mnras, 412, 1913

\bibitem[{{Ivison} {et~al.}(2013)}]{Ivison:2013fk}
{Ivison}, R.~J. {et~al.} 2013, \apj, 772, 137

\bibitem[{{Karim} {et~al.}(2013)}]{Karim:2013lr}
{Karim}, A. {et~al.} 2013, \mnras, 432, 2

\bibitem[{{Keeton}(2001)}]{2001astro.ph..2340K}
{Keeton}, C.~R. 2001, ArXiv Astrophysics e-prints

\bibitem[{{Keeton} {et~al.}(1997){Keeton}, {Kochanek}, \&
  {Seljak}}]{Keeton:1997ys}
{Keeton}, C.~R., {Kochanek}, C.~S., \& {Seljak}, U. 1997, \apj, 482, 604

\bibitem[{{Kennicutt}(1998)}]{1998ARA&A..36..189K}
{Kennicutt}, Jr., R.~C. 1998, \araa, 36, 189

\bibitem[{{Kere{\v s}} {et~al.}(2009){Kere{\v s}}, {Katz}, {Fardal},
  {Dav{\'e}}, \& {Weinberg}}]{Keres:2009kx}
{Kere{\v s}}, D., {Katz}, N., {Fardal}, M., {Dav{\'e}}, R., \& {Weinberg},
  D.~H. 2009, \mnras, 395, 160

\bibitem[{{King} \& {Browne}(1996)}]{King:1996fk}
{King}, L.~J. \& {Browne}, I.~W.~A. 1996, \mnras, 282, 67

\bibitem[{{Kreysa} {et~al.}(1998)}]{Kreysa:1998uq}
{Kreysa}, E. {et~al.} 1998, in Society of Photo-Optical Instrumentation
  Engineers (SPIE) Conference Series, Vol. 3357, Society of Photo-Optical
  Instrumentation Engineers (SPIE) Conference Series, ed. T.~G. {Phillips},
  319--325

\bibitem[{{Lapi} {et~al.}(2012){Lapi}, {Negrello}, {Gonz{\'a}lez-Nuevo}, {Cai},
  {De Zotti}, \& {Danese}}]{Lapi:2012kx}
{Lapi}, A., {Negrello}, M., {Gonz{\'a}lez-Nuevo}, J., {Cai}, Z.-Y., {De Zotti},
  G., \& {Danese}, L. 2012, \apj, 755, 46

\bibitem[{{Le Floc'h} {et~al.}(2005)}]{2005ApJ...632..169L}
{Le Floc'h}, E. {et~al.} 2005, \apj, 632, 169

\bibitem[{{Lupu} {et~al.}(2012)}]{Lupu:2012ly}
{Lupu}, R.~E. {et~al.} 2012, \apj, 757, 135

\bibitem[{{Magnelli} {et~al.}(2011){Magnelli}, {Elbaz}, {Chary}, {Dickinson},
  {Le Borgne}, {Frayer}, \& {Willmer}}]{Magnelli:2011ul}
{Magnelli}, B., {Elbaz}, D., {Chary}, R.~R., {Dickinson}, M., {Le Borgne}, D.,
  {Frayer}, D.~T., \& {Willmer}, C.~N.~A. 2011, \aap, 528, A35+

\bibitem[{{Magnelli} {et~al.}(2012)}]{Magnelli:2012lr}
{Magnelli}, B. {et~al.} 2012, \aap, 539, A155

\bibitem[{{Marsden} {et~al.}(2013)}]{Marsden:2013lr}
{Marsden}, D. {et~al.} 2013, ArXiv e-prints, 1306.2288

\bibitem[{{Mihos} \& {Hernquist}(1996)}]{1996ApJ...464..641M}
{Mihos}, J.~C. \& {Hernquist}, L. 1996, \apj, 464, 641

\bibitem[{{Mocanu} {et~al.}(2013)}]{Mocanu:2013fk}
{Mocanu}, L.~M. {et~al.} 2013, ArXiv e-prints

\bibitem[{{Mu{\~n}oz} {et~al.}(1998){Mu{\~n}oz}, {Falco}, {Kochanek},
  {Leh{\'a}r}, {McLeod}, {Impey}, {Rix}, \& {Peng}}]{Munoz:1998mz}
{Mu{\~n}oz}, J.~A., {Falco}, E.~E., {Kochanek}, C.~S., {Leh{\'a}r}, J.,
  {McLeod}, B.~A., {Impey}, C.~D., {Rix}, H.-W., \& {Peng}, C.~Y. 1998, \apss,
  263, 51

\bibitem[{{Murphy} {et~al.}(1996){Murphy}, {Armus}, {Matthews}, {Soifer},
  {Mazzarella}, {Shupe}, {Strauss}, \& {Neugebauer}}]{1996AJ....111.1025M}
{Murphy}, Jr., T.~W., {Armus}, L., {Matthews}, K., {Soifer}, B.~T.,
  {Mazzarella}, J.~M., {Shupe}, D.~L., {Strauss}, M.~A., \& {Neugebauer}, G.
  1996, \aj, 111, 1025

\bibitem[{{Muxlow} {et~al.}(2005)}]{Muxlow:2005fj}
{Muxlow}, T.~W.~B. {et~al.} 2005, \mnras, 358, 1159

\bibitem[{{Myers} {et~al.}(2003)}]{Myers:2003lr}
{Myers}, S.~T. {et~al.} 2003, \mnras, 341, 1

\bibitem[{{Navarro} {et~al.}(1997){Navarro}, {Frenk}, \&
  {White}}]{Navarro:1997ys}
{Navarro}, J.~F., {Frenk}, C.~S., \& {White}, S.~D.~M. 1997, \apj, 490, 493

\bibitem[{{Negrello} {et~al.}(2007){Negrello}, {Perrotta},
  {Gonz{\'a}lez-Nuevo}, {Silva}, {de Zotti}, {Granato}, {Baccigalupi}, \&
  {Danese}}]{2007MNRAS.377.1557N}
{Negrello}, M., {Perrotta}, F., {Gonz{\'a}lez-Nuevo}, J., {Silva}, L., {de
  Zotti}, G., {Granato}, G.~L., {Baccigalupi}, C., \& {Danese}, L. 2007,
  \mnras, 377, 1557

\bibitem[{{Negrello} {et~al.}(2010)}]{Negrello:2010fk}
{Negrello}, M. {et~al.} 2010, Science, 330, 800

\bibitem[{{Nguyen} {et~al.}(2010)}]{Nguyen:2010fk}
{Nguyen}, H.~T. {et~al.} 2010, \aap, 518, L5

\bibitem[{{Oliver} {et~al.}(2010)}]{2010A&A...518L..21O}
{Oliver}, S.~J. {et~al.} 2010, \aap, 518, L21+

\bibitem[{{Oliver} {et~al.}(2012)}]{Oliver:2012lr}
---. 2012, \mnras, 424, 1614

\bibitem[{{Omont} {et~al.}(2011)}]{Omont:2011fk}
{Omont}, A. {et~al.} 2011, \aap, 530, L3

\bibitem[{{Paciga} {et~al.}(2009){Paciga}, {Scott}, \&
  {Chapin}}]{Paciga:2009uq}
{Paciga}, G., {Scott}, D., \& {Chapin}, E.~L. 2009, \mnras, 395, 1153

\bibitem[{{Pascale} {et~al.}(2011)}]{2011MNRAS.415..911P}
{Pascale}, E. {et~al.} 2011, \mnras, 415, 911

\bibitem[{{Perrotta} {et~al.}(2002){Perrotta}, {Baccigalupi}, {Bartelmann}, {De
  Zotti}, \& {Granato}}]{2002MNRAS.329..445P}
{Perrotta}, F., {Baccigalupi}, C., {Bartelmann}, M., {De Zotti}, G., \&
  {Granato}, G.~L. 2002, \mnras, 329, 445

\bibitem[{{Pilbratt} {et~al.}(2010)}]{Pilbratt:2010fk}
{Pilbratt}, G.~L. {et~al.} 2010, \aap, 518, L1

\bibitem[{{Planck Collaboration} {et~al.}(2011)}]{Planck-Collaboration:2011qy}
{Planck Collaboration} {et~al.} 2011, \aap, 536, A16

\bibitem[{{Riechers} {et~al.}(2011{\natexlab{a}})}]{Riechers:2011uq}
{Riechers}, D.~A. {et~al.} 2011{\natexlab{a}}, \apjl, 733, L12

\bibitem[{{Riechers} {et~al.}(2011{\natexlab{b}})}]{2011ApJ...733L..11R}
---. 2011{\natexlab{b}}, \apjl, 733, L11+

\bibitem[{{Rigby} {et~al.}(2011)}]{2011MNRAS.415.2336R}
{Rigby}, E.~E. {et~al.} 2011, \mnras, 415, 2336

\bibitem[{{Roseboom} {et~al.}(2010)}]{Roseboom:2010lk}
{Roseboom}, I.~G. {et~al.} 2010, \mnras, 409, 48

\bibitem[{{Rujopakarn} {et~al.}(2011){Rujopakarn}, {Rieke}, {Eisenstein}, \&
  {Juneau}}]{Rujopakarn:2011dq}
{Rujopakarn}, W., {Rieke}, G.~H., {Eisenstein}, D.~J., \& {Juneau}, S. 2011,
  \apj, 726, 93

\bibitem[{{Sault} {et~al.}(1995){Sault}, {Teuben}, \&
  {Wright}}]{1995ASPC...77..433S}
{Sault}, R.~J., {Teuben}, P.~J., \& {Wright}, M.~C.~H. 1995, in Astronomical
  Society of the Pacific Conference Series, Vol.~77, Astronomical Data Analysis
  Software and Systems IV, ed. R.~A. {Shaw}, H.~E. {Payne}, \& J.~J.~E.
  {Hayes}, 433--+

\bibitem[{{Schmidt} {et~al.}(1989){Schmidt}, {Weymann}, \&
  {Foltz}}]{Schmidt:1989fk}
{Schmidt}, G.~D., {Weymann}, R.~J., \& {Foltz}, C.~B. 1989, \pasp, 101, 713

\bibitem[{{Schneider} {et~al.}(1992){Schneider}, {Ehlers}, \&
  {Falco}}]{Schneider:1992fk}
{Schneider}, P., {Ehlers}, J., \& {Falco}, E.~E. 1992, {Gravitational Lenses}
  (Springer)

\bibitem[{{Schneider} {et~al.}(2006){Schneider}, {Kochanek}, \&
  {Wambsganss}}]{Schneider:2006ab}
{Schneider}, P., {Kochanek}, C.~S., \& {Wambsganss}, J. 2006, Gravitational
  Lensing: Strong, Weak and Micro: Saas-Fee Advanced Courses 33 (Berlin:
  Springer-Verlag)

\bibitem[{{Scott} {et~al.}(2011)}]{2011ApJ...733...29S}
{Scott}, K.~S. {et~al.} 2011, \apj, 733, 29

\bibitem[{{Serjeant}(2012)}]{Serjeant:2012lr}
{Serjeant}, S. 2012, \mnras, 424, 2429

\bibitem[{{Serjeant} \& {Harrison}(2005)}]{2005MNRAS.356..192S}
{Serjeant}, S. \& {Harrison}, D. 2005, \mnras, 356, 192

\bibitem[{{Sersic}(1968)}]{1968adga.book.....S}
{Sersic}, J.~L. 1968, {Atlas de galaxias australes} (Cordoba, Argentina:
  Observatorio Astronomico, 1968)

\bibitem[{{Soifer} {et~al.}(1986){Soifer}, {Sanders}, {Neugebauer},
  {Danielson}, {Lonsdale}, {Madore}, \& {Persson}}]{1986ApJ...303L..41S}
{Soifer}, B.~T., {Sanders}, D.~B., {Neugebauer}, G., {Danielson}, G.~E.,
  {Lonsdale}, C.~J., {Madore}, B.~F., \& {Persson}, S.~E. 1986, \apjl, 303, L41

\bibitem[{{Solomon} {et~al.}(1997){Solomon}, {Downes}, {Radford}, \&
  {Barrett}}]{Solomon:1997qy}
{Solomon}, P.~M., {Downes}, D., {Radford}, S.~J.~E., \& {Barrett}, J.~W. 1997,
  \apj, 478, 144

\bibitem[{{Sonnenfeld} {et~al.}(2013){Sonnenfeld}, {Gavazzi}, {Suyu}, {Treu},
  \& {Marshall}}]{Sonnenfeld:2013fj}
{Sonnenfeld}, A., {Gavazzi}, R., {Suyu}, S.~H., {Treu}, T., \& {Marshall},
  P.~J. 2013, ArXiv e-prints

\bibitem[{{Swetz} {et~al.}(2011)}]{Swetz:2011qy}
{Swetz}, D.~S. {et~al.} 2011, \apjs, 194, 41

\bibitem[{{Swinbank} {et~al.}(2010)}]{Swinbank:2010lr}
{Swinbank}, A.~M. {et~al.} 2010, \nat, 464, 733

\bibitem[{{Tacconi} {et~al.}(2006)}]{Tacconi:2006lr}
{Tacconi}, L.~J. {et~al.} 2006, \apj, 640, 228

\bibitem[{{Tacconi} {et~al.}(2008)}]{2008ApJ...680..246T}
---. 2008, \apj, 680, 246

\bibitem[{{Tacconi} {et~al.}(2010)}]{Tacconi:2010vn}
---. 2010, \nat, 463, 781

\bibitem[{{Thompson} {et~al.}(2005){Thompson}, {Quataert}, \&
  {Murray}}]{Thompson:2005vn}
{Thompson}, T.~A., {Quataert}, E., \& {Murray}, N. 2005, \apj, 630, 167

\bibitem[{{Treu}(2010)}]{Treu:2010fk}
{Treu}, T. 2010, \araa, 48, 87

\bibitem[{{Vernet} {et~al.}(2011)}]{Vernet:2011lr}
{Vernet}, J. {et~al.} 2011, \aap, 536, A105

\bibitem[{{Vieira} {et~al.}(2010)}]{Vieira:2010rr}
{Vieira}, J.~D. {et~al.} 2010, \apj, 719, 763

\bibitem[{{Vieira} {et~al.}(2013)}]{Vieira:2013fk}
---. 2013, \nat, 495, 344

\bibitem[{{Wardlow} {et~al.}(2013)}]{Wardlow:2013lr}
{Wardlow}, J.~L. {et~al.} 2013, \apj, 762, 59

\bibitem[{{Wei{\ss}} {et~al.}(2013)}]{Weis:2013qy}
{Wei{\ss}}, A. {et~al.} 2013, \apj, 767, 88

\end{thebibliography}

\end{document}